\documentclass[a4paper,11pt]{article}
\pdfoutput=1

\usepackage{jheppub}
\usepackage{latexsym}
\usepackage{amsfonts}
\usepackage{float,slashed,indentfirst}
\usepackage{amsmath,amssymb,xcolor}

\def\ifmath#1{\relax\ifmmode #1\else $#1$\fi}
\def\ls#1{\ifmath{_{\lower1.5pt\hbox{$\scriptstyle #1$}}}}
\def\lss#1{\ifmath{^{\,\lower2.5pt\hbox{$\scriptstyle #1$}}}}

\def\beq{\begin{equation}}
\def\eeq{\end{equation}}
\def\bea{\begin{eqnarray}}
\def\eea{\end{eqnarray}}

\def\lsim{\mathrel{\raise.3ex\hbox{$<$\kern-.75em\lower1ex\hbox{$\sim$}}}}
\def\gsim{\mathrel{\raise.3ex\hbox{$>$\kern-.75em\lower1ex\hbox{$\sim$}}}}

\def\gev{~{\rm GeV}}

\def\dg{\dagger}
\def\lam{\lambda}

\def\eq#1{Eq.~(\ref{#1})}

\def\tth  {t_{\theta}}
\def\sth  {s_{\theta}}
\def\cth  {c_{\theta}}

\def\tb  {t_{\beta}}
\def\sb  {s_{\beta}}
\def\stwob  {s_{2\beta}}
\def\cb  {c_{\beta}}

\def\cosba {\cos(\beta-\alpha)}
\def\sinba {\sin(\beta-\alpha)}

\def\hm{{\hat m}}

\def\typei{Type~I}
\def\typeii{Type~II}

\usepackage{ulem,fancyvrb}

\title{A new insight into the phase transition in the early Universe with two Higgs doublets}
\author[c]{J\'er\'emy Bernon}
\author[a,b]{\!\!, Ligong Bian}
\author[d]{and Yun Jiang}

\affiliation[a]{Department of Physics, Chongqing University, Chongqing 401331, China}
\affiliation[b]{Department of Physics, Chung-Ang University, Seoul 06974, Korea}
 \emailAdd{lgbycl@cqu.edu.cn}
\affiliation[c]{Institute for Advanced Studies, The Hong Kong University of Science and Technology, Clear
Water Bay, Kowloon, Hong Kong S.A.R, China }
\emailAdd{iasbernon@ust.hk}
\affiliation[d]{NBIA and Discovery Center, Niels Bohr Institute, University of Copenhagen,
Blegdamsvej 17, DK-2100, Copenhagen, Denmark}
\emailAdd{yunjiang@nbi.ku.dk}

\date{\today}

\abstract{
\\[-3mm]
\indent

We study the electroweak phase transition in the alignment limit of the CP-conserving two-Higgs-doublet model  (2HDM) of \typei~and \typeii.
The effective potential is evaluated at one-loop, where the thermal potential includes Daisy corrections and is reliably approximated by means of a sum of Bessel functions.
Both 1-stage and 2-stage electroweak phase transitions are shown to be possible, depending on the pattern of the vacuum development as the Universe cools down.  For the 1-stage case focused on in this paper, we analyze the properties of phase transition and discover that the field value of the electroweak symmetry breaking vacuum at the critical temperature at which the first order phase transition occurs is largely correlated with the vacuum depth of the 1-loop potential at zero temperature.

We demonstrate that a strong first order electroweak phase transition (SFOEWPT) in the 2HDM is achievable and establish benchmark scenarios leading to different testable signatures at colliders. 
In addition, we verify that an enhanced triple Higgs coupling (including loop corrections) is a typical feature of the SFOPT driven by the additional doublet. As a result, SFOEWPT might be able to be probed at the LHC and future lepton colliders through Higgs pair production.

}

\keywords{Electroweak phase transition, Beyond the Standard Model, Multi-scalar sector}
\arxivnumber{arXiv:1712.08430}

\begin{document}

\maketitle 
\setcounter{page}{2}
 \newpage

\section{Introduction}
\label{sec:intro}

After the discovery of the 125 GeV Higgs boson~\cite{Aad:2012tfa,Chatrchyan:2012xdj} and the accumulation of LHC data, no evidence of the new physics has been observed yet. Therefore, it is time to inquire whether the Standard Model (SM) of particle physics is actually complete to describe the physics at the electroweak scale. In the meantime, the origin of the baryon asymmetry of the Universe (BAU) 
is still one of the important open puzzles in particle physics and cosmology. To explain the BAU, the three Sakharov conditions~\cite{Sakharov:1967dj} must be fulfilled. 
The electroweak baryogenesis (EWBG)~\cite{Kuzmin:1985mm} is a possible mean to account for the generation of an asymmetry (imbalance) between baryons and antibaryons produced in the very early Universe. The success of EWBG requires two crucial ingredients: CP violation and strong first order phase transition (SFOPT), neither of which however can be addressed in the SM framework.
First, the SM fails to produce a sufficiently large baryon number due to a shortage of CP violation in the Cabibbo-Kobayashi-Maskawa (CKM) matrix.
The other shortcoming of the SM is the absence of departure from thermal equilibrium which could have been realized by a SFOPT: for the observed value of the SM-like Higgs mass this is not accomplished. The phase transition in the early Universe from the symmetric phase to the electroweak symmetry breaking (EWSB) phase actually belongs to a smooth crossover type~\cite{DOnofrio:2014rug}. It has been derived using lattice computation that the phase transition in the SM can only be strong first order when the Higgs mass is around 70-80 GeV~\cite{Kajantie:1996mn,Kajantie:1996qd,Csikor:1998eu,Aoki:1999fi}.
Therefore, a successful EWBG invokes new physics at the electroweak scale~\cite{Morrissey:2012db}. Theories that go beyond the SM typically have an extended Higgs sector, which may contain the ingredients for a SFOPT as well as new CP-violating interactions as needed for EWBG, usually also producing new signatures at colliders.

The two-Higgs-doublet model (2HDM) is the simplest renormalizable framework to realize EWBG.~\footnote{Though the electroweak phase transition has been extensively studied in the singlet extended model, the BAU generation cannot be addressed without extra CP-violation sources.~\cite{Jiang:2015cwa}} 
In this model the scalar potential is extended by an additional $SU(2)_L$  doublet, where a charged Higgs together with two additional neutral scalars are introduced. Through their portal interactions with the SM-like Higgs, the finite temperature potential can develop a potential barrier during the Universe cooling down, leading to strengthen the phase transition at the critical temperature.
On the other hand, the CP violation can exist either explicitly in these portal couplings or spontaneously via a relative phase between the vacuum expectation values (VEVs) of two doublets~\cite{Lee:1974jb}. Interestingly, the CP violation (beyond the SM) can be detected indirectly at high precision electric dipole moments (EDMs) experiments. With the recent improvement of the EDMs measurements, the CP violation phases that are needed for the baryon asymmetry are severely constrained~\cite{Inoue:2014nva}. In order to evade the bound  one may expect a cancellation arising from the different contributions of EDMs predictions, see Ref.~\cite{Bian:2017jpt,Bian:2017wfv,Bian:2016zba,Bian:2014zka,Chupp:2017rkp}.

The electroweak phase transition (EWPT) in the 2HDM context has been extensively studied for both the CP conserving case~\cite{Dorsch:2013wja,Dorsch:2014qja,Basler:2016obg} and with the source of CP-violation~\cite{Cline:1996mga,Fromme:2006cm,Cline:2011mm,Dorsch:2016nrg,Haarr:2016qzq,Basler:2017uxn}. 
While the CP phase at zero temperature is supposed to play an insignificant  in the EWPT process~\cite{Fromme:2006cm,Dorsch:2013wja,Dorsch:2014qja}, the CP-violating phase at finite temperature is found to be important in a recent study~\cite{Basler:2017uxn} where the analysis was performed after taking into account the LHC run-2 constraints.
In general, none of the scalar states of the 2HDM resembles a SM-Higgs boson that was observed at the LHC. 
However, such a SM-like Higgs boson $h$ can arise in the alignment limit, a particularly interesting limit of this model where only one Higgs doublet acquires the total electroweak vev, namely the couplings of $H$ to gauge boson pairs vanish while $h$ possesses SM-like couplings~\cite{Bernon:2015qea,Bernon:2015wef,Grzadkowski:2016szj}.
In terms of the model parameters, this limit corresponds to $\sin(\beta-\alpha)$ (always positive in our convention) to be close to 1.
Driven by the LHC Higgs data, in this paper we focus on the alignment limit (here $\sin(\beta-\alpha)\geq 0.99$) of the CP-conserving 2HDMs of \typei~and \typeii~models.
We consider the lightest CP-even state $h$ to be the 125 GeV SM-like Higgs observed at the LHC. 
To proceed the numerical analysis we take the points passing all existing experimental bounds (by the time of paper publication) generated from extensive scans in Ref.~\cite{Bernon:2015qea} and additionally employ the 1-loop improved theoretical constraints, the updated measurements coming from flavor physics and the recent LHC run-2 bounds searching for heavy resonances.
Our aim is to identity the parameter space of the 2HDM that can lead to a SFOPT and investigate the implications of a SFOPT required by baryogenesis on the LHC Higgs phenomenology.

It is inspiring to note that the cosmological EWPT can leave signatures of gravitational waves (GW) after the nucleation of the true vacuum bubbles, with typical red-shifted spectrum frequency around $\mathcal{O}(10^{-4}-10^{-2})~\text{Hz}$ \cite{Kamionkowski:1993fg}, which are detectable in the Evolved Laser Interferometer Space Antenna (eLISA)~\cite{Caprini:2015zlo}, DECi-hertz Interferometer Gravitational wave Observatory (DECIGO), UltimateDECIGO and Big Bang Observer (BBO)~\cite{Kudoh:2005as}. 
However, these two effects might be quite incompatible due to an opposite preference occurring in the bubble wall velocity.
The baryon asymmetry generation process within EWBG demands a relatively low bubble wall velocity in order to have enough time for the chiral asymmetry generation process to take place, this will later be transformed to the baryon asymmetry by the sphelaron process~\cite{Morrissey:2012db}. Of course, when performing the computation of the BAU in the EWBG mechanism, one should keep in mind that in addition to being subject to large theoretical uncertainties, the detailed calculations of the baryon asymmetry rely on the wall velocity of the bubble generated during the EWPT, see Ref.~\cite{Lee:2004we, Riotto:1998zb,Moore:2000wx,Konstandin:2004gy,Li:2008ez,Cline:1997vk}.
On the contrary, a testable GW signal requires a higher strength of the FOPT and a larger wall velocity. 
Very intriguingly, the recent development~\cite{Dorsch:2016nrg} shows that it is possible, although difficult in the 2HDM, to simultaneously accomplish the EWBG and produce the detectable GW signals generated during the EWPT especially through acoustic waves~\cite{Hindmarsh:2013xza,Hindmarsh:2015qta}.

This paper is organized as follows. In Sec.~\ref{sec:model} we first briefly review the CP-conserving 2HDMs of \typei~and \typeii~and discuss the status in view of the existing experimental bounds.
Next, we describe in Sec.~\ref{sec:effV} the details of the finite temperature potential and provide a fast numerical handle for the thermal potential.
In Sec.~\ref{sec:ptc}, the one-stage and two-stage PT are demonstrated and classified.
Subsequently, we present in Sec.~\ref{sec:Tceval} a useful computational scheme used to single out the one-stage PT and, more importantly, to evaluate the critical temperature $T_c$ for the one-stage PT.
Having studied the theoretical issues of the model and built the computational tools, 
we then proceed with the numerical analysis and investigate the properties of the phase transition which are presented in Sec.~\ref{sec:res}.
In particular, the relations between $T_c$ and extra Higgs masses as well as the influence of the effective potential at zero temperature on the field value of the electroweak symmetry breaking vacuum are analyzed.
In Sec.~\ref{sec:1stcol} benchmark scenarios leading to the SFOEWPT are established and their implications for future measurements at colliders are also discussed. Finally, Sec.~\ref{sec:Con} contains our conclusions and outlook for future studies. 
In Appendix~\ref{sec:thermass}, explicit formulas for the thermal mass corrections of the SM gauge bosons are given.

%-------------------------------------------------------------------------------
\section{The two-Higgs-doublet model } 
\label{sec:model}
%-------------------------------------------------------------------------------

Let us start with a brief review of the tree-level 2HDM at zero temperature. 
The general 2HDM is obtained by doubling the scalar sector of the SM, two doublets with identical quantum numbers are present. In general, CP violation may be present in the scalar sector and the Yukawa sector contains generic tree-level flavor changing neutral currents (FCNCs) mediated by the neutral scalar states. Here we consider a CP conserving Higgs sector and the absence of tree-level FCNCs. The first condition is obtained by imposing a reality condition on the parameters of the potential, and the second requirement is achieved by imposing a \typei~or \typeii~structure on the Yukawa sector, this is achieved by imposing a softly-broken $\mathbb{Z}_{2}$ symmetry~\cite{Glashow:1976nt,Paschos:1976ay}.

Denoting by $\Phi_1,\Phi_2$ the two Higgs doublets, the tree-level potential of this model is expressed as,
\beq
\begin{split}
V_{0}(\Phi_1, \Phi_2)&=m_{11}^2 \Phi_1^{\dg} \Phi_1+m_{22}^2 \Phi_2^{\dg} \Phi_2-\left[m_{12}^2 \Phi_1^{\dg} \Phi_2 + h.c.\right] +\frac{\lam_1}{2}(\Phi_1^{\dg} \Phi_1)^2+\frac{\lam_2}{2}(\Phi_2^{\dg} \Phi_2)^2
 \\
&\quad +\lam_3 (\Phi_1^{\dg} \Phi_1)(\Phi_2^{\dg} \Phi_2) +\lam_4 |\Phi_1^{\dg} \Phi_2|^2 +\left[ \frac{\lam_5}{2}(\Phi_1^{\dg} \Phi_2)^2 + h.c.\right] \ .
\label{potsym}
\end{split}
\eeq
In this basis, the $\mathbb{Z}_{2}$ symmetry under which $\Phi_2 \to - \Phi_2$ is manifest in the quartic terms, while it is softly broken by the introduction of the $m_{12}^2$ term.  
In general, $m_{12}^{2}$ and $\lambda_{5}$ are complex. We consider in this work a CP conserving Higgs sector and set all $\lam_i$ and $m^2_{12}$ as real parameters, see \cite{Basler:2017uxn} for a CP violating study. In this basis, both Higgs doublets have a non-zero vacuum expectation value (vev). We parametrize the degrees of freedom contained in the Higgs doublets as,
\beq
\Phi_i =
	\begin{pmatrix}
		\phi_i^+ \\ (v_i + \rho_i + i \eta_i)/\sqrt2
	\end{pmatrix}, \quad i=1,2 \ .
\eeq
where $v_{i}$ are the vevs of the two Higgs doublets. At zero temperature one has the relation $v^2_1+v^2_2=v_{\rm 0T}^2 \simeq (246 \gev)^2$. For convenience, we use the shorthand notation $v \equiv v_{\rm 0T}$ from now on and define $v_1 =v \cos\beta$ and $v_2=v \sin \beta$, $\tan\beta$ is therefore the ratio of the two vevs at $T=0$. 

The mass parameters $m^{2}_{11}$ and $m^{2}_{22}$ in the potential \eq{potsym} are determined by the potential minimization conditions,
\beq
\begin{split}
&\quad m_{11}^2 = m_{12}^2 \tb - \frac{1}{2} v^2 \left( \lambda_1 \cb^2 + \lambda_{345}\sb^2 \right)\,,\\
& \quad m_{22}^2 =  m_{12}^2 / \tb - \frac{1}{2} v^2 \left( \lambda_2 \sb^2 + \lambda_{345}\cb^2 \right)\,,
\end{split}
\label{min_cond}
\eeq
here the shorthand notations $\sb\equiv \sin\beta$, $\cb \equiv \cos\beta$ and $\tb\equiv \tan\beta$ and $\lambda_{345} \equiv \lambda_3+\lambda_4+\lambda_5$ are employed.

Though $\tan\beta$ is a physical parameter here, it is still possible to redefine the two doublets and go to a  basis where the full vev resides entirely in one of the two doublets: the so-called Higgs basis $(H_1,H_2)$~\cite{Davidson:2005cw}. This change of basis is generally possible as observed by the invariance of the gauge kinetic terms of the two doublets under a U(2) Higgs flavor transformation. In the Higgs basis $H_1$ has the full vev $v$ and thus is precisely the SM Higgs doublet. In general however both neutral components mix upon EWSB and a SM-like CP-even mass eigenstate is not automatic. On the contrary, if one of the CP-even eigenstates is parallel to the neutral $H_1$ direction, this realizes the alignment limit of the 2HDM~\cite{Gunion:2002zf} which the LHC Higgs data appears to favor. In general, in a basis-independent manner, the alignment limit is defined as the presence of a CP-even eigenstate in the vev direction in the scalar field space.

In the electroweak vacuum, the squared mass matrices in the neutral CP-even, CP-odd and charged scalar sectors are respectively given by, 
\begin{align}
\mathcal{M}^2_P & = \begin{pmatrix}
m_{12}^2 \tb+ \lambda_1 v^2 \cb^2  &\quad - m^2_{12} + {\lambda_{345} \over 2} v^2 \stwob \\[5pt]
- m^2_{12} + {\lambda_{345} \over 2} v^2  & \quad m_{12}^2/ \tb+ \lambda_2 v^2 \sb^2  \\[5pt]
\end{pmatrix} \label{eq:CPEvenMass} \ ,\\
\mathcal{M}^2_A & = \left[ m_{12}^2 - {1\over 2} \lambda_5 v^2 \stwob \right] \begin{pmatrix}
\tb & \quad -1 \\[5pt]
-1 & \quad  1/\tb
\end{pmatrix} \label{eq:CPOddMass} \ ,\\
\mathcal{M}^2_\pm &= \left[ m_{12}^2 - {1\over 4} (\lambda_4+ \lambda_5) v^2 \stwob \right] \begin{pmatrix}
\tb & \quad -1 \\[5pt]
-1 & \quad  1/\tb
\end{pmatrix}.
\label{eq:ChargedMass}
\end{align}
The CP-even mass eigenstates $h$ and $H$, with $m_h \leq m_H$, are obtained through the diagonalization of $\mathcal{M}^2_P$, they are expressed in terms of the neutral components of the doublets as,
\begin{equation}
\label{Z2_basis_diag}
\begin{pmatrix}H\\ h\end{pmatrix}= \begin{pmatrix} c_\alpha & s_\alpha \\  -s_\alpha &  c_\alpha \end{pmatrix} \begin{pmatrix}\rho_1\\ \rho_2\end{pmatrix}\,,
\end{equation}
where the mixing angle $\alpha$ is introduced and is expressed in terms of the entries of the mass matrix. Diagonalization of $\mathcal{M}^2_A$ leads to a massive CP-odd scalar $A$ and a massless Goldstone boson $G^{0}$, while $\mathcal{M}^2_\pm$ leads to a charged state $H\pm$ and a charged massless Goldstone boson $G^{\pm}$. 
Their tree-level masses read~\footnote{We point out that in the review article~\cite{Branco:2011iw} a factor of 2 is missing in front of $\lam_{5}$ and $\lam_{4}+\lam_{5}$ terms in the formula of $m^{2}_{A}$ and $m^{2}_{+}$, respectively.} 
\begin{align}
m^2_{H,h} &= \frac{1}{2}\left[\mathcal{M}^2_{P, 11} + \mathcal{M}^2_{P,22}\pm \sqrt{(\mathcal{M}^2_{P,11}-\mathcal{M}^2_{P,22})^2+4 (\mathcal{M}^2_{P,12})^2 } \right]  \label{eq:hmass} \ ,\\
m_A^2 &= \frac{m_{12}^2}{s_\beta c_\beta} - \lambda_5 v^2 \label{eq:Amass} \ , \\
m_{H^{\pm}}^2 &= \frac{m_{12}^2}{s_\beta c_\beta} - \frac{1}{2} (\lambda_4+\lambda_5) v^2 \ .\label{eq:Hcmass}
\end{align}

Using Eqs. (\ref{eq:hmass})-(\ref{eq:Hcmass}) one can inversely solve the potential parameters, $\lam_1, \dots, \lam_5$ in terms of four physical Higgs masses and the CP-even Higgs mixing angle $\alpha$, supplemented by the $\mathbb{Z}_2$ soft-breaking parameter $m^{2}_{12}$~\cite{Gunion:2002zf}. 
This means that the scalar potential can be entirely determined by these seven parameters and therefore allows us to choose them as a set of complete free inputs for the numerical analysis.

As mentioned previously, we imposed a $\mathbb{Z}_{2}$ symmetry on the potential Eq.~\eqref{potsym} in order to forbid Higgs-mediated tree-level FCNCs. Out of the four independent realizations of this symmetry in the fermion sector, we study two of them: the so-called \typei~model where only $\Phi_1$ couples to fermions and the \typeii~model where $\Phi_1$ couples to down-type fermions and $\Phi_2$ to up-type fermions, see~\cite{Aoki:2009ha} for details. These particular structures redefine multiplicatively the Higgs couplings to fermions as compared to the SM predictions, we denote as $C_{U,D,V}$ theses multiplicative factors for the up-type fermions, down-type fermions and massive gauge bosons, respectively. The Higgs couplings to massive gauge bosons do not depend on the $\mathbb{Z}_{2}$ symmetry charges but are directly obtained from gauge symmetry alone. In Table~\ref{2hdmcouplings} we present these factors for the three physical scalar states of the theory. Important intuition can be gained by re-expressing these factors in terms of $(\beta-\alpha)$ and $\beta$, in particular to understand their behavior in the alignment limit $\sinba\approx1$:
\begin{eqnarray}
\label{CFh}
C_F^{h, \text{I}}= C_U^{h, \text{II}}=& {\cos\alpha}/{\sin\beta}&=\ \sinba +\cosba\cot\beta,\\
\label{CDh}
C_D^{h, \text{II}}=& -{\sin\alpha}/{\cos\beta}&=\ \sinba -\cosba\tan\beta,\\
\label{CFHH}
C_F^{H, \text{I}}= C_U^{H, \text{II}}=& {\sin\alpha}/{\sin\beta}&=\ \cosba- \sinba\cot\beta,\\
\label{CDHH}
C_D^{H, \text{II}}=& {\cos\alpha}/{\cos\beta}&=\ \cosba +\sinba\tan\beta.
\end{eqnarray}

\begin{table}[t!]
\begin{center}
\caption{Tree-level vector boson couplings $C_V$ ($V=W,Z$) and fermionic couplings $C_{U}$ and $C_D$ to up-type and down-type fermions respectively, normalized to their SM values for the two scalars $h,H$ and the pseudoscalar $A$ in \typei~and \typeii~models.}
\label{2hdmcouplings}
\vspace{5mm}
\begin{tabular}{|c|c|c|c|c|c|}
\hline
\ & Type~I, II  & \multicolumn{2}{c|}  {Type~I} & \multicolumn{2}{c|}{Type~II} \\
\hline
Higgs & $C_V$ & $C_U$ & $C_D$  & $C_U$ & $C_D$  \\
\hline
 $h$ & $\sin(\beta-\alpha)$ & $\cos\alpha/ \sin\beta$ & $\cos\alpha/ \sin\beta$  &  $\cos\alpha/\sin\beta$ & $-{\sin\alpha/\cos\beta}$   \\
\hline
 $H$ & $\cos(\beta-\alpha)$ & $\sin\alpha/ \sin\beta$ &  $\sin\alpha/ \sin\beta$ &  $\sin\alpha/ \sin\beta$ & $\cos\alpha/\cos\beta$ \\
\hline
 $A$ & 0 & $\cot\beta$ & $-\cot\beta$ & $\cot\beta$  & $\tan\beta$ \\
\hline
\end{tabular}
\end{center}
\end{table}

%%%%%%%%%%%%%%%%%%
\subsection{Theoretical constraints}
\label{subsec:theoconst}

For a viable 2HDM scenario, we require here tree-level stability of the potential, which means that Eq.~(\ref{potsym}) has to be bounded from below, requiring 
\bea
& \lam_1, \lam_2>0, \quad\lam_3+\lam_4-|\lam_5|>-\sqrt{\lam_1 \lam_2}, \quad
\lam_3>-\sqrt{\lam_1 \lam_2} & \label{first3}\ .
\eea
In this work we improve the bounds supplemented by the radiative corrected potential, as will be shown in Sec.~\ref{sec:cw}.
Additional theoretical constraints from S-matrix unitarity and perturbativity are required. Tree-level unitarity~\footnote{For a recent one-loop analysis, leading to slightly more stringent bounds, see~\cite{Grinstein:2015rtl}.} imposes bounds on the size of the quartic couplings $\lambda_i$ or various combinations of them~\cite{Akeroyd:2000wc,Ginzburg:2005dt}.
Similarly (often less stringent) bounds on $\lambda_i$ may be obtained from perturbativity arguments.

%%%%%%%%%%%%%%%%%%
\subsection{The experimental constraints}
\label{sec:LHCbounds}

Next, we briefly describe the impact of the experimental bounds on the parameter space of the model.
First, electroweak precision data (EWPD), essentially the T parameter, constrains the mass difference between $m_{H^\pm}$ and $m_{A}$ or $m_{H}$, one of the two neutral states should indeed be approximatively paired with the charged state in order to restore a custodial symmetry of the Higgs sector~\cite{Gerard:2007kn,Haber:2010bw}.
Second, the recent measurement on $BR(B \rightarrow X_{s} \gamma)$~\cite{Belle:2016ufb} excludes low values of $m_{H^\pm}\lesssim 580$ GeV in the \typeii~model~\cite{Misiak:2017bgg}. As a consequence, the preferred ranges for the scalar masses are pushed above $\sim 400$~GeV. 
Third, LHC measurements of the 125~GeV signal rates put large constraints on the 2HDM parameter space, in particular they tend to favor the alignment limit where the Higgs couplings are similar to the SM ones. To evaluate these constraints, we use \texttt{Lilith-1.1.3}~\cite{Bernon:2015hsa}.

\begin{figure}[t]
\begin{center}
\includegraphics[width=0.55\textwidth]{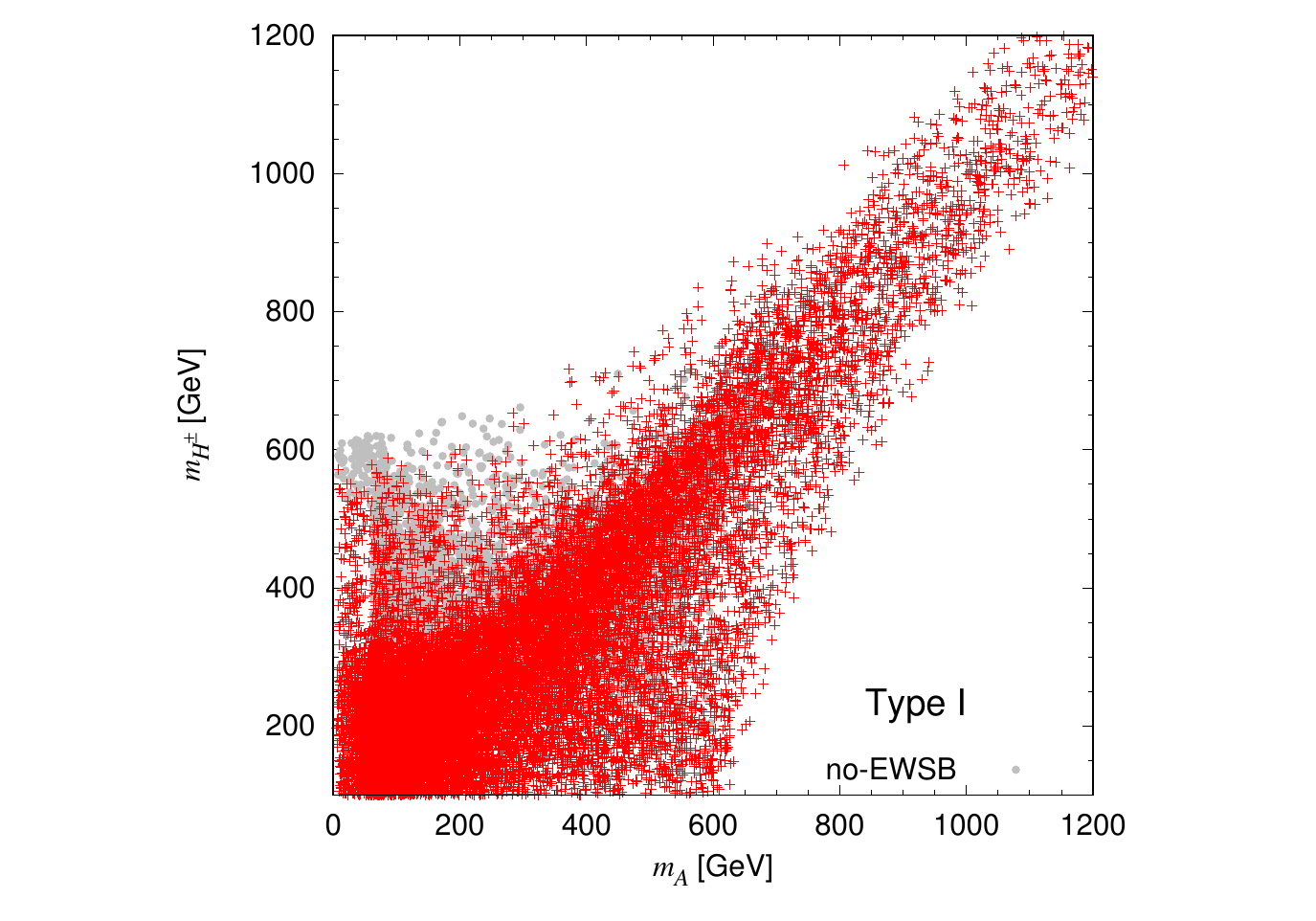}
\hspace{-18mm}
\includegraphics[width=0.55\textwidth]{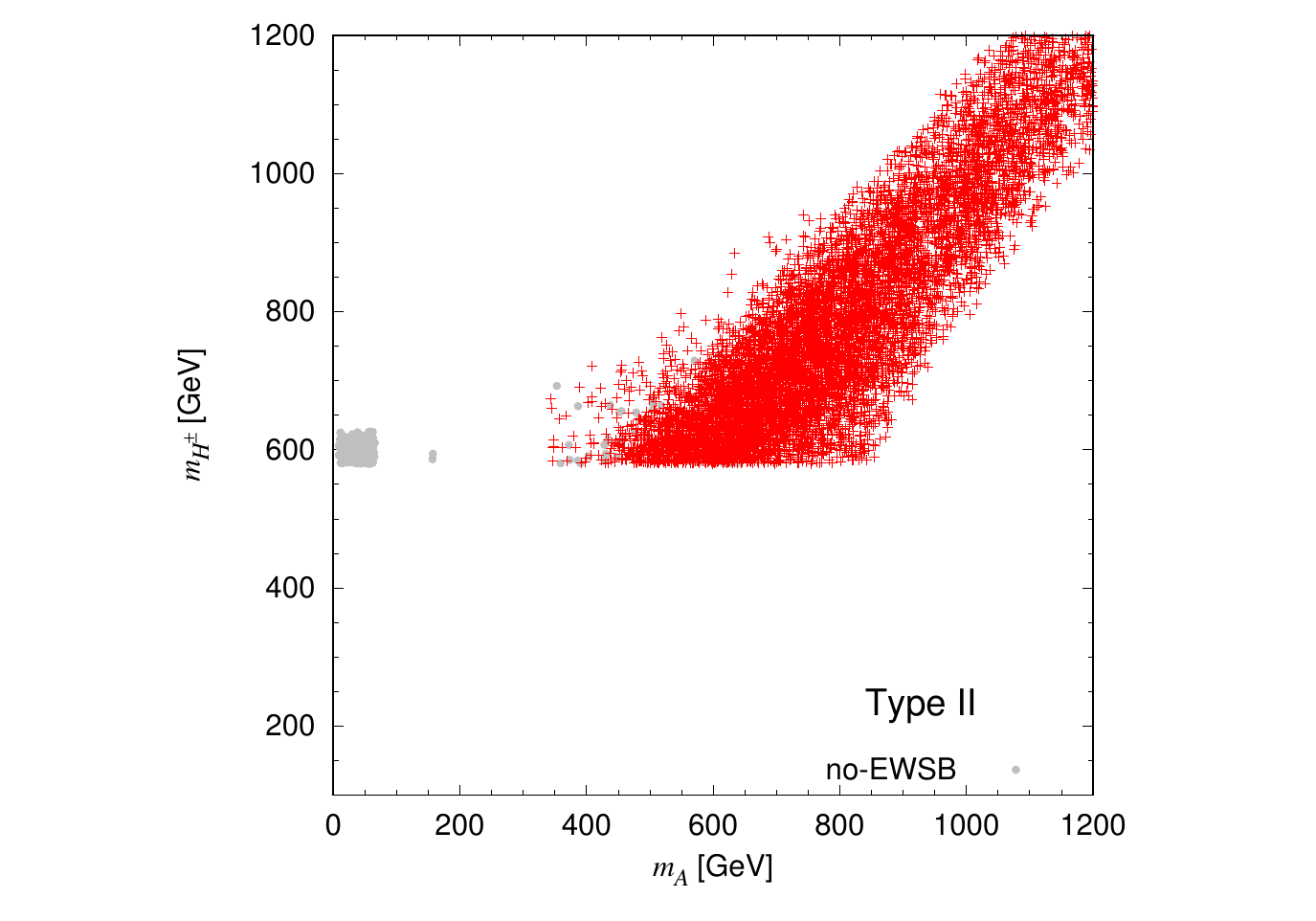}\\
\includegraphics[width=0.55\textwidth]{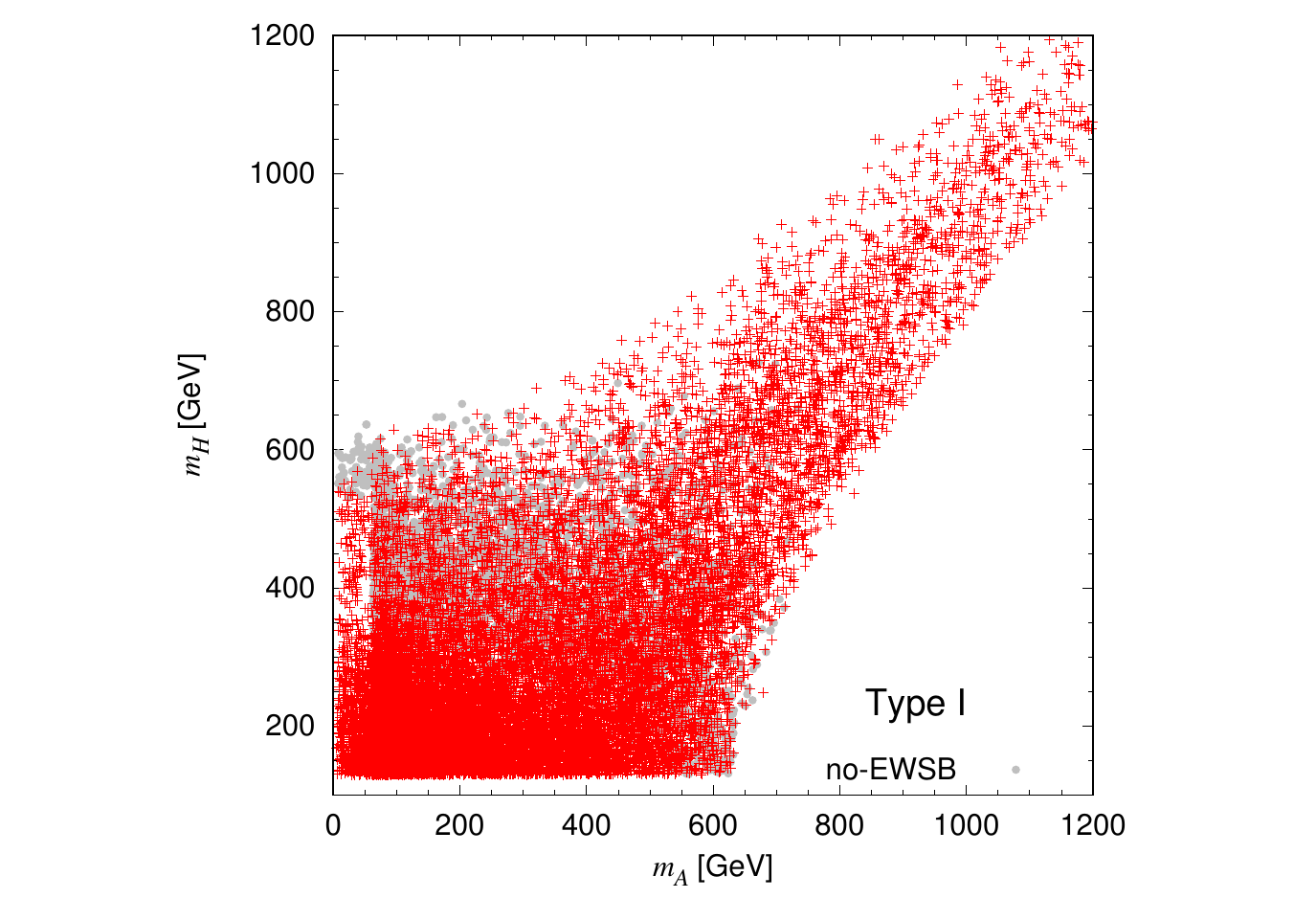}
\hspace{-18mm}
\includegraphics[width=0.55\textwidth]{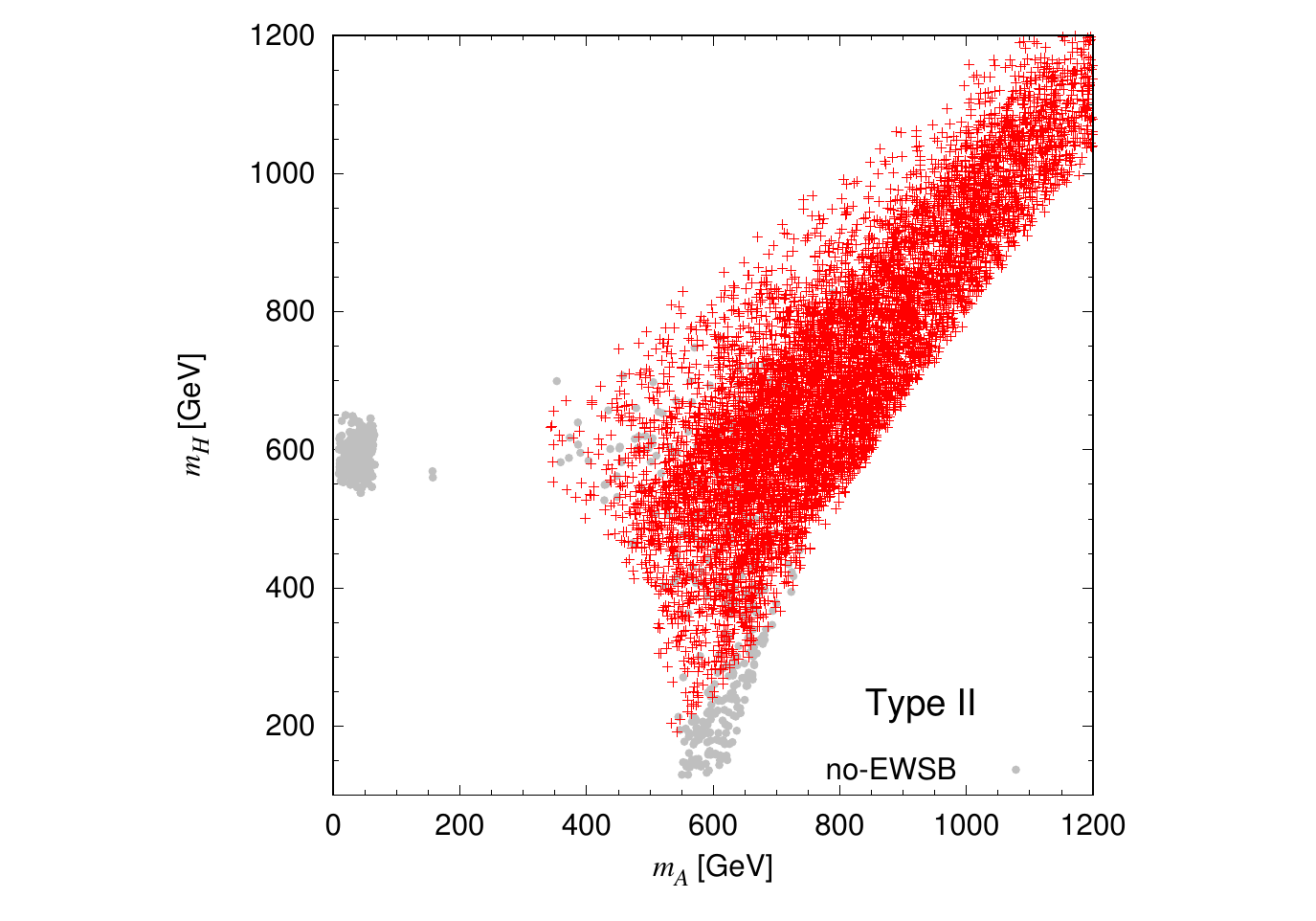}
\caption{The allowed mass spectra (up to 1.2 TeV) of the BSM Higgs bosons confronting with LHC Run-2. In \typeii~(right), the B-physics constraint on the charged Higgs $m_{H^\pm} \ge580$ GeV~\cite{Misiak:2017bgg} is imposed and nearly excludes the low $m_A$ points. Gray points indicate EWSB is not ensured at zero temperature when one-loop effect is included in the Higgs potential.}
\label{fig:LHCbounds}
\end{center}
\end{figure}

Finally, regarding direct searches, we implement the Run-1 and LEP constraints as performed in \cite{Bernon:2015qea}. 
A very important search for the \typeii~model is in the $A,H\to\tau\tau$ channel, either through gluon-fusion or $b\bar b$-associated production~\cite{Aaboud:2017sjh,CMS:2016pkt}. The ATLAS Run-2 constraint is much stronger than the corresponding Run-1 searches, eliminating larger portion of the parameter space at large $\tan\beta$ in particular. 
For $m_A \lsim 350$~GeV we only find few scenarios compatible with the experimental constraints in the \typeii~model~\footnote{This result is not fully consistent with Ref.~\cite{Basler:2016obg} where the authors claimed the experimental constraints are less severe for $m_A \lsim 120$~GeV.}. This is both coming from the aforementioned $\tau\tau$ search, as well as the $H\to ZA$ searches for CP-odd state down to 60~GeV.
This final state has been searched for by the CMS collaboration during Run-1~\cite{Khachatryan:2016are}, and leads to severe constrains of the parameter region. The $A\to Zh$ channel has been searched for during both LHC Run-1~\cite{Aad:2015wra,Khachatryan:2016are} and Run-2~\cite{ATLAS:2017nxi} but the resulting constraints have little impact. 
In Fig.~\ref{fig:LHCbounds} we show the allowed spectra (red pluses) for the two types of models considered here. 
The points labeled `no-EWSB' comes from the requirement of proper EWSB at the 1-loop level, which will be extensively discussed in Sec.~\ref{sec:cw}.
Due to the severe constraints on the mass spectrum of the extra Higgs bosons, these experimental constraints have significant influence on the requirement of a SFOPT as we will see in Sec.~\ref{sec:res}.

We now move to investigate the possibility of having a first-order phase transition for the surviving sample points.
The interesting question is whether the parameter space that LHC Higgs data favors, simultaneously satisfying both theoretical constraints and experimental bounds, can lead to a favorable prediction for a strong first-order phase transition.

%%%%%%% Fig.1 %%%

%%%%%%%%%%%%%%%%%%%%%%%%%%%%%%%
\section{The effective potential at finite temperature}
\label{sec:effV}

To study the phase transition we consider the scalar potential of the model at finite temperature.
In the standard analysis, the effective potential $V_{\rm eff} (h_{1},h_{2},T)$ is 
\begin{align}
\label{potVeff}
V_{\rm eff}(h_{1},h_{2},T) &= V_0(h_1,h_2) +V_{\rm CW}(h_1,h_2) +V_{\rm CT}(h_1,h_2) + V_{\rm th}(h_{1},h_{2},T)\; ,
\end{align}
which is composed of the tree-level potential at zero temperature $V_0(h_{1},h_{2})$ derived in \eq{eq:V0tree}, the Coleman-Weinberg one-loop effective potential $V_{\rm CW}(h_{1},h_{2})$
at $T=0$ given in \eq{eq:CWpot}, the counter-terms  $V_{\rm CT}$ given in \eq{eq:Vct} being chosen to maintain the tree level relations of the parameters in $V_0$, and the leading thermal corrections being denoted by $V_{\rm th}(h_{1},h_{2},T)$. We discuss these terms separately now.

%-------------------------------------------------------------------------------
\subsection{The tree level potential}

Since our model is CP conserving, the classical value of the CP-odd field $A$ is zero and so are the ones for the neutral Goldstone fields. We assume the charged fields do not get VEV during the EWPT process, by taking the classical values for the charged fields to be zero, to strictly respect the U(1) electromagnetic symmetry and therefore ensure the photon massless
~\cite{Branco:2011iw}.~\footnote{
The charge breaking vacuum in multi-Higgs doublet models has been studied in Refs.~\cite{DiazCruz:1992uw,Ferreira:2004yd,Barroso:2005sm,Barroso:2006pa,Barroso:2007rr,Ginzburg:2007jn,Ginzburg:2010wa}.
Once the U(1) electromagnetic symmetry is broken during the EWPT, the photon acquires mass, which may change the thermal history of the Universe~\cite{Ginzburg:2010wa}. We leave it to future work.
Also, we do not expect the presence of color-breaking vacuum in the process of EWPT since the bosons which actively participate into the evolution of Higgs scalar potential are color neutral. As of our knowledge, color-breaking baryogenesis is achievable in the model with the inclusion of colored bosons (i.e. scalar leptoquarks)~\cite{Patel:2013zla,Ramsey-Musolf:2017tgh}.}
The relevant tree level potential $V_0$ in terms of their classical fields $(h_1,h_2)$~\footnote{To avoid confusion we distinguish the dynamical fields and EW vev in this paper. The classical fields $(h_{1}, h_{2})$ approach the EW vacuum $(v_{1}, v_{2})$ at zero temperature.} derived from Eq.~(\ref{potsym}) is
\begin{align}
\label{eq:V0tree}
V_0(h_{1},h_{2}) &=  {1\over 2} m^{2}_{12} \tb \left(h_{1}  - h_{2} \tb^{-1}\right)^{2}
 - {v^{2}\over 4} {\lam_{1} h^{2}_{1} + \lam_{2} h^{2}_{2} \tb^{2} \over 1+\tb^{2}}  - {v^{2}\over 4} {\lam_{345} (h^{2}_{1} \tb^{2}  + h^{2}_{2} ) \over 1+\tb^{2}} \nonumber\\
& \quad + {1\over 8} \lam_{1} h^{4}_{1} + {1\over 8} \lam_{2} h^{4}_{2} + {1 \over 4} \lam_{345} h^{2}_{1} h^{2}_{2}
\end{align}
here we have eliminated $m^{2}_{11}$ and $m^{2}_{22}$ by using the minimization conditions \eq{min_cond}.  

%-------------------------------------------------------------------------------
\subsection{The Coleman-Weinberg potential at zero temperature}
\label{sec:cw}
To obtain the radiative corrections of the potential at one-loop level, we use Coleman and Weinberg method~\cite{Coleman:1973jx}.
The Coleman-Weinberg (CW) potential in the $\overline{\rm MS}$ scheme and Landau gauge\footnote{As noted in \cite{Camargo-Molina:2013qva}, the VEVs are slightly different in various gauges and the recent study~\cite{Haarr:2016qzq} find this effect to be numerically small in the physically interesting regions of parameter space.} at 1-loop level has the form:
\beq
\label{eq:CWpot}
V_{\rm CW}(h_{1},h_{2}) = \sum_{i} (-1)^{2s_i} n_i\frac{\hm_i^4 (h_{1},h_{2})}{64\pi^2}\left[\ln \frac{\hm_i^2 (h_{1},h_{2})}{Q^2}-C_i\right] \; \ .
\end{equation}
The sum $i$ runs over the contributions from the top fermion, massive $W^\pm,Z$ bosons, all Higgs bosons and Goldstone bosons~\footnote{We ignore the light SM fermions because of the smallness of their masses. In contrast, the inclusion of Goldstone modes is necessary as their masses are non-vanishing for  field configurations outside the electroweak vacuum. The photon at zero temperature is strictly massless due to gauge invariance.}; in the sum $s_i$ and $n_i$ are the spin and the numbers of degree of freedom for the $i$-th particle listed in Table~\ref{tab:nicount};
$Q$ is a renormalization scale which we fix to $Q=v$
and $C_i$ are constants depending on the renormalization scheme.
In the $\overline{\text{MS}}$ on-shell scheme employed, $C_i=\frac{1}{2}$ ($\frac{3}{2}$) for the transverse (longitudinal) polarizations of gauge bosons~\footnote{In most literature $C_i=5/6$ is taken for gauge bosons without the distinction between transverse and longitudinal modes. In fact, these two ways of counting are equivalent as the field-dependent mass are identical for both transverse and longitudinal modes at zero temperature. For instance, $n_Z C_Z=2\times 1/2 + 1 \times 3/2 = 3 \times 5/6$ and $n_W C_W=2 n_Z C_Z$. The mass difference between transverse and longitudinal modes arises from thermal corrections as will see later.} and $C_i=3/2$ for the particles of other species~\cite{Quiros:1999jp}.
Finally, the field-dependent squared masses $\hm^2_i$ for SM particles include~\footnote{We notice typos occurring in the thermal mass of SM fermions (c.f. Eqs.~(A.19) and (A.20)) in Ref.~\cite{Haarr:2016qzq}.}
\begin{align}
%\hm^2_t &= {1\over 2} y^2_t \left(h^{2}_{1} +h^{2}_{2} \right) \ , \\
\hm^2_t &= {1\over 2} y^2_t h^{2}_{2}/{s_\beta^2} \ , \\
\hm^2_{W^\pm} &= {1\over 4} g^2_t \left(h^{2}_{1} +h^{2}_{2} \right), \quad \hm^2_{Z} = {1\over 4} (g^2+g'^2)  \left(h^{2}_{1} +h^{2}_{2} \right), \quad \hm^2_{\gamma}=0 \ ,
\end{align}
with the corresponding SM Yukawa and gauge couplings being defined $g=2M_W/v, g'=2\sqrt{M_Z^2-M_W^2}/v$, $y_t=\sqrt{2}m_t/v$
and the ones for scalar bosons are given by
\begin{align}
\hm^2_{h,H} &=\rm{eigenvalues} ( \widehat{\mathcal{M}^2_P} ) \ , \\
\hm^2_{G,A} &=\rm{eigenvalues} ( \widehat{\mathcal{M}^2_A}) \ , \\
\hm^2_{G^\pm,H^\pm} &=\rm{eigenvalues}  (\widehat{\mathcal{M}^2_\pm})  \ ,
\end{align}
where the corresponding matrices $\widehat{\mathcal{M}}^2_X\,(X=P,A,\pm)$ are
\begin{align}
\widehat{\mathcal{M}_X^2} & = \begin{pmatrix}
{\lam_{1}\over 2}   h^{2}_{1}+m^{2}_{12} \tb - {\lam_{1} \over 2} {v^{2} \over 1+ \tb^{2}} -  {\lam_{345} \over 2} {v^{2} \tb^{2} \over 1+ \tb^{2}} +  \Theta^{X}_{11} & \quad  -m^{2}_{12} +  \Theta^{X}_{12} \\[5pt]
 -m^{2}_{12} +  \Theta^{X}_{12} &  {\lam_{2} \over 2}  h^{2}_{2}+m^{2}_{12} \tb^{-1} - {\lam_{2} \over 2} {v^{2} \tb^{2} \over 1+ \tb^{2}} -  {\lam_{345} \over 2} {v^{2}\over 1+ \tb^{2}} +  \Theta^{X}_{22}
\end{pmatrix} \ .
\label{eq:FieldDepMass}
\end{align}
Here the $\Theta^{X}_{ij}$ terms listed below are different for $X=P,A,\pm$
\beq
\begin{array}{llll}
&\Theta^{P}_{11} =\lam_{1} h^{2}_{1} + {1\over 2} \lam_{345} h^{2}_{2}, \quad & \Theta^{A}_{11} ={1\over 2} \bar{\lam}_{345} h^{2}_{2}, \quad & \Theta^{\pm}_{11} ={1\over 2} \lam_{3} h^{2}_{2} \ , \\[10pt]
&\Theta^{P}_{12} =\lam_{345} h_{1} h_{2}, \quad & \Theta^{A}_{12} =\lam_{5} h_{1} h_{2}, \quad & \Theta^{\pm}_{12} ={1 \over 2} (\lam_{4} +\lam_{5}) h_{1} h_{2} \ , \\[10pt]
&\Theta^{P}_{22} =\lam_{2} h^{2}_{1} + {1\over 2} \lam_{345} h^{2}_{1}, \quad & \Theta^{A}_{22} ={1\over 2} \bar{\lam}_{345} h^{2}_{1}, \quad & \Theta^{\pm}_{22} ={1\over 2} \lam_{3} h^{2}_{1} \ ,
\end{array}
\eeq
in which $\bar{\lam}_{345}\equiv\lam_{3}+\lam_{4}-\lam_{5}$.

\begin{table}[t]
\caption{The number of d.o.f. from SM particles of different species contributing to the thermal potential. The fermions except the top quark are neglected due to their small masses.}
\begin{center}
\begin{tabular}{|c|c|c|c|c|c|c|}
\hline
$i$ & $t$ & $W^\pm$ & $Z$ & $\mathcal{H}=\{h,H,A\}$ &  $H^\pm$ & $G^{0},G^{\pm}$ \\
\hline
$n_{i}$ & $2 \times 2  \times 3$ & $3\times 2$ & 3 & $1\times 3$ & 2 & 1+2\\
\hline
\end{tabular}
\end{center}
\label{tab:nicount}
\end{table}%

With $V_{\rm CW}$ being included in the potential, the minimum of the Higgs potential  will be slightly shifted, and hence the minimization conditions Eq.~(\ref{min_cond}) no longer hold. 
To maintain these relations, we add the so-called ``counter-terms" (CT)\cite{Cline:2011mm}~\footnote{In addition, we do not include more complicate terms to compensate the shift of mass matrix of $h$, because these shift effects are estimated to be negligible in our scenario.},
\begin{equation}
\label{eq:Vct}
V_{\rm CT}=\delta m_1^2 h_{1}^2+\delta m_2^2 h_{2}^2 +\delta \lam_1 h_{1}^4+\delta \lam_{12} h_{1}^2 h_{2}^2 +\delta \lam_2 h_{2}^4\; ,
\end{equation}
where the relevant coefficients are determined by,
\beq
\label{eq:V1der}
\frac{\partial V_{\rm CT}}{\partial h_{1}} = -\frac{\partial V_{\rm CW}}{\partial h_{1}}\;, \quad \frac{\partial V_{\rm CT}}{\partial h_{2}} = -\frac{\partial V_{\rm CW}}{\partial h_{2}}\;, 
\eeq
\beq
\label{eq:V2der}
\frac{\partial^{2} V_{\rm CT}}{\partial h_{1}\partial h_{1}} = - \frac{\partial^{2} V_{\rm CW}}{\partial h_{1}\partial h_{1}}\;, \quad
\frac{\partial^{2} V_{\rm CT}}{\partial h_{1}\partial h_{2}} = - \frac{\partial^{2} V_{\rm CW}}{\partial h_{1}\partial h_{2}}\;, \quad
\frac{\partial^{2} V_{\rm CT}}{\partial h_{2}\partial h_{2}} = - \frac{\partial^{2} V_{\rm CW}}{\partial h_{2}\partial h_{2}}\;, 
\eeq
which are evaluated at the EW minimum of $\{ h_{1}=v_{1}, h_{2}=v_{2}, A=0 \}$ on both sides.
As a result, the vevs of $h_{1}$, $h_{2}$ and the CP-even mass matrix will not be shifted.

One technical difficulty involved at this step arises from the inclusion of the Goldstone bosons in the CW potential. Due to the variation of the scalar field configuration with temperature (which we will see shortly), the Goldstone boson may acquire a non-zero mass at finite temperature, enforcing the inclusion of Goldstone modes in the sum. Nonetheless, in the electroweak vacuum at zero temperature the masses of the Goldstone bosons are vanishing in the Landau gauge, which leads to an infrared (IR) divergence due to the second derivative present in our renormalization conditions \eq{eq:V2der}. This means that renormalizing the Higgs mass at the IR limit is ill-defined~\cite{Delaunay:2007wb}. To overcome this divergence, we take a straightforward  treatment developed in~\cite{Cline:2011mm} and impose for Goldstone bosons an IR cut-off at SM Higgs mass, $m^{2}_{IR} = m^{2}_{h}$.
Although a rigorous prescription used to deal with the Goldstone's IR divergence was developed in~\cite{Cline:1996mga}, Ref.~\cite{Cline:2011mm} argued that this simple approach can give a good approximation to the exact on-shell renormalization. 
Practically, in evaluating the derivatives for the CW potential, we remove the Goldstone modes from the sum and add instead the following Goldstone contribution to the right hand of \eq{eq:V2der} 
\beq
{1\over 32\pi^{2}} \ln {m^{2}_{G} (h_{1},h_{2}) \over Q^{2}} \left( \frac{\partial^{2} m^{2}_{G}(h_{1},h_{2})} {\partial h_{1}\partial h_{1}} ,
2\frac{\partial^{2} m^{2}_{G}(h_{1},h_{2})} {\partial h_{1}\partial h_{2}} , \frac{\partial^{2} m^{2}_{G}(h_{1},h_{2})} {\partial h_{2}\partial h_{2}} \right)\Bigg|_{\text{vev}}
\eeq
with the replacement for the singular term $m^{2}_{G} (h_{1},h_{2})|_{\text{VEV}} \to m^{2}_{IR}$ in the logarithm. Note that the Goldstone bosons have a vanishing contribution to the first derivative evaluated at the vev.

%%%%%%% Fig.2 %%%
\begin{figure}[t]
\begin{center}
\includegraphics[width=0.56\textwidth]{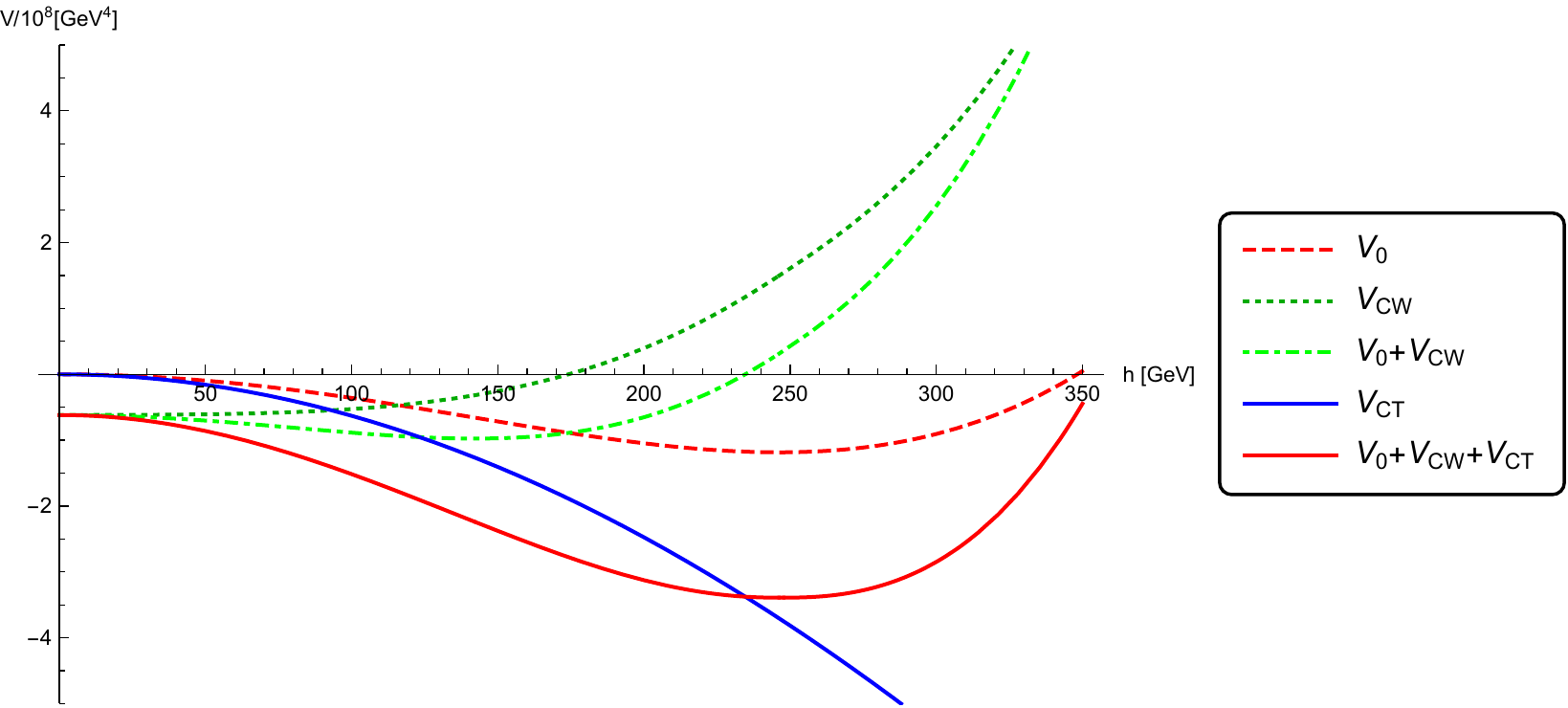}
\includegraphics[width=0.43\textwidth]{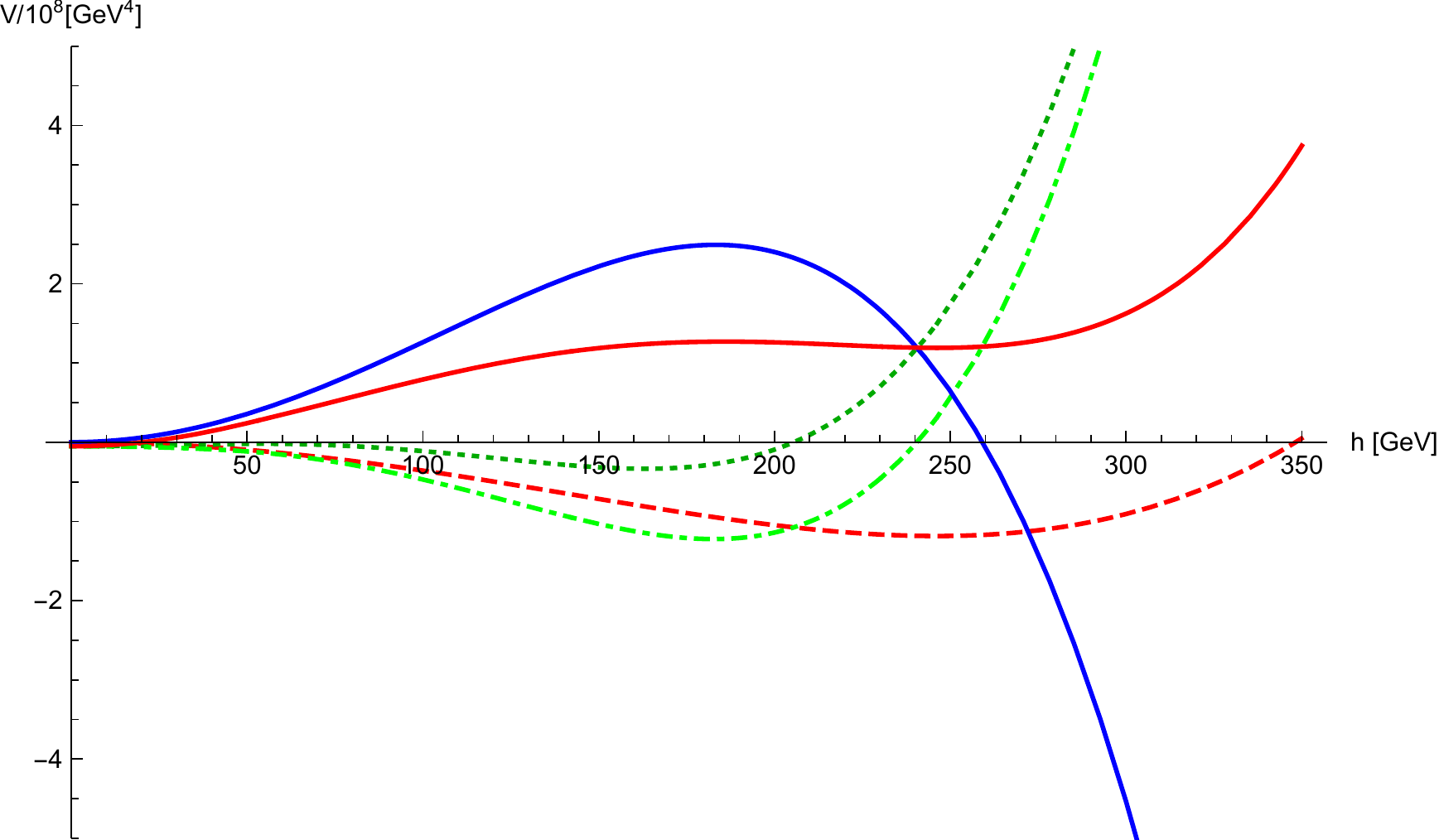}
\caption{
Tree level and loop-correction contributions to the potential at zero temperature for two model points with $\tan\beta=1, \sin(\beta-\alpha)=1$. The remaining parameters corresponding to the point shown in the left (right) plot are $\lam_{1}=\lam_{2}=\lam_{3}=2.9~(6.36), \lam_{4}=-8.5~(-12), \lam_{5}=3.3~(-0.2), m_{12}=315~(-70)\gev$. Clearly, the point shown in the left plot has a true EW vacuum while the one on the right plot has only a local minimum at $v$.}
\label{fig:CWCTplot}
\end{center}
\end{figure}

%%%%%%% Fig.3 %%%
\begin{figure}[t]
\begin{center}
\includegraphics[width=0.7\textwidth]{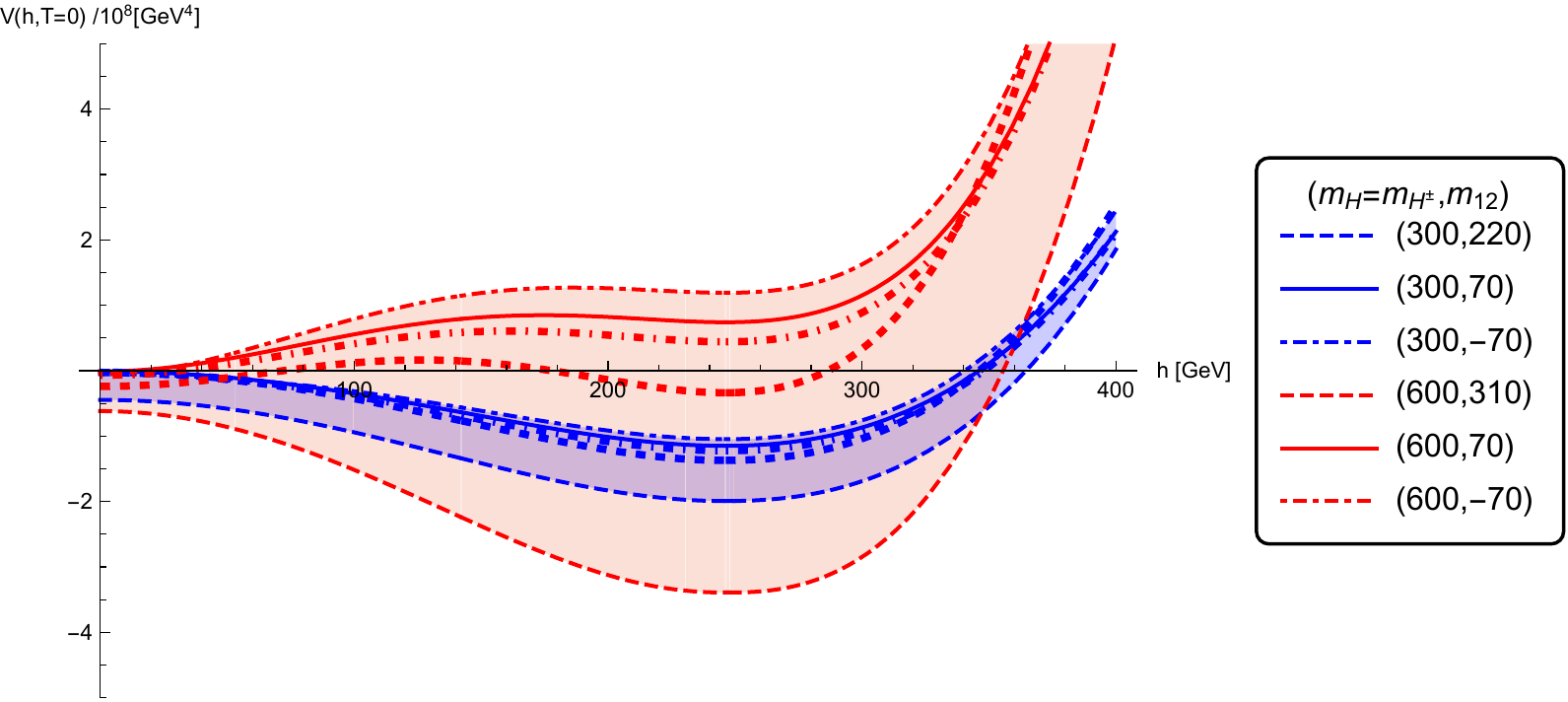}
\caption{The one-loop potential at zero temperature (including the CW potential and counter-terms). $m_H = m_{H^\pm}$ is assumed at two common scales 300 (blue) and 600 (red) GeV, $m_{A}=50\gev$ is taken. The picture is negligibly modified for $m_{A}=200\gev$. Three values of $\tan\beta$, $\tan\beta=1$ (solid and dashed lines), $\tan\beta=10$ (thick dashed lines) and $\tan\beta=20$ (thick dotdashed lines) are shown for which $m^{2}_{12}$ is chosen such that none of quartic couplings exceeds the pertubativity bound.}
\label{fig:zeoTpotent}
\end{center}
\end{figure}

Beyond tree level the true EW vacuum must be preserved when the one-loop corrections are taken into account. 
This demands that the potential after the inclusion of the CW and counter-terms still form a global minimum at the EW vacuum.
As seen in Fig.~\ref{fig:CWCTplot}, the CW term (green dotted) often lifts up the potential at the EW vacuum, resulting the local minimum shifting inward or even leading to a false vacuum. On the other hand, the CT effect (blue) drags down the potential at the EW vacuum and thus helps to accomplish a true EW vacuum.
As a result, the competition between these two opposite effects determines the existence of a global minimum at the EW vacuum. 
We present in Fig.~\ref{fig:CWCTplot} two examples where the left one accomplishes a true EW vacuum, while the potential in the right plot has only a local minimum at $v$. 
The latter example is phrased `no-EWSB' in our terminology and such type of points are displayed in Fig.~\ref{fig:LHCbounds}. 
This is an additional important constraint that excludes about 10\% (5\%) points in the \typei~(II) model, in particular for the points with $m_A \leq300\gev$. In the $m_A<m_h/2$ regime (termed low-$m_A$ scenario), it turns out that EWSB at zero temperature can be achieved as long as at least one lighter $H$ or $H^\pm$ is present in the spectrum. 
For the case where both $H$ and $H^\pm$ are heavier than $\sim 550\gev$,  EWSB would be hardly successful. 
To understand this, we display in Fig.~\ref{fig:zeoTpotent} the one-loop potential at zero temperature (including the CW potential and counter-terms).
For simplicity, we assume $m_H = m_{H^\pm}$, which is typical mass spectra required by the T parameter~\footnote{The lighter the CP-odd state $A$ is, the stronger the degeneracy between $H$ and $H^{\pm}$ should be.}.
The authors of Ref.~\cite{Bernon:2014nxa} have shown that low-$m_{A}$ scenario can be phenomenologically alive in the parameter space where the SM-like Higgs $h$ has very small coupling to $AA$, which leads to $\tan\beta \lesssim 2$ or $\tan\beta \gtrsim 12$ for $m_{H}=600\gev$ in the deep alignment limit. The low $\tan\beta$ solution requires a severe tuning in the parameter $m^{2}_{12}$, and in the allowed range $m^{2}_{12}\simeq 5000\gev^{2}$ the zero temperature potential
(c.f. the red solid line) at EW vacuum $v$ is higher than the one at the origin. Moreover, a proper EW vacuum can be developed as the symmetry soft-breaking parameter $m^{2}_{12}$ increases. This can be achievable for the case of $m_{A} \geq m_{h}/2$ where the $h\to AA$ decay is kinematically suppressed. 
On the other hand, the large $\tan\beta$ solution, though possible in \typei~model, strongly constrains $m^{2}_{12}$ and tends to lift the potential. Hence, the importance of this class of solution is very marginal and no points were found in our numerical analysis. In addition, $\tan\beta \gtrsim 5$ in \typeii~model was already excluded by the CMS bound searching for a light pseudoscalar scalar in the mass range of 20-80~GeV through the bottom-quark associated production and decaying into $\tau\tau$ final states during Run-1~\cite{Khachatryan:2015baw}.
As a comparison, we also exhibit the potential at a lower common scale $m_H = m_{H^\pm}=300\gev$. This example is only applicable in \typei~model. One can observe that the potential generically reaches a global minimum at the EW vacuum and the depth of this minimum is less sensitive to $\tan\beta$. This implies that when the new scalars introduced are not heavy, the loop effect is not substantial and thus the potential is largely governed by the tree-level.

We conclude that the requirement of proper EWSB at zero temperature, in synergy with $m_{H^\pm}\geq580\gev$ required by $B$-physics measurements~\cite{Belle:2016ufb}, entirely exclude the scenario of existing a light pseudoscalar $A$ in \typeii~model that was delicately studied in Ref.~\cite{Bernon:2014nxa}.
As will show shortly, these theoretical constraints will play an important role in achieving a strong first-order phase transition.

%%%%%%%%%%%%%%%%
\subsection{The thermal effective potential}

The finite temperature corrections to the effective potential at one-loop are given by~\cite{Dolan:1973qd}
\beq
\label{potVth}
 V_{\rm th}(h_{1},h_{2}, T) = \frac{T^4}{2\pi^2}\, \sum_i n_i J_{B,F}\left( \frac{ m_i^2(h_{1},h_{2})}{T^2}\right)\;,
\eeq
where the functions $J_{B,F}$ are 
\beq
\label{eq:jfunc}
J_{B,F}(y) = \pm \int_0^\infty\, dx\, x^2\, \ln\left[1\mp {\rm exp}\left(-\sqrt{x^2+y}\right)\right]\; ,
\eeq
with $y\equiv m_{i}^2(h_{1},h_{2})/T^2$ and the upper (lower) sign corresponds to bosonic (fermionic) contributions.
The numerical evaluation of this exact integral is very time-consuming (notably for the $y<0$ case present for the bosonic degrees of freedom).
Thus, computational techniques to reduce the computation time are welcome.
A widely used solution is to consider the asymptotic expansions of $J_{B,F}$.
At small $y$ ($y \ll 1$)~\footnote{The high/low T approximations do not necessarily lead to small/large $y$, which also depends on the field-dependent mass in the numerator.}, Eq.~(\ref{eq:jfunc}) can be approximated by
\begin{align}
J^{y\ll 1}_{B} (y)   & \simeq -\frac{\pi ^{4}}{45}+\frac{\pi ^{2}}{12}y -\frac{\pi }{6}y ^{3/2}-\frac{y ^{2}}{32}\ln \frac{y }{a_{B}},  \label{eq:JBhighT} \\
J^{y\ll 1}_{F} (y) & \simeq -\frac{7\pi ^{4}}{360}+\frac{\pi ^{2}}{24}y +\frac{y ^{2}}{32}\ln \frac{y}{a_{F}},  \label{eq:JFhighT}
\end{align}%
where $a_{F}=\pi ^{2}\exp
({3 \over 2}-2\gamma _{E})$ and $a_{B}= 16 a_{F}$ with the Euler constant $\gamma _{E}=0.5772156649$.
Whereas at large $y$,
\beq
\label{eq:JlowT}
J^{y\gg 1}_{B,F} (y) \simeq - \left(\frac{\pi}{2}\right)^{1/2}y^{3/4} \exp\left(-y^{1/2}\right)\left(1+\frac{15}{8}y^{-1/2}\right).
\eeq

\begin{figure}[t]
\begin{center}
%\hspace*{-3.5mm}
\includegraphics[width=0.34\textwidth]{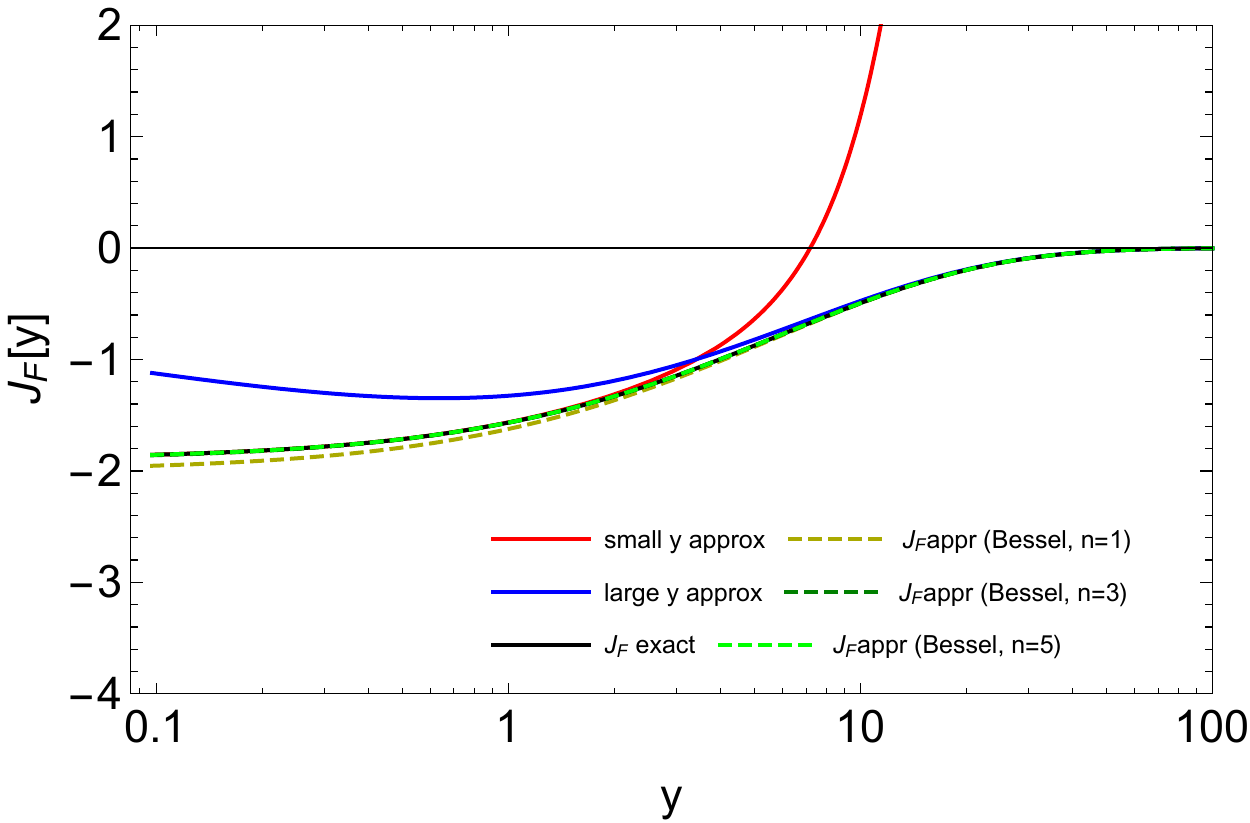}
\hspace{-4mm}
\includegraphics[width=0.34\textwidth]{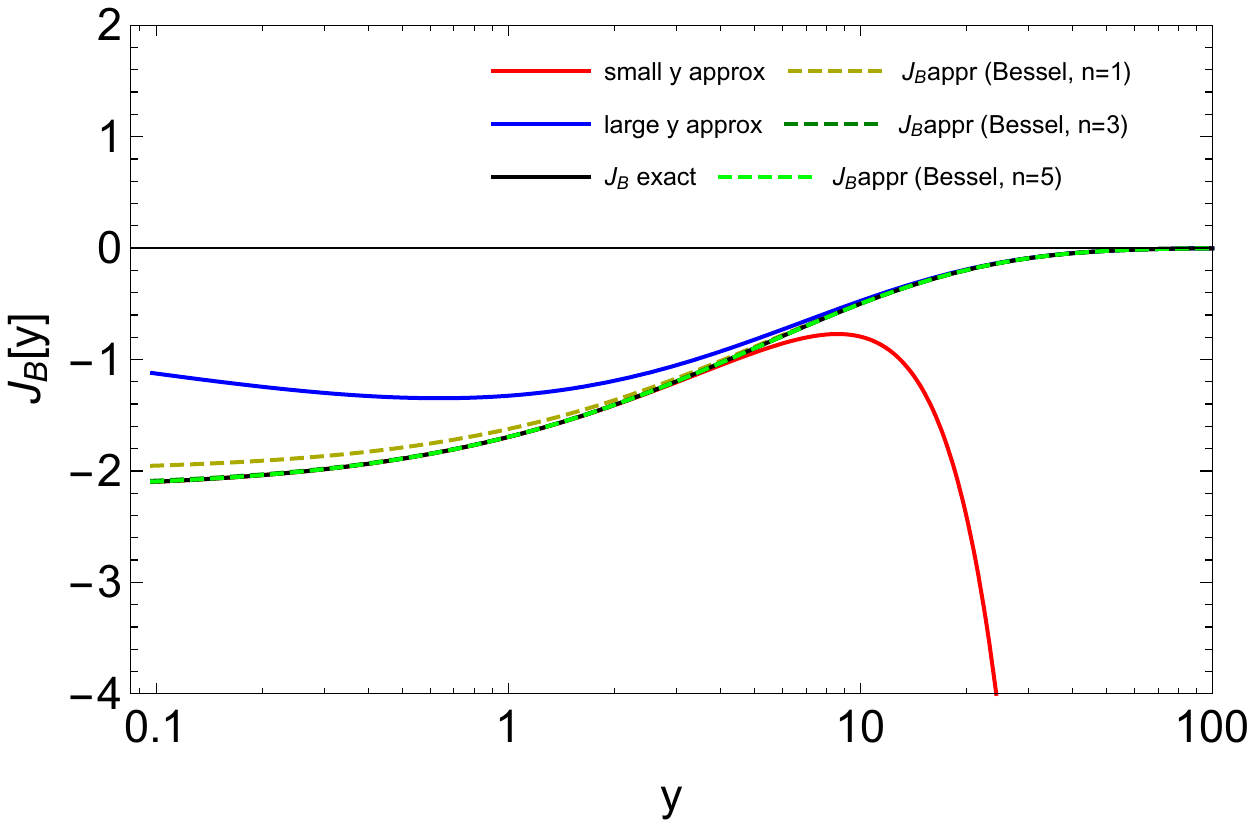}
\hspace{-4mm}
\includegraphics[width=0.33\textwidth]{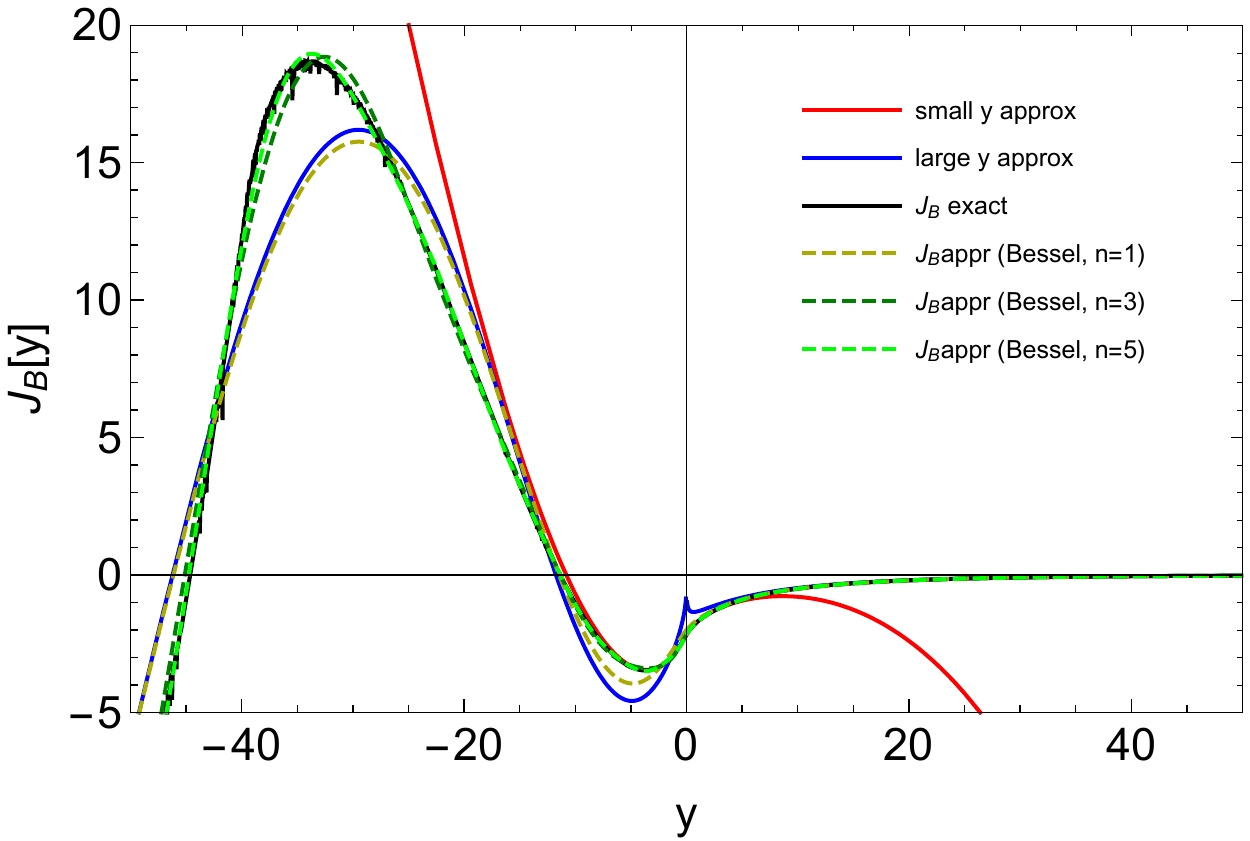}

\caption{Thermal function for fermionic (left) and bosonic (middle) states for positive $y$. For bosonic states, we additionally present the negative $y$ range since their thermal mass can be negative at $T\ne0$.
In each plot the result of the exact integral is shown in solid black curve. Red and blue curves give the small and large $y$ approximations, respectively. Three dashed lines illustrates the result evaluated by summing over the Bessel functions at different order. 
}
\label{fig:JBJFcurve}
\end{center}
\end{figure}

In order to make a quantitive assessment of the approximation precision we plot in Fig.~\ref{fig:JBJFcurve} the small/large $y$ approximations as well as the direct numerical evaluation of the integral. (For the evaluation of the latter one we use the \textsf{NIntegrate} function built in \textsc{Mathematica}.) 
It is clearly seen that the small $y$ approximation (red curve) is valid in the ranges $y \in (-5,5)$ for bosons and $y \in (0,5)$ for fermions, while the large $y$ expansion (blue curve) converges to the exact integral for $y>10$ for both functions. A gap is then present between the small and large $y$ approximations in the transition range $y\in(5,10)$. In this situation an interpolation can be introduced to connect smoothly the two approximations. Even though this reduces the deviation of the approximate results from the exact integral to less than 2\%, there are still two serious shortcomings. 
First, this requires a conditional judgement for each state at temperature $T$  to know which approximation should be applied, this largely increases the evaluation time.
Second, the above approximations Eqs.~(\ref{eq:JBhighT})-(\ref{eq:JlowT}) are only valid for $y>0$ as shown in Fig.~\ref{fig:JBJFcurve}. 
However, the eigenvalues of the mass matrix of the neutral scalar states can become negative depending on the field configuration.~\footnote{For instance, in the SM the field-dependent mass for Higgs field is $m^{2}_{h}=3 \lambda h^2 - \mu^2$ and turns negative at low field configuration. Similarly for the Goldstone bosons.}
If this happens, Ref.~\cite{Basler:2016obg} suggests that only the real part of the integral $J_{B}$ should be chosen in the evaluation as the imaginary part is irrelevant in extracting the global minimum.~\footnote{Tachyonic mass configurations generate a negative local curvature of the potential, leading to a local maximum rather than a minimum.}

%%%% Fig.4 %%%%%

The thermal integrals $J_{B,F}$ given by \eq{eq:jfunc} can be expressed as an infinite sum of modified Bessel functions of the second kind $K_{n} (x)$ with $n=2$~\cite{Anderson:1991zb}, 
\beq
\label{Bessel_JFJB}
J_{B,F}(y) = \lim_{N \to +\infty} \mp \sum_{l=1}^{N} {(\pm1)^{l}  y \over l^{2}} K_{2} (\sqrt{y} l),
\eeq
with the upper (lower) sign corresponds to bosonic (fermionic) contributions. Our numerical results show that the leading order $l=1$ does not provide a good approximation of the full integrals. Instead, inclusion up to $l=5$ order in the expansion can match the exact integral very well for both positive and negative $y$ values. Therefore, in this work we take $N=5$ in the evaluation of the thermal integrals Eq.~\eqref{Bessel_JFJB}.~\footnote{A similar numerical analysis taking $N=50$ was performed in a recent study~\cite{Jain:2017sqm}.}
Fig.~\ref{fig:JBJFcurve} also shows that the thermal function is negative for positive $y$ thus dragging the potential down and leading to the formation of two degenerate vacua. As expected, this dragging effect arising from the temperature corrections diminishes as $y$ approaches to the infinity, which corresponds to zero temperature or the decoupling limit.  

Finally, there is another important part of the thermal corrections to the scalar masses coming from the resummation of \textit{ring }(or\textit{\ daisy}) 
diagrams~\cite{Carrington:1991hz,Arnold:1992rz}, 
\beq
V_{\rm daisy}\left(h_{1},h_{2},T\right) =-\frac{T}{12\pi }\sum_{i} n_{i}\left[ \left( M_{i}^{2}\left(h_{1},h_{2},T\right) \right)^{\frac{3}{2}}-\left( m_{i}^{2}\left( h_{1},h_{2}\right) \right)^{\frac{3}{2}}\right] ,
\label{eq:daisy}
\eeq
where $M_{i}^{2}\left( h_{1},h_{2},T\right) $ are the thermal Debye masses of the
bosons corresponding to the eigenvalues of the full mass matrix 
\begin{equation}
\label{eq:thermalmass}
M_{i}^{2}\left( h_{1},h_{2},T\right) ={\rm eigenvalues} \left[\widehat{\mathcal{M}_{X}^2}\left( h_{1},h_{2}\right) +\Pi _{X}(T)\right]  ,
\end{equation}%
which consists of the field dependent mass matrices at $T=0$ Eq.~\eqref{eq:FieldDepMass} and the finite temperature correction to the mass function $\Pi_X, (X=P,A,\pm)$ given by 
\begin{align}
\Pi_X & = \begin{pmatrix}
\Pi^{X}_{11} &  \quad \Pi^{X}_{12}  \\[5pt]
\Pi^{X}_{12} &  \quad \Pi^{X}_{22}  
\end{pmatrix} {T^{2}\over 24} \ ,
\end{align}
with the diagonal terms being
\beq
\begin{array}{ll}
\label{eq:thermalmass2}
\Pi^{P}_{11} =  \Pi^{A}_{11} =  \Pi^{\pm}_{11} = c_{\rm SM}- 6 y_t^2 + 6 \lambda_1 +4 \lambda_3 + 2 \lambda_4, \\[10pt]
\Pi^{P}_{22} = \Pi^{A}_{22} =  \Pi^{\pm}_{22}  = c_{\rm SM} + 6 \lambda_2 +4 \lambda_3 + 2 \lambda_4      \, ,
\end{array}
\eeq
here the subscripts $\{1,2\}$ denote the states $\{h_{1}, h_{2}\}$ and
\begin{eqnarray}	
c_{\rm SM} = \frac{9}{2}g^2 + \frac{3}{2}g'^2 + 6y_t^2 ,
\end{eqnarray}
is the known SM contribution from the SU(2)$_{\mathrm{L}}$ and U(1)$_Y$ gauge fields and the top quark~\cite{Carrington:1991hz}. It is important to note that the temperature corrections are independent of $\lam_{5}$ where a possible CP phase can reside.
On the other hand, the leading correction to off-diagonal thermal mass is vanishingly small due to $\mathbb{Z}_{2}$ symmetry imposed in the scalar sector. Moreover, it was argued by~\cite{Blinov:2015vma} that subleading thermal corrections to off-diagonal self-energies are suppressed by additional powers of coupling constants and EW vevs which are usually neglected.
Therefore, we shall treat the thermal mass correction $\Pi_{i}$ as diagonal matrices in the following numerical analysis. The thermal mass corrections of the SM gauge bosons are given in Appendix~\ref{sec:thermass}. 

Historically, there was an alternative algorithm proposed by Parwani in dealing with the thermal corrections~\cite{Parwani:1991gq}. He included the effect of thermal correction from Daisy diagrams by means of substituting $m^2_i (h_{1},h_{2})$ by $M^2_i (h_{1},h_{2},T)$ in the  $V_{\rm th}(h_{1},h_{2}, T)$, Eq.~(\ref{potVth}).
It is important to note that these two approaches are not physically equivalent and the results produced are quantitively incompatible~\cite{Basler:2016obg}. They differ in the organization of the perturbative expansion and consistent implementation of higher order terms. The method formulated in \eq{eq:daisy} 
restricts the corrections to the thermal masses at one-loop level, whereas Parwani's method inconsistently blends higher-order contributions. Because of this dangerous artifact unrealistically large values of the phase transition strength $\xi$ (defined in \eq{eq:xidef}) would be obtained.  Therefore, we will adopt the former consistent method in the following analysis.

%%%%%%%%%%%%%%%
\section{Phase transition: classification}
\label{sec:ptc}

In general, a system may transit from one symmetry phase to another one. Here the electroweak symmetry is broken as the Universe cools down, this is singled as a change in the nature of the global 0-vacuum at high temperature that gets replaced by an electroweak breaking global vacuum at lower temperature.
At any given set of parameters, the full effective potential \eq{potVeff} can have several extrema. Our major interest is the global minimum vacuum state, the deepest minimum of the potential. The other extrema can be either saddle points or maxima or local minima of the potential. In studying the thermal phase transition, it is useful to trace the evolution of the extrema as well as calculate the difference in potential depth between the global minimum (called true electroweak (EW) vacuum) and a secondary local minimum.

%%%% Fig.5 %%%%%
\begin{figure}[t]
\includegraphics[width=1.00\textwidth]{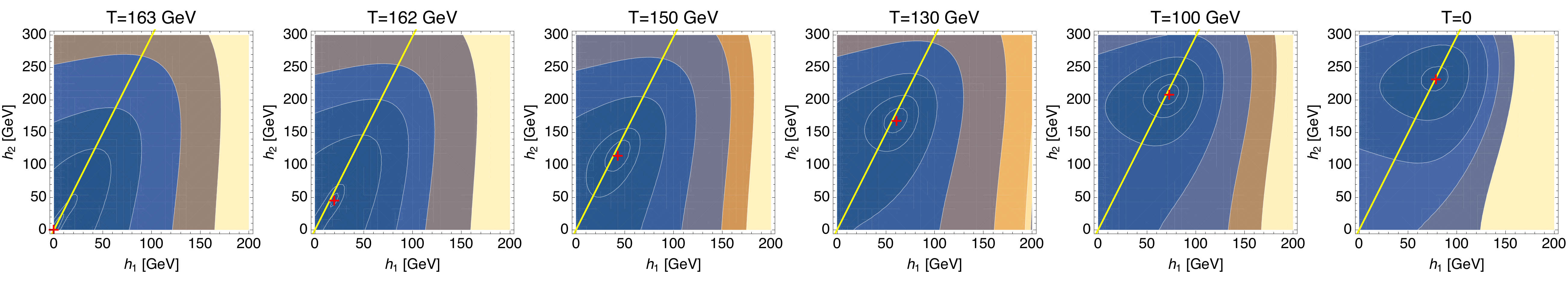}\\
\includegraphics[width=1.00\textwidth]{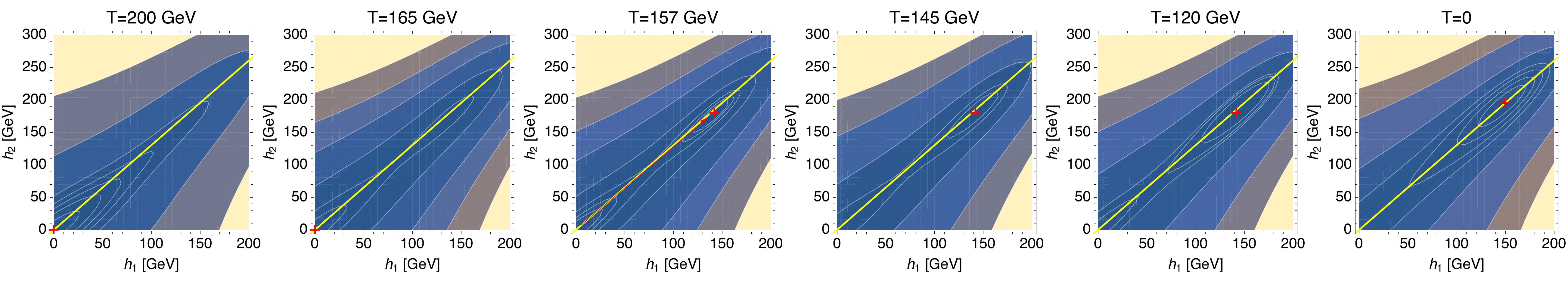}\\
\includegraphics[width=1.00\textwidth]{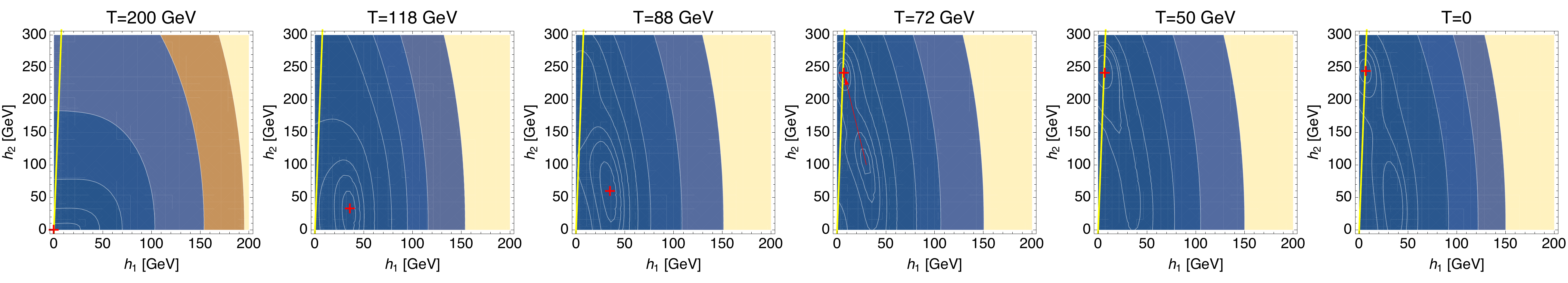}
\caption{Temperature evolution of the Higgs potential on the $(h_{1}, h_{2})$ plane. As temperature cools down, the EW vacuum shifts away from the 0-vacuum. Depending on the way in which the vacuum (marked by the red plus) develops, three types of phase transition presented are possible in the 2HDM: 1-stage second order (top panel), 1-stage first order (middle panel) and 2-stage transition (bottom panel). Red arrows indicate a jump between two degenerate vacuum in the first order phase transition while the vacuum transitions smoothly in the second phase transition.}
\label{fig:potevol}
\end{figure}

First, since at very high temperatures electroweak symmetry is not broken, the effective potential has one global minimum, which tends towards the point $(h_{1}, h_{2}) = (0,0)$. We refer to this minimum as the 0-vacuum.
As the Universe is cooling down, the parameters that characterize the thermal effects of the model evolve with temperature. This leads to a change of the classical values of $h_{1}, h_{2}$ fields~\footnote{It may have resulted not only in variation of the absolute values of particle masses, but also in rearrangement of the particle mass spectrum, which can have interesting cosmological consequences.} and thermal phase transition takes place. In general one can classify the thermal phase transition according to the behavior of the vacuum development during the cooling down. For instance, the phase transition may be of first or second order, one-stage or two-stage process. In Fig.~\ref{fig:potevol} we show three examples that illustrate the different behavior of the vacuum development with temperature, where the temperature decreases from left to right and the true vacuum is marked as a red plus in each graph. The model parameters corresponding to each point are summarized in Table~\ref{tab:threepts}.
For the case shown in the top panel, the vacuum starts to depart from the origin at $T=163\gev$, and then moves closely along the yellow line until reaching the EW vacuum at zero temperature. 
This phase transition is called of second order, because the potential minimum shifts continuously while no potential barrier develops during the cooling down. In contrast, the vacuum of the potential displayed in the middle panel is localized in the vicinity of the origin point at high temperature. When the temperature decreases to $T\simeq 157\gev$ an EW vacuum located away from the origin appears, forming two degenerate vacua separated by an energy barrier. This gives rise to the first-order phase transition from the origin to the EW vacuum, which is indicated by the red arrow. 
In these two examples, the phase transition is termed one-stage.
In addition to experiencing only one standard EWSB phase transition, the 2HDM can undergo a two-stage phase transition as the temperature falls as shown in the lower panel. In this mechanism the first stage is a conventional second order PT in which the symmetry is broken, shortly thereafter follows a first order PT. 
Another remarkable thing is that the ratio of the classical value between the two fields $h_2/h_1$ shown in the upper and middle panels has very little dependence on the temperature. However, in general, the value of $h_2(T)/h_1(T)$ is a temperature-dependent parameter and the change in the temperature growth can even be large in magnitude. In particular, the lower panel displays a peculiar behavior of the ratio $h_2(T)/h_1(T)$ as temperature decreases: at first it monotonically increases, resulting in a deviation of the vacuum from the yellow line, then jumps to its zero temperature value (that is $\tan\beta$) at the transition point and maintains unchanged in the remaining process.

\begin{table}[t]
\caption{Parameters for three benchmark points (with all mass parameters in units of GeV) that lead to different types of EWPT.}
\label{tab:threepts}
\begin{center}
\begin{small}
\vspace*{-5mm}
\hspace*{-5mm}
\begin{tabular}{|c|c|c|c|c|c|c|c|c|c|c|c|c|}
\hline
Points & Properties & $\tan\beta$ & $\sin\alpha$ & $\lambda_{1}$ & $\lambda_{2}$ & $\lambda_{3}$ & $\lambda_{4}$ & $\lambda_{5}$ & $m_{12}$ & $m_{H}$ & $m_{A}$ & $m_{H^\pm}$ \\
\hline
Top & 2nd order PT & 2.98 & -0.24 & 3.99 & 0.29 & 0.86 & -1.06 & 0.11 & 49 & 181 & 35 & 192 \cr
%Middle & 1st order PT & 1.34 & -0.67 & 6.57 & 3.14 & 4.75 & -8.89 & 0.14 & 57 & 508 & 34 & 524\cr
Middle & 1st order PT & 1.30 & -0.59 & 6.05 & 2.00 & 6.63 & -8.27 & -1.27 & 176 & 510 & 376 & 594\cr
Bottom & 2 stage PT & 40 & 0.06 & 0.16 & 0.27 & 4.25 & 1.04 & 1.80 & 59 & 375 & 176 & 232 \cr
\hline
\end{tabular}
\end{small}
\end{center}
\end{table}%

%%%%%%%%%%%%%%%
\section{Numerical procedures: $T_c$ evaluation scheme}
\label{sec:Tceval}

The dynamics of the EWPT is governed by the effective potential at finite temperature \eq{potVeff} in our model.
For purposes of analyzing the temperature evolution of the potential involving both $h_{1}$ and $h_{2}$, it is convenient to work with a polar coordinate representation of the classical fields $h_{1}(T)$ and $h_{2}(T)$. To that end, we define $h(T)$ and $\theta(T)$ via
\begin{eqnarray}
h_{1}(T) &\equiv &h (T)\cos\theta (T), \\
h_{2}(T) &\equiv &h (T) \sin\theta (T).
\end{eqnarray}
The tree-level potential \eq{eq:V0tree} in the $(h,\theta)$ plane becomes 
\begin{align}
\label{eq:potential2}
V_{0}(h,\theta)&= {1\over 8} \left( \lam_{1} \cth^{4} +\lam_{2} \sth^{4} + 2 \lam_{345} \sth^{2} \cth^{2} \right) h^{4} \\
&\quad + {h^{2} \over 4} \left[ {2m^{2}_{12} (\tb -\tth)^{2} \over \tb (1+\tth^{2})^{2}} -v^{2} \left( \lam_{1} \cth^{2} \cb^{2} + \lam_{2} \sth^{2} \sb^{2} + \lam_{345} (\sth^{2}\cb^{2}+\cth^{2} \sb^{2}) \right) \right],
\end{align}
and the remaining parts of the effective potential are much more involved and hence not shown here. 

When a first order PT takes place, a local minimum with $\langle h \rangle\neq 0$ develops and becomes degenerate with the symmetric minimum $\langle h \rangle = 0$ as the temperature decreases, this defines the critical temperature $T_c$, and the two minima are separated by a potential barrier. Therefore, the evaluation of $T_{c}$ is of great importance in studying the EWPT and its cosmological consequences. 
A straightforward approach is to decrease the temperature by small steps and make a potential plot (like Fig.~\ref{fig:potevol}) at each step. Then the global minimum of the potential (starting from the EW vacuum at zero temperature) can be followed step-by-step and the critical point is found once the minimum displays a jump rather than a smooth transition. Obviously, this graphic method is feasible only for benchmark points but is barely applicable for extensive scan due to its non-numerical nature. 
To date several numerical methods have been developed. In Ref.~\cite{Dorsch:2013wja}, going from zero temperature to higher temperature, the critical point is taken to be the last one for which the minimum lied below the origin. This approach is no able to resolve the 2-step phase transitions where the potential experiences a second order PT prior to the first order PT, giving rise to a vacuum shift from the origin at higher temperature.
To overcome this problem, the authors of~\cite{Basler:2016obg} used advanced numerical algorithms to search for the global minimum of the effective potential in the $(h, \theta)$ plane for each temperature.
We employ a method consisting of the following procedures:

First, we deal with points for which the ratio $h_2/h_1$ is (approximately) temperature-independent, that is $\theta (T) = \beta$. The effective potential \eq{potVeff} reduces to a function of two parameter -- temperature $T$ and $T$-dependent field norm $h(T)$. In this case, one can easily determine the critical temperature $T_c$ and the field norm $v_{c}$ at which the potential reaches a minimum by solving the equations 
\beq
\label{eq:Tceval}
\left.  \frac{\partial}{\partial h}V_{\mathrm{eff}}(h
,T_{c})\right\vert _{h=v_{c}}=0, \ \ \ V_{\mathrm{eff}}(h=v_{c}
,T_{c})=V_{\mathrm{eff}}(h=0,T_{c}).
\eeq
In searching for the solution of the above equations, we require a difference between the potential at the minimum and its value at the origin smaller than $10^{-10}$GeV$^4$.
As a consequence, the solution for $v_{c}$ would be a value close to zero if there were no degenerate minima of the effective potential present in the process of temperature drop. This means that below a certain small value of $v_{c}$ we do not expect a decent probability of achieving a first-order phase transition.
Instead, very likely such points lead to a second-order phase transition. For this reason we employ a technical cut $v_{c} > 1\gev$ in order to remove these points.

Next, we are going to deal with the points that exhibit an explicit temperature dependence for the ratio of two fields. 
This type of points often lead to a 2-stage phase transition~\cite{Land:1992sm,Hammerschmitt:1994fn}, as illustrated in the last row of Fig.~\ref{fig:potevol}.
There must exist a global minimum for which $\tan\theta(T) = \tan\beta$ is not obeyed at a certain temperature or within a small temperature interval. 
In this situation, $(v_{c},T_{c})$ obtained as a solution of \eq{eq:Tceval} is not the critical vev and temperature where the phase transition occurs because the true vacuum is no longer located at the origin. Searching for the global minimum should be performed not along the $\tan\beta$ line but on a two-dimensional $(h,\theta)$ space. We employ an algorithm which uses the steepest descent method to find the global minimum of the effective potential. 
At $T_{c}$ the searched minimum is then compared with the value of the effective potential evaluated at $v_{c}$ and the one with the lower value is chosen as the candidate for the global minimum. 
For the general 2-stage PT, tracing the (temperature) evolution of the global minimum on a 2D plane is inevitable.
Here, we discard points leading to a 2-stage PT and focus on the scenarios featuring a 1-stage PT.
In the following analysis, we only retain parameter points with $T_c \leq 300\gev$.\footnote{It appears possible that the potential has a global minimum at large value of $h$. However, the probability of having a strong phase transition for these points is quite low, unless high scale phase transition is considered.}

%%%%%%%%%%%%%%%
\section{Properties of the first order EWPT}
\label{sec:res}

The strength of the phase transition is quantified as the ratio of the norm of the neutral fields to the temperature at the critical point,
\beq
\label{eq:xidef}
\xi = \frac{v_c}{T_c}.
\eeq
Here $v_c=\sqrt{ \langle h_{1}\rangle ^2+ \langle h_{2}\rangle^2+\langle A\rangle^2}$,~\footnote{Since we restrict ourselves to a CP-conserving model, the global minimum has $\langle A\rangle =0$. In general there may exist local minima that are CP-violating, while a recent study~\cite{Basler:2016obg} found that it is always vanishes up to numerical fluctuations at both $T=0$ and $T=T_c$.} in general, represents the value of the norm of all scalar fields involved at the broken vacuum at critical temperature $T_c$.
Note that when interpreting the ratio as the strength of the electroweak phase transition, one should be aware of its gauge dependence~\cite{Dolan:1973qd,Patel:2011th,Wainwright:2011qy,Garny:2012cg}.
In order to ensure that a baryon number generated during the phase transition is not washed out, 
a strong first-order phase transition is demanded and occurs if $\xi \geq1$~\cite{Moore:1998swa}.~\footnote{The choice of the washout factor is subject to additional uncertainties. It was argued that the EW sphaleron is not affected much if extra degrees of freedom are SM-gauge singlets~\cite{Patel:2014} but the situation in the presence of an additional doublet is unclear yet. As a more conservative choice,  other criterion such as $\xi \geq 0.7$ was also taken in other works.}

Before presenting the main results, we discuss the specific features of the parameter space compatible with the theoretical and experimental constraints and at the same time leads to first order and second order phase transition. 
We will show results for both \typei~and \typeii~models.

%%%%%%%%%%%%%%%
\subsection{first order vs. second order phase transition}

\begin{figure}[t]
\begin{center}
\includegraphics[width=0.38\textwidth]{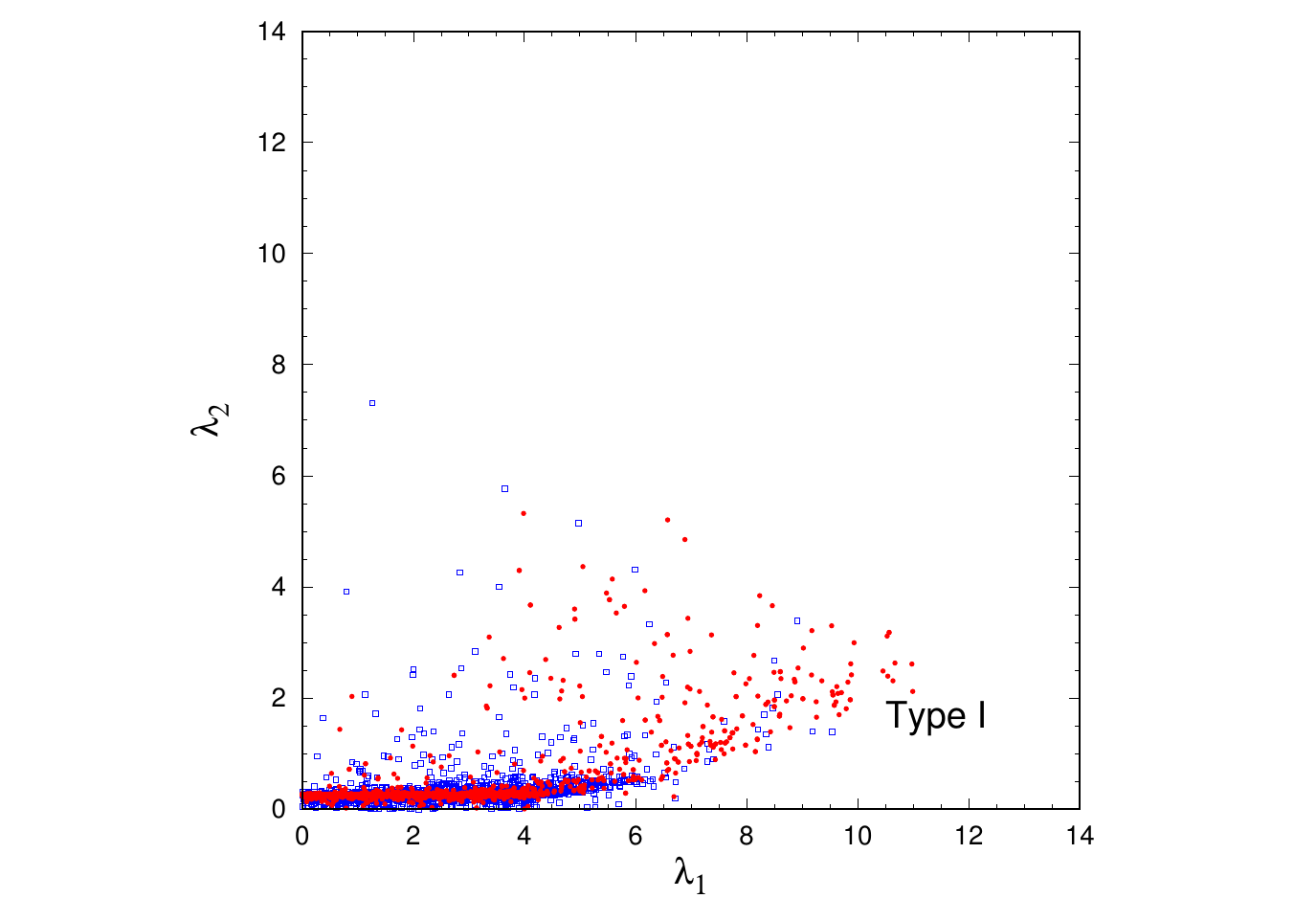}
\hspace{-15mm}
\includegraphics[width=0.39\textwidth]{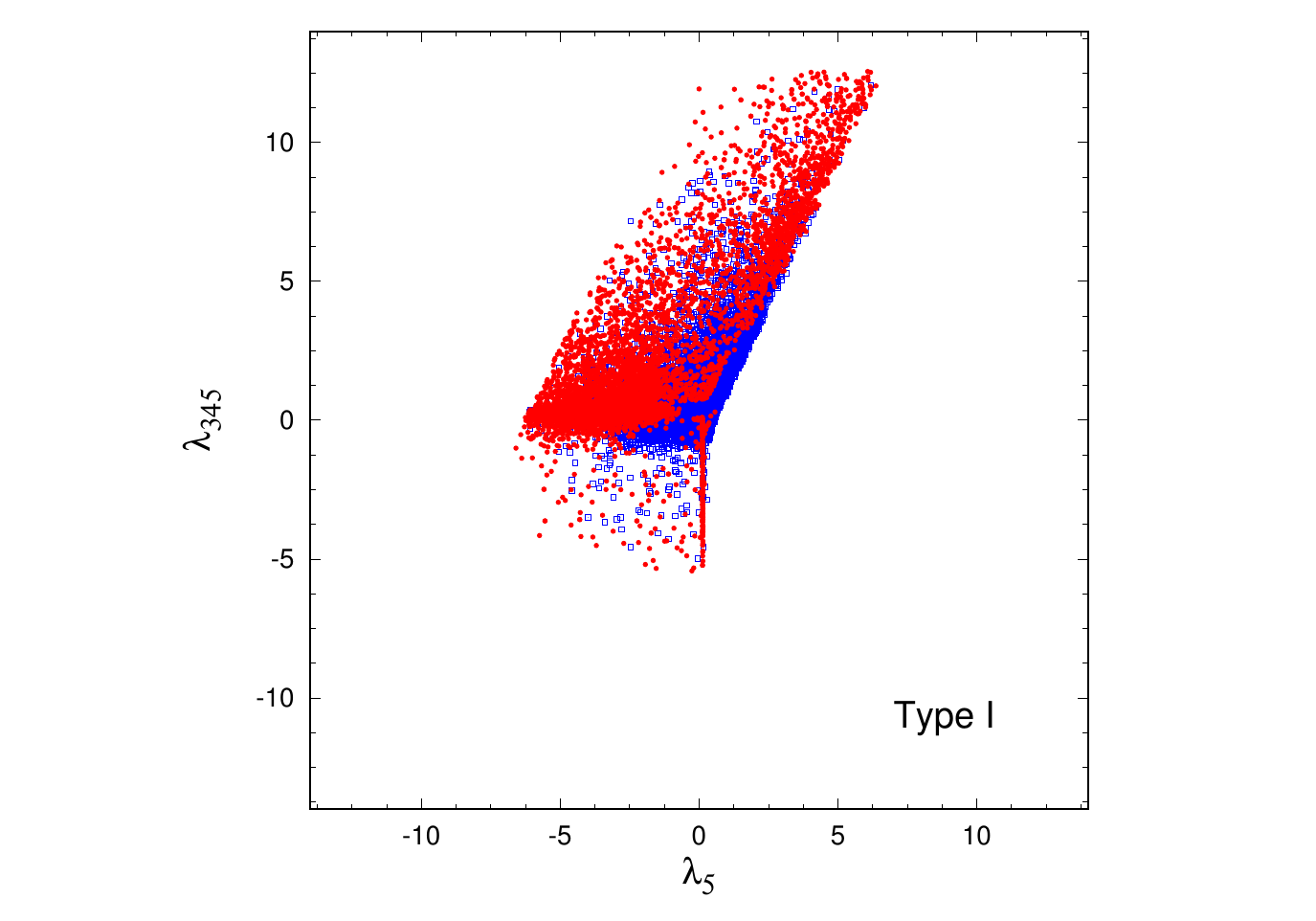}
\hspace{-15mm}
\includegraphics[width=0.38\textwidth]{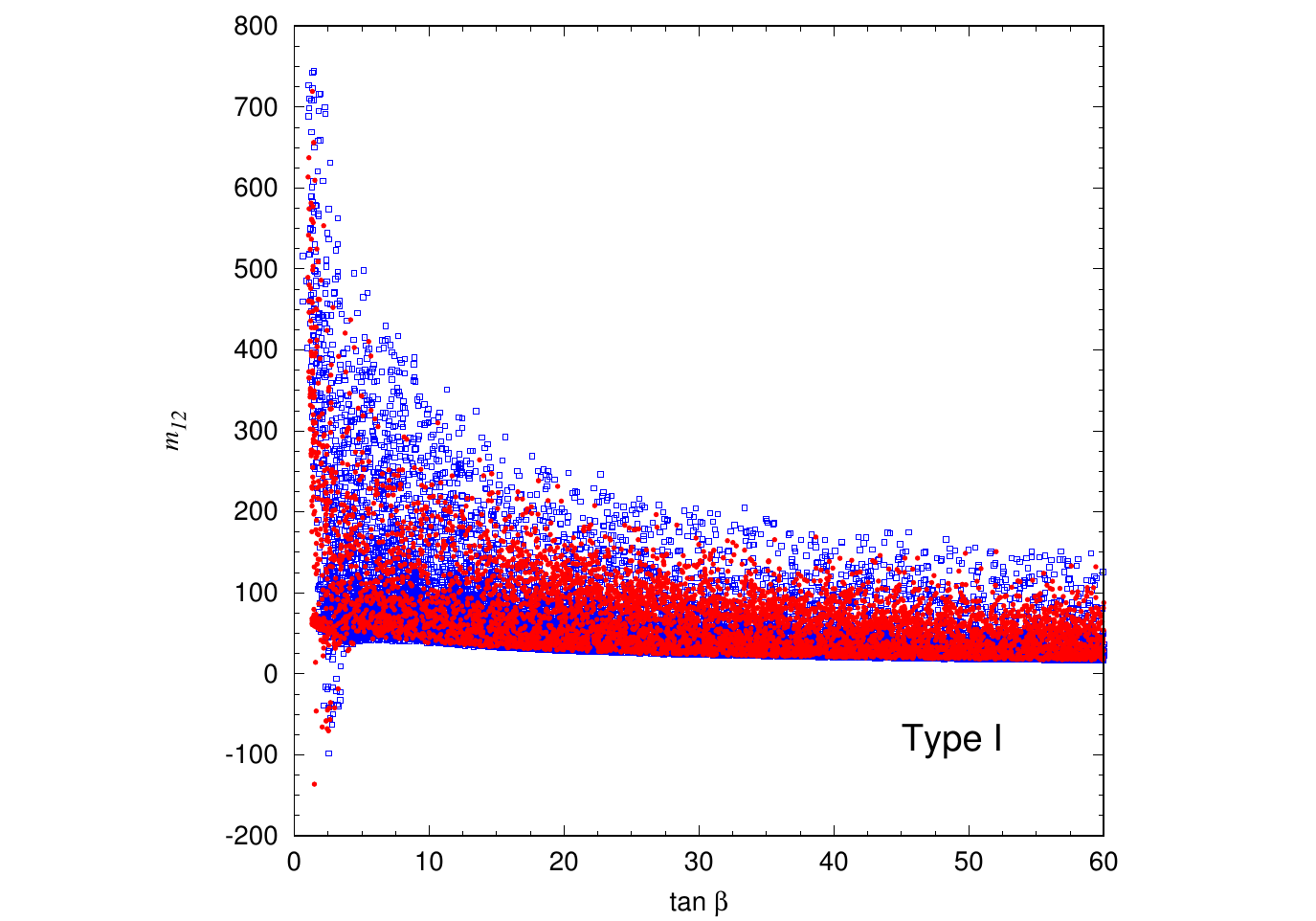}\\
\includegraphics[width=0.38\textwidth]{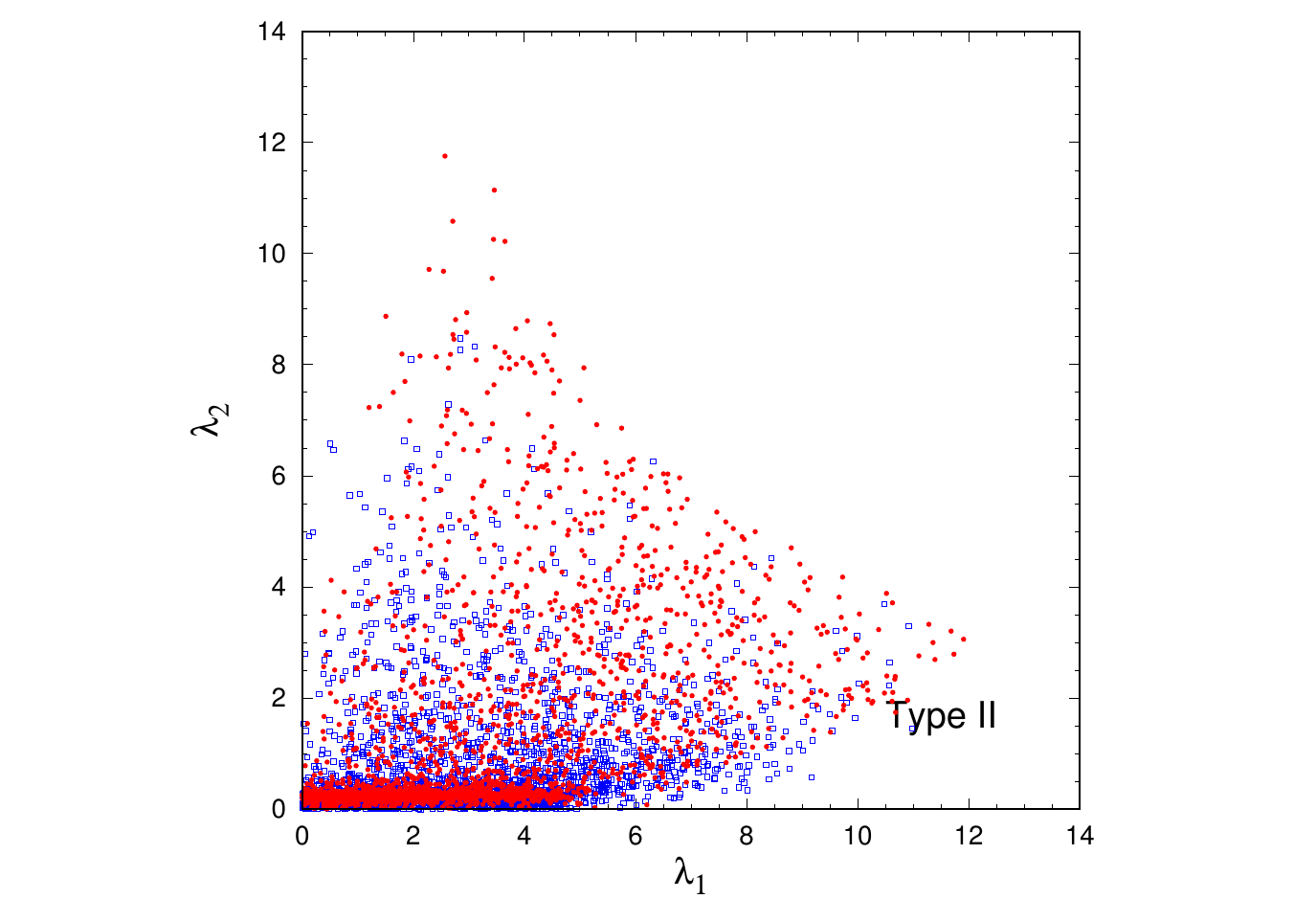}
\hspace{-15mm}
\includegraphics[width=0.39\textwidth]{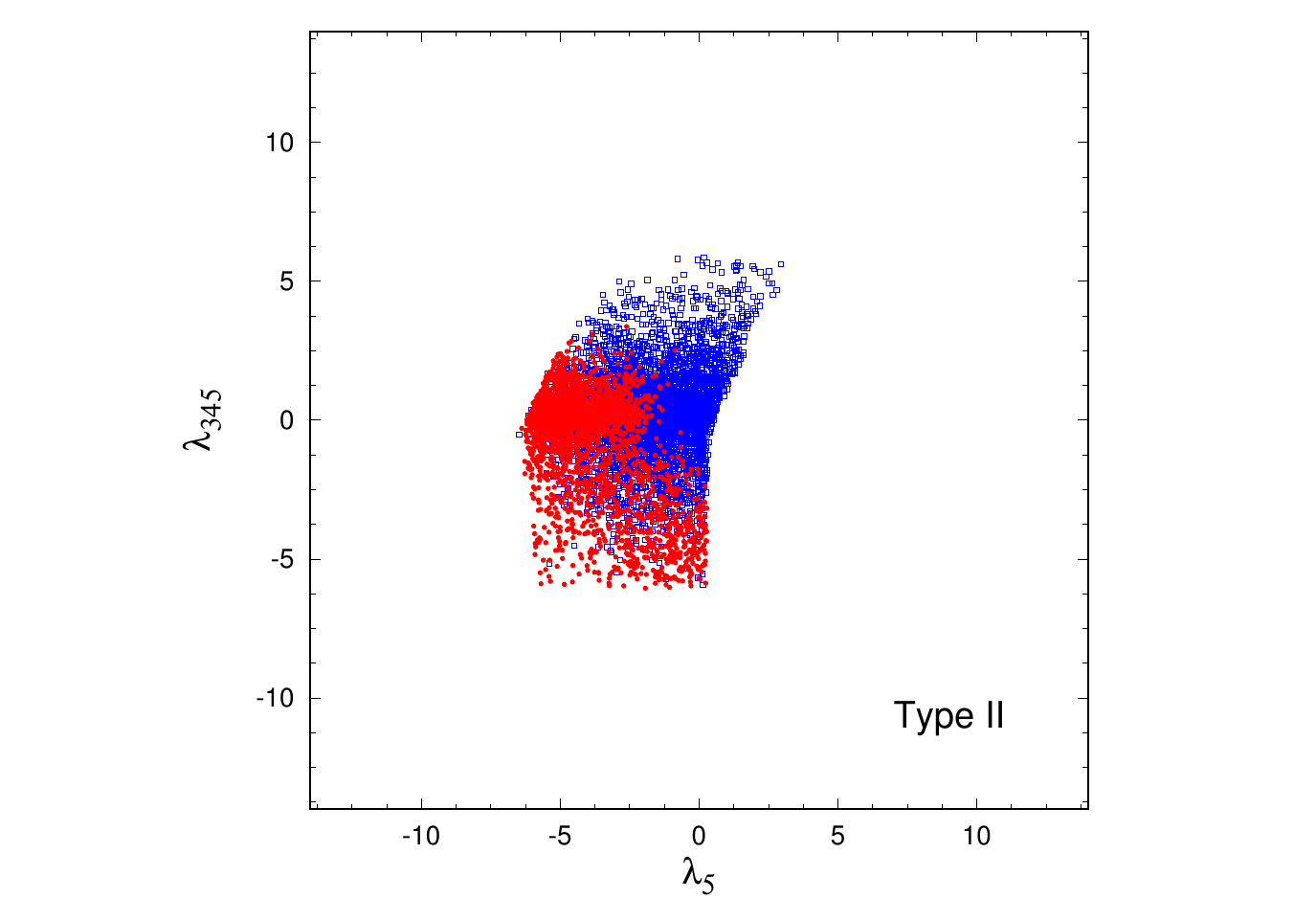}
\hspace{-15mm}
\includegraphics[width=0.38\textwidth]{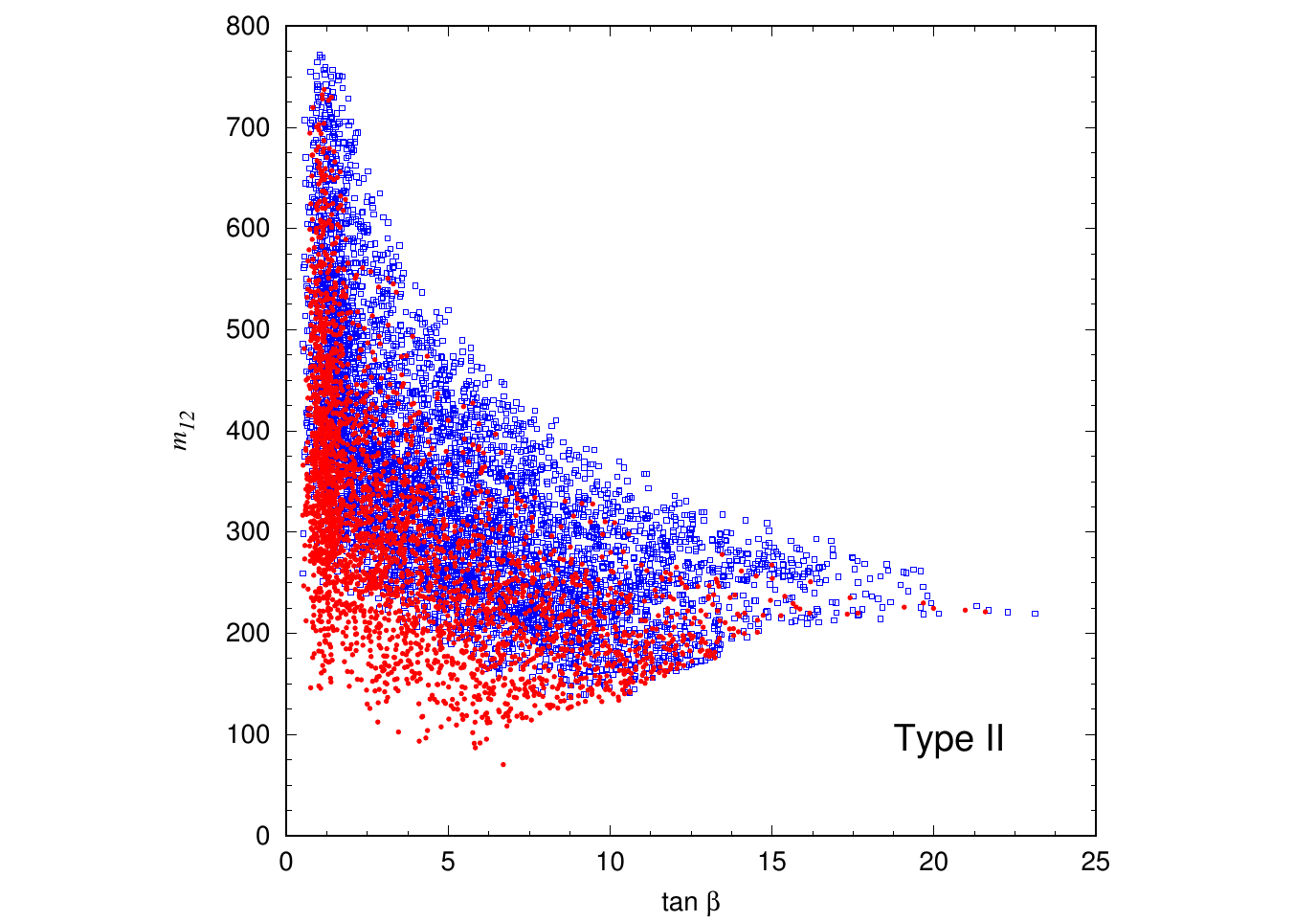}
\caption{The mapping of the first order (red circles) and second order (blue boxes) PT points on the 2D space of model parameters: $\lam_{1}$ vs. $\lam_{2}$ (left), $\lam_{345} $ vs. $\lam_{5}$ (middle) and $m_{12}$ vs. $\tan \beta$ (right). 
Only $T_c \leq 300$ GeV points are retained. Note that all the points with $\tan\beta >25$ in \typeii~model have been excluded by the $H,A \to \tau\tau$ bounds~\cite{Aaboud:2017sjh,CMS:2016pkt}.}
\label{fig:PTcomp}
\end{center}
\end{figure}

It has been shown in Fig.~\ref{fig:potevol} that both first order and second order phase transition can take place in the 2HDM. Whether first order or second order PT is developed depends on the mass spectrum among the three extra Higgs bosons, which is directly related to the five quartic couplings $\lam_i$ and the soft symmetry breaking parameter $m^2_{12}$ through Eqs.~(\ref{eq:hmass})-(\ref{eq:Hcmass}). Thus, it would be very interesting and useful if one can divide the entire model parameter space into different sectors where distinct dynamics of vacuum evolution leading to first order and second order PT take place.
An initial attempt along this direction was made in~\cite{Ginzburg:2009dp} in accordance with the general geometric analysis of~\cite{Ginzburg:2004vp}. In~\cite{Ginzburg:2009dp} the authors introduced several discriminators in terms of certain combinations of $\lam_{i}$ and succeeded in dividing into four sectors which do not overlap. However, the analysis conducted in~\cite{Ginzburg:2009dp} is oversimplified, only the effect of thermal mass corrections was included. When considering the full effective potential \eq{potVeff}, such division may be highly difficult or even impossible, which is reflected in Fig.~\ref{fig:PTcomp} where we map our first order (red) and second order (blue) PT points in the 2D space of model parameters and none of the parameters exclusively distinguish the two types of PT points.~\footnote{We examined that none of the discriminators defined in~\cite{Ginzburg:2009dp} can effectively isolate the first order (red) and second order (blue) PT points when considering the full effective potential.}
As expected, $\lam_{1}$ and $\lam_{2}$ have marginal influence since both of them only enter into the masses of two neutral CP-even scalars. On the contrary, $\lam_{3,4,5}$ can be potentially used as discriminators as they affect the masses of three extra Higgs bosons simultaneously. For instance, $\lam_5$ is bounded from -6 to 6~(2) in \typei~(II) model and a small value of $\lam_{5}$ tends to induce a second order PT unless the sum $\lam_{345}$ is negatively large. 

%%% Fig.6 

%%%% Fig.7 %%%%%
\begin{figure}[t]
\begin{center}
\includegraphics[width=0.52\textwidth]{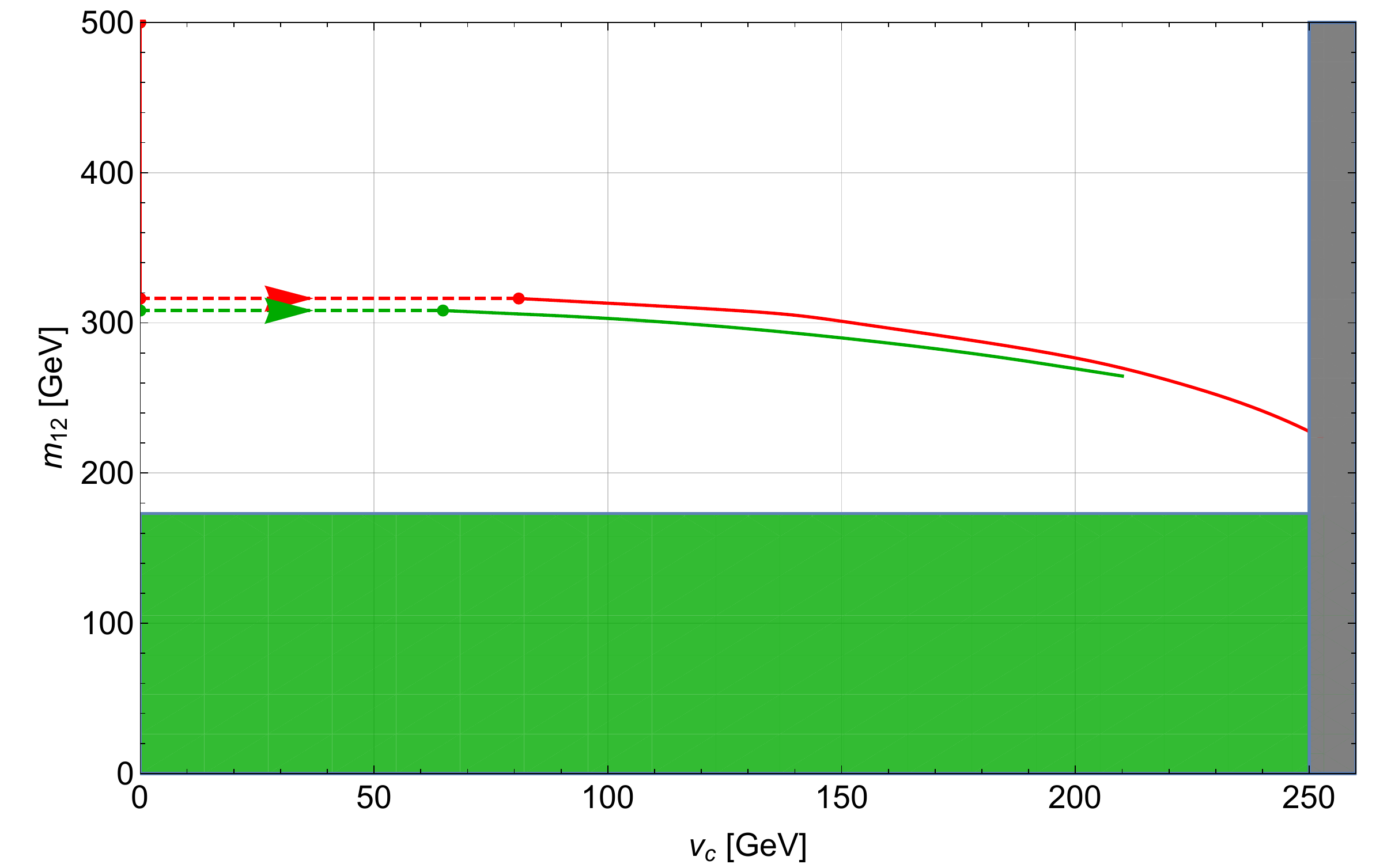}
\caption{The evolution of the critical vev $v_{c}$ as function of $m_{12}$. The alignment limit and a common mass scale among the three BSM states $M=600\gev$ are assumed. The dashed line with an arrow indicates the jump from the second order PT to the first order PT. Red and green curves represent $\tan\beta=1$ and $1.5$, respectively and terminate at which a proper EWSB at one loop level does not happen at zero temperature. In the gray-shaded region $v_{c}$ exceeds the EW vacuum $v=246\gev$ and in the green-shade region at least one of the $\lambda$'s (mostly $|\lam_{1}|$ or $|\lam_{2}|$) exceeds the perturbativity bound (i.e. $4\pi$) for the $\tan\beta=1.5$ case.}
\label{fig:m12hcplot}
\end{center}
\end{figure}

\begin{figure}[t]
\begin{center}
\includegraphics[width=0.44\textwidth]{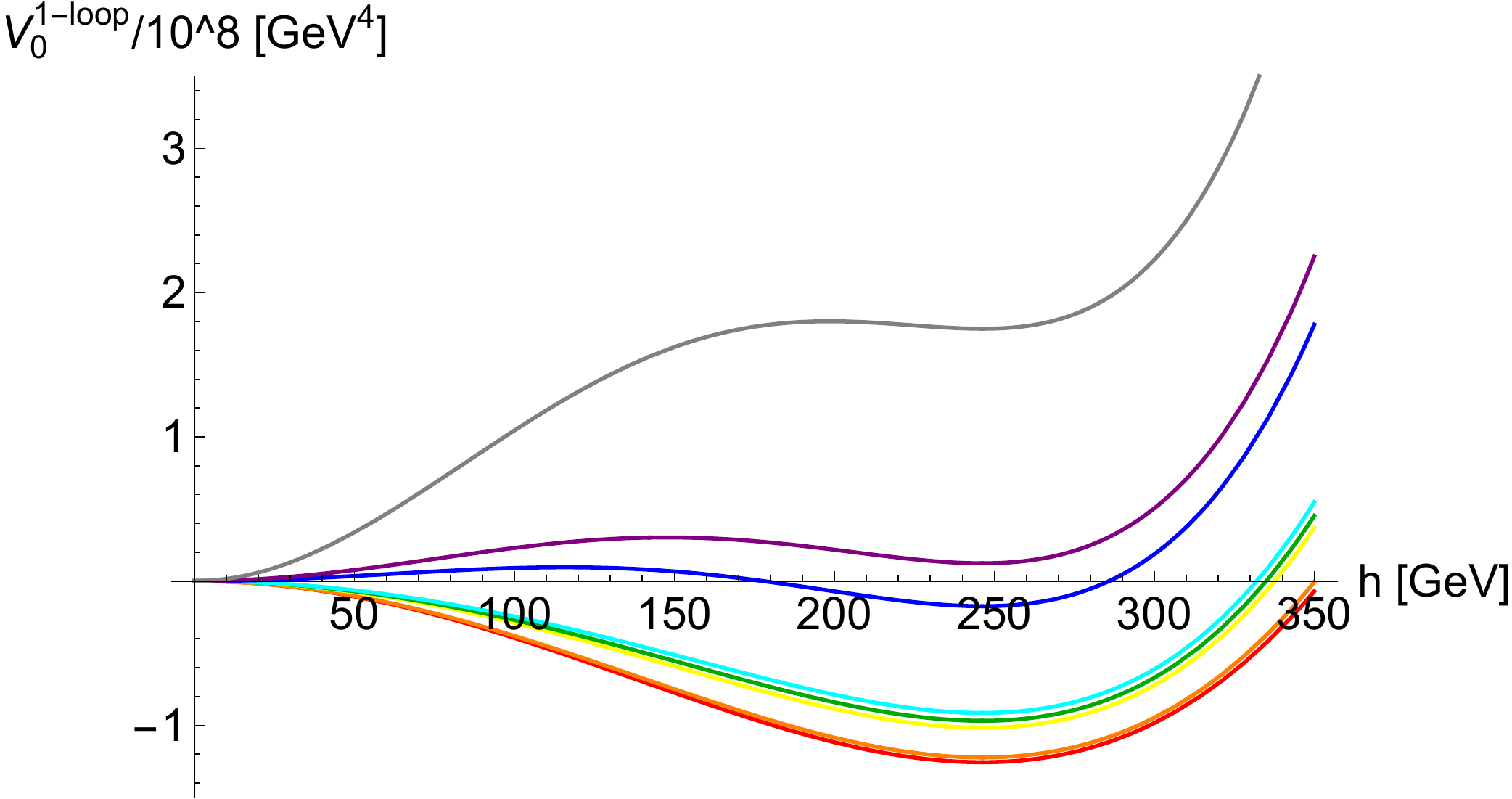}
\hspace*{-5mm}
\includegraphics[width=0.52\textwidth]{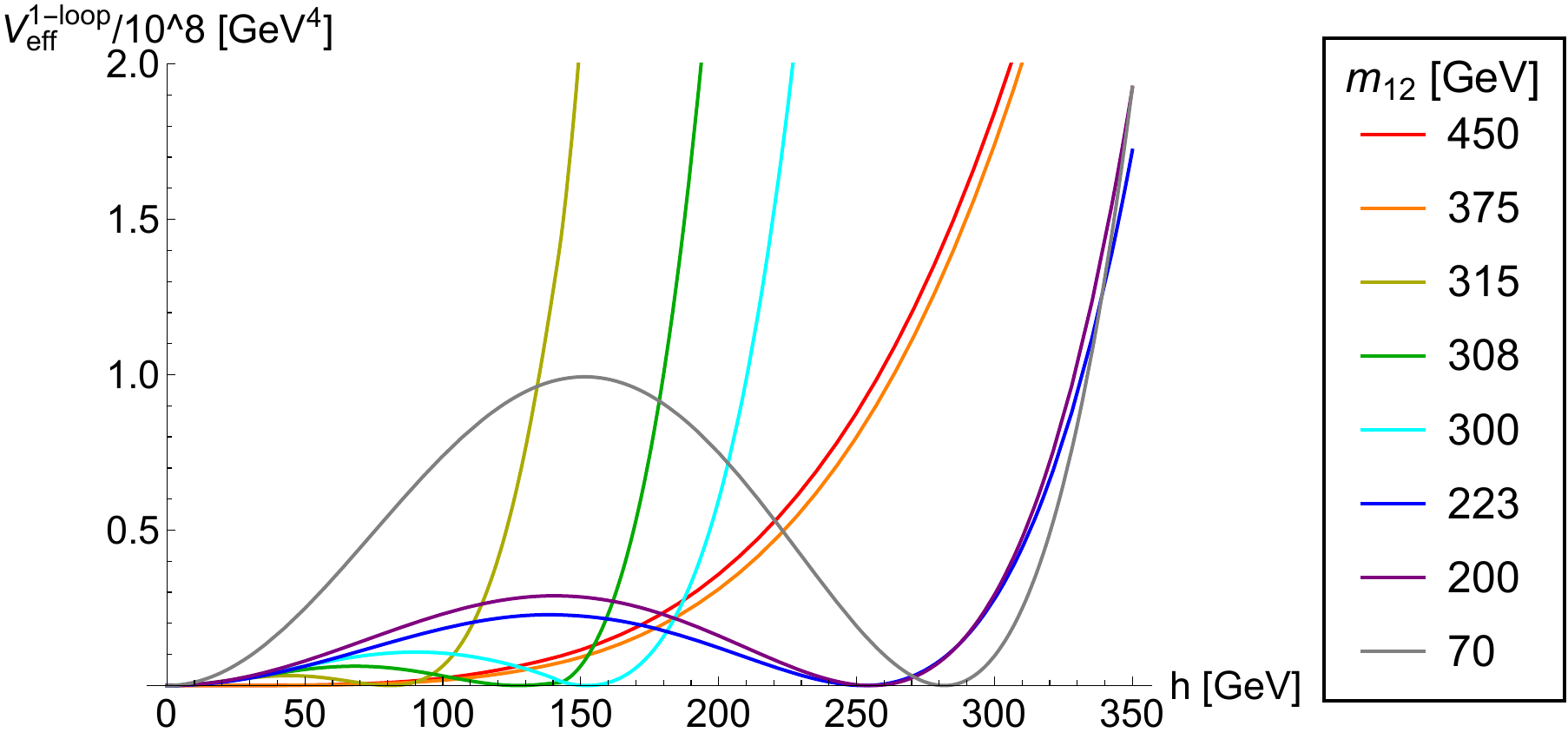}\\[5mm]
\caption{The 1-loop potential curve at zero temperature (left) and finite temperature effective potential evaluated at the critical temperature (right) for $\tan\beta=1$ and various values of $m_{12}$ given in the legend. As Fig.~\ref{fig:m12hcplot}, the mass of three BSM Higgs states are commonly fixed at $600\gev$ and $\sin(\beta-\alpha)=1$ is assumed.}
\label{fig:V0Vfullplot}
\end{center}
\end{figure}

Another important observation is that for a given value of $\tan\beta$, larger $m_{12}$, allowing for larger $m_{H}$, favours a second order PT. 
This points to the fact that the phase transition in the theory degrades to the SM case when the new scalars reside in a decoupled sector, as expected intuitively. 
In reverse, it has the implication that first order PT is more probable for a small or modest value of $m_{12}$ when $m_{H}$ is fixed. To illustrate this, we evaluate the phase transition properties in the process of slowly varying $m_{12}$, assuming the alignment limit and a common mass scale among the three BSM states $M=600\gev$ for simplicity. 
The situation is shown in Fig.~\ref{fig:m12hcplot}, where red and green curves represent $\tan\beta=1$ and $1.5$, respectively.
This plot can be used to track the evolution of the critical vev $v_{c}$: it starts from zero (in the second order PT stage) at large $m_{12}$ to a non-zero value (in the first order PT stage). The jump from the second order PT to the first order PT is indicated by a dashed line with an arrow. 
Notably, a severe fine-tuning on $m_{12}$ is required for a successful first order PT and the $v_{c}$ value approaches the EW vacuum at smaller $m_{12}$.
This interesting behavior is explicitly illustrated in Fig~\ref{fig:V0Vfullplot} which gives, for $\tan\beta=1$, the 1-loop potential curve at zero temperature (left) and the finite temperature effective potential evaluated at the critical temperature (right) for various values of $m_{12}$.
As $m_{12}$ decreases, thermal effects generate a higher potential barrier and simultaneously push the degenerate vacuum towards the EW vev $v$, giving rise to a growth in $\xi$ (owing to the small fluctuation on $T_{c}$ in the stage of the first order PT). On the other hand, a smaller contribution from the $m_{12}^{2}$ term to the tree-level and 1-loop potential at zero temperature will remove the potential barrier. For example, the SM potential $V\sim \lam h^{4}$ when the mass term $\mu^{2}\to 0$. Consequently, the desired vacuum disappears, resulting in a terminal value of $v_c$ near $v$, as we will also see in Fig.~\ref{fig:vcTc}.
Furthermore, the effect of increasing $\tan\beta$ on the phase properties is also visible by comparing the red and green curves. For a larger $\tan\beta$ and the same mass spectrum, the first order PT is realised at a lower value of $m_{12}$ and in the meanwhile the `no-EWSB' situation takes place at a smaller value of $v_{c}$.

%%%% Fig.8 %%%%%

As also seen in Fig.~\ref{fig:PTcomp}, most of our points have $\tan\beta$ close to one, which agree well with the findings of previous studies~\cite{Dorsch:2013wja}. Yet we would like to clarify that such preference is absolutely not the consequence of requiring a (strong) first order PT. The underlying reason is that in the vicinity of $\tan\beta \simeq 1$, a large range of $m^{2}_{12}$ satisfying the theoretical constraints outlined in Sec.~\ref{subsec:theoconst} is allowed.~\footnote{The correlation between $\tan\beta$ and $m^{2}_{12}$ were discussed in details in Ref.~\cite{Jiang:2017}.} Oppositely, $m^{2}_{12}$ is strongly constrained in the high $\tan\beta$ region and a fine-tuning is required, which will greatly increase the difficulty of accumulating the points by means of random scan. Numerically, very limited range of $\tan\beta$ is allowed for large $m^{2}_{12}$.

%%%%%%%%%%
\subsection{Properties of the first order EWPT}

%%%% Fig.9 %%%%%
\begin{figure}[t]
\begin{center}
\includegraphics[width=0.49\textwidth]{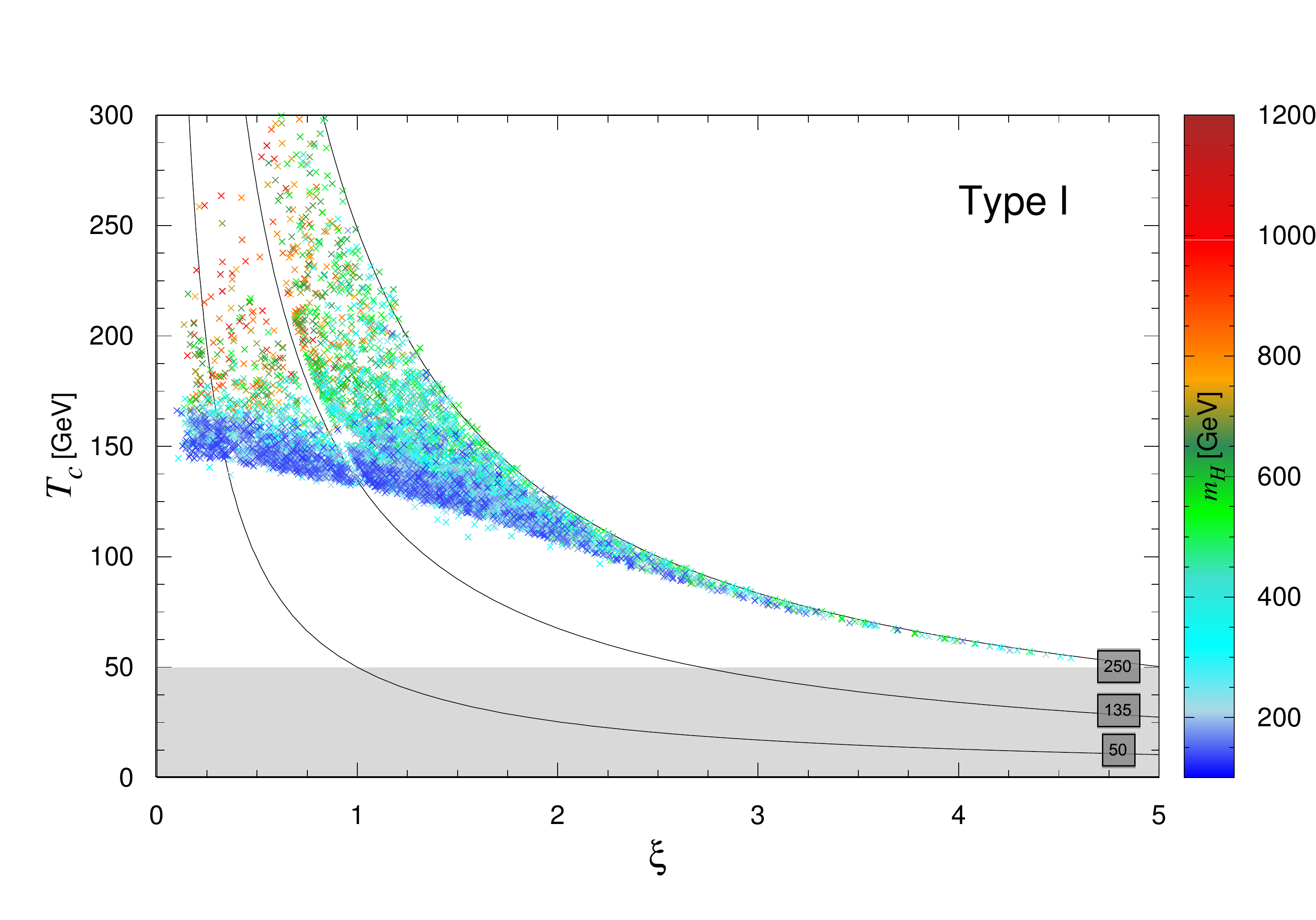}
\hspace*{-4mm}
\includegraphics[width=0.49\textwidth]{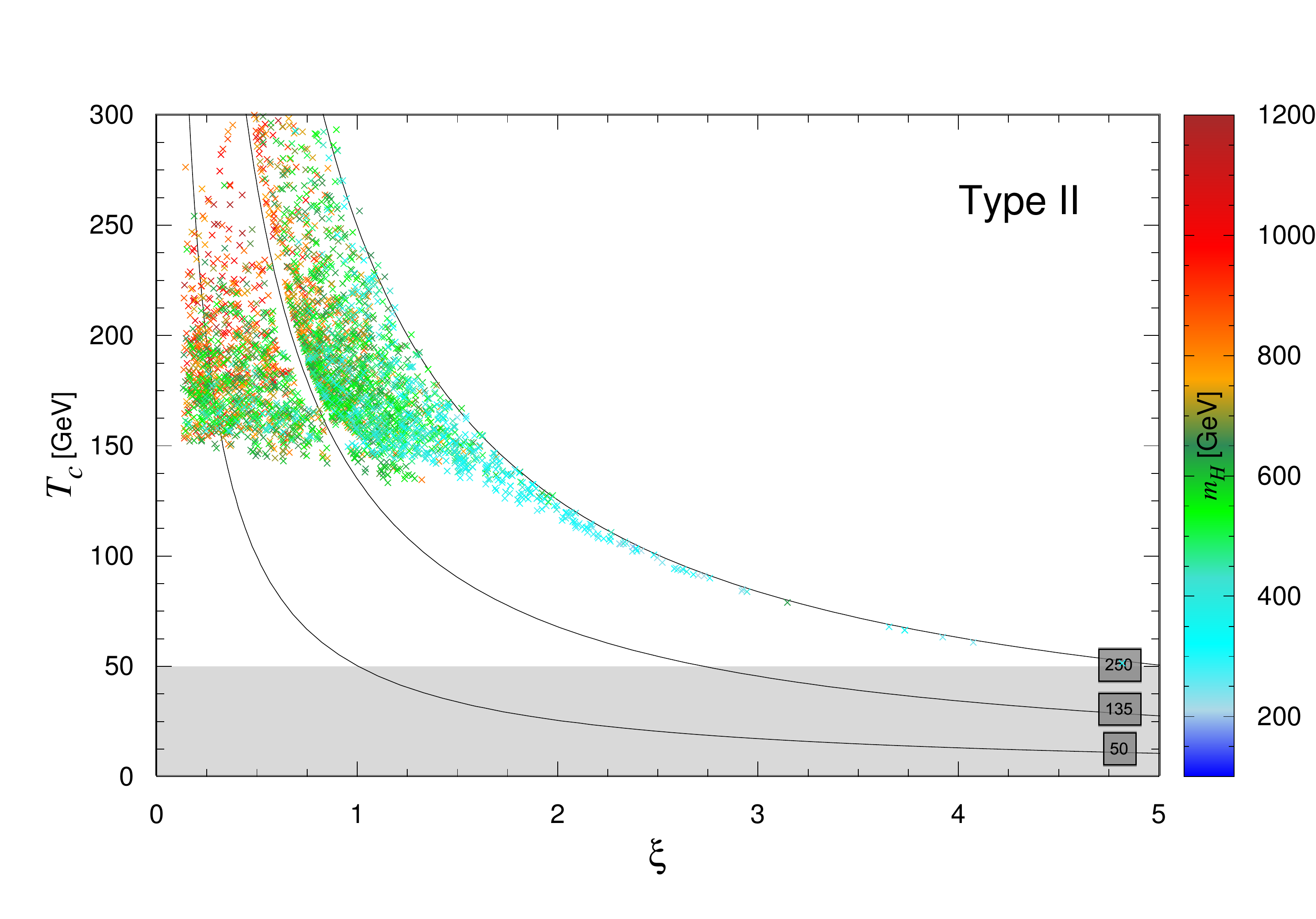}\\
\hspace*{0.5mm}
\includegraphics[width=0.49\textwidth]{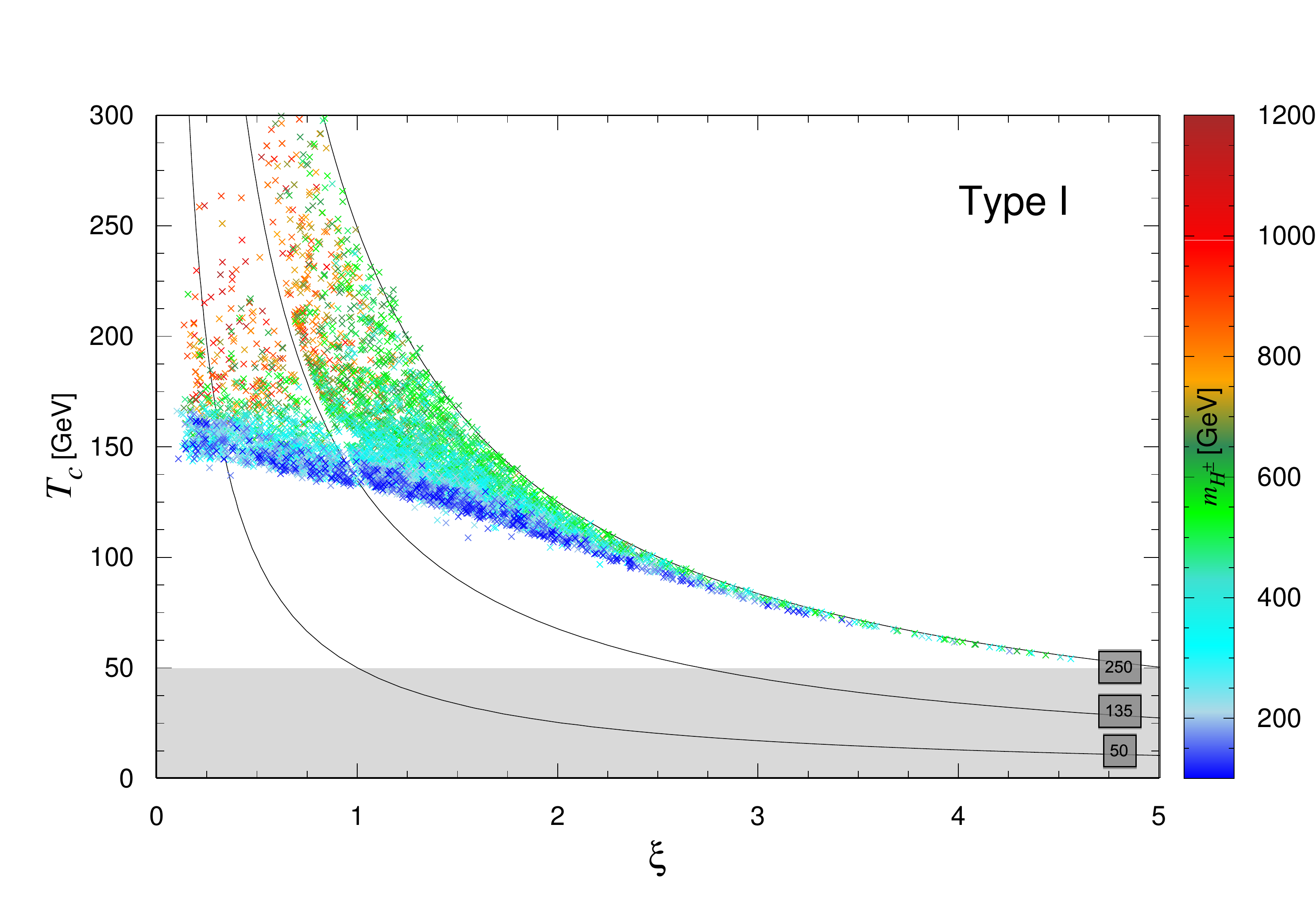}
\hspace*{-4mm}
\includegraphics[width=0.49\textwidth]{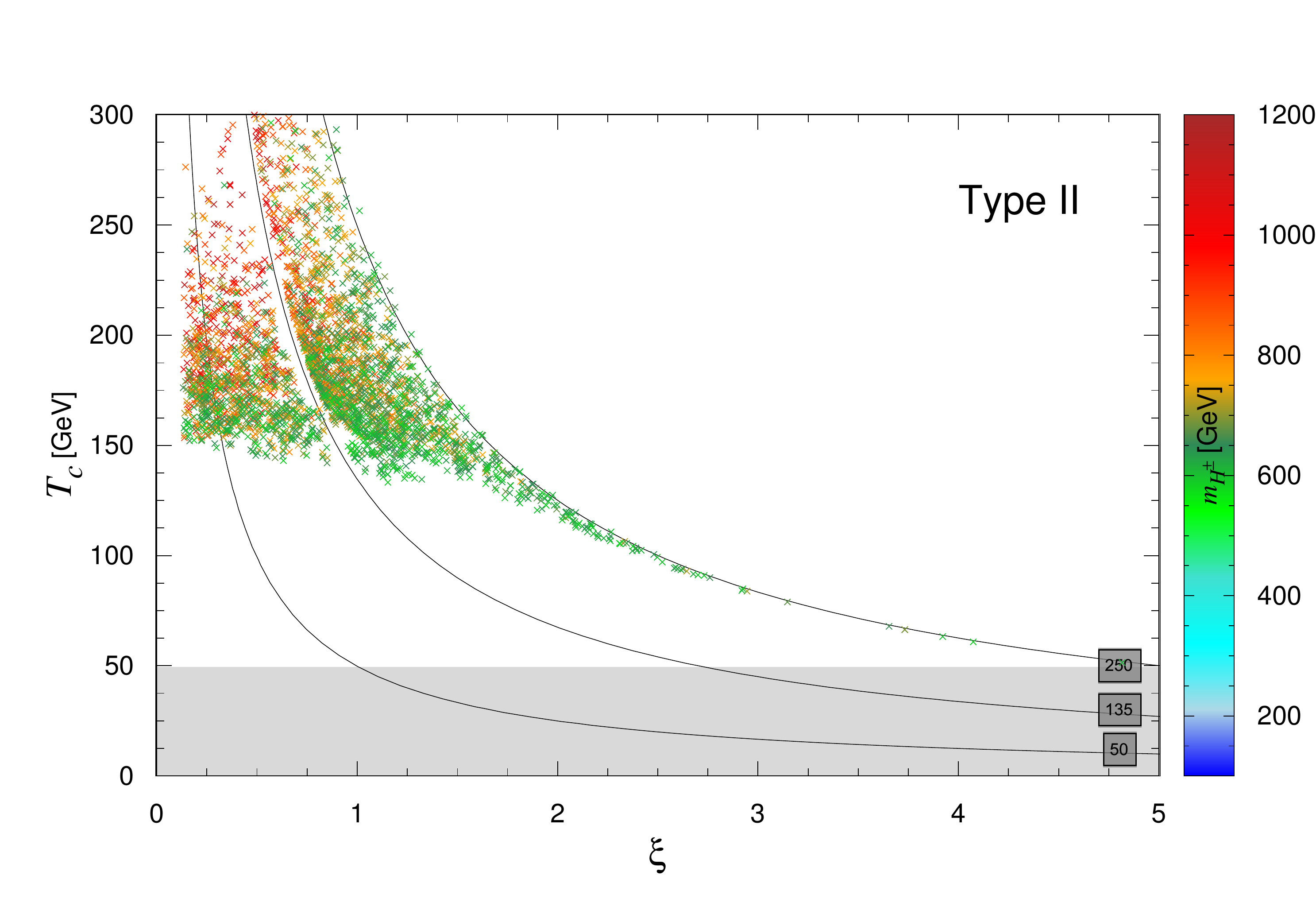}
\caption{Properties of first order PT in the 2HDM. Only $T_c \lesssim 300\gev$ points are retained. Three black contours from top to down correspond to $v_{c}=250, 135, 50\gev$. The value of $m_{H}, m_{H^{\pm}}$ is color coded as indicated by the scales on the right of the plots in the upper and lower panel, respectively.}
\label{fig:vcTc}
\end{center}
\end{figure}

We now turn to discuss the general properties of the first-order PT accomplished in the 2HDM.
The crucial parameters of the phase transition include the critical temperature $T_c$, the field value $v_c$ at $T_c$ and their ratio $\xi = v_{c}/T_{c}$ which is used as a measure of the strength of the EWPT. 
In Fig.~\ref{fig:vcTc} we display in the $(T_{c}, \xi)$ plane the points consistent with all theoretical constraints on the potential and up-to-date LHC limits at Run-2. Three black contours from top to down correspond to $v_{c}=250, 135, 50\gev$. 
We first discuss the impact of extra scalars in the spectrum. Suppose all extra scalars are heavy (\textit{i.e}, above $800 \gev$) and thus their masses are highly degenerate required by the EWPD (see Fig.~\ref{fig:LHCbounds}), then the sector consisting of the new scalars decouple from the SM Higgs and the dynamics of phase transition behaves like the SM. Of course, the strength of EWPT is not closely related to the masses of any of additional Higgs bosons but more directly linked to the mass splittings among them, which can be explicitly visualized in Fig.~\ref{fig:vcTcMass} presented later. 

A general tendency observed is that $v_{c}$ is more constrained as $T_c$ decreases. 
In the extreme case of $T_c \lesssim 100\gev$, the thermal effect, while still playing the role of lifting the effective potential and forming two degenerate minima, is too weak to compete with the zero-temperature loop corrections to the potential. As a result, the critical classical field value is mostly localized around $v$, which makes it slowly vary with respect to the temperature change. Nonetheless, $v_{c}$ shown in Fig.~\ref{fig:vcTc} does not exceed the zero temperature EW vacuum value $v$ owing to the EW vacuum run away (`no-EWSB' bound) as sketched in Fig.~\ref{fig:m12hcplot}, implying that the PT strength $\xi$ necessarily improves at low $T_{c}$. More quantitively, this leads to a maximum PT strength $\xi \simeq5$ at $T_{c} = 50\gev$, and, on the other hand, implies an upper bound on $T_c$ at $250 (350)\gev$ for $\xi \geq 1 (0.7)$~\footnote{This result supports us to efficiently place a cut $T_c \lesssim 300\gev$ in the analysis.}. 
In addition, we observe that a lower bound on $T_{c}$ for each value of $v_{c}$. 
For the value of the critical classical field $v_{c}$ being slightly away from the EW vacuum $v$, 
the lower bounds on the critical temperature would be around $T_{c} \gtrsim 100\gev$ in \typei~and the lower bound on $T_{c}$ in \typeii~model is slightly raised due to the lack of $m_{A}\leq 350\gev$ points.
We stress that this is an useful finding that one can utilize to greatly optimize the algorithm for the evaluation of $T_{c}$.
Last, we point out that the extremum, if coexisting in the vicinity of $v_c \simeq 135\gev$, often develops to a local maximum (corresponding to a barrier) rather than a local minimum of the potential, which causes a narrow gap dividing the displayed points into two parts. 

\begin{figure}[t]
\begin{center}
\includegraphics[width=0.57\textwidth]{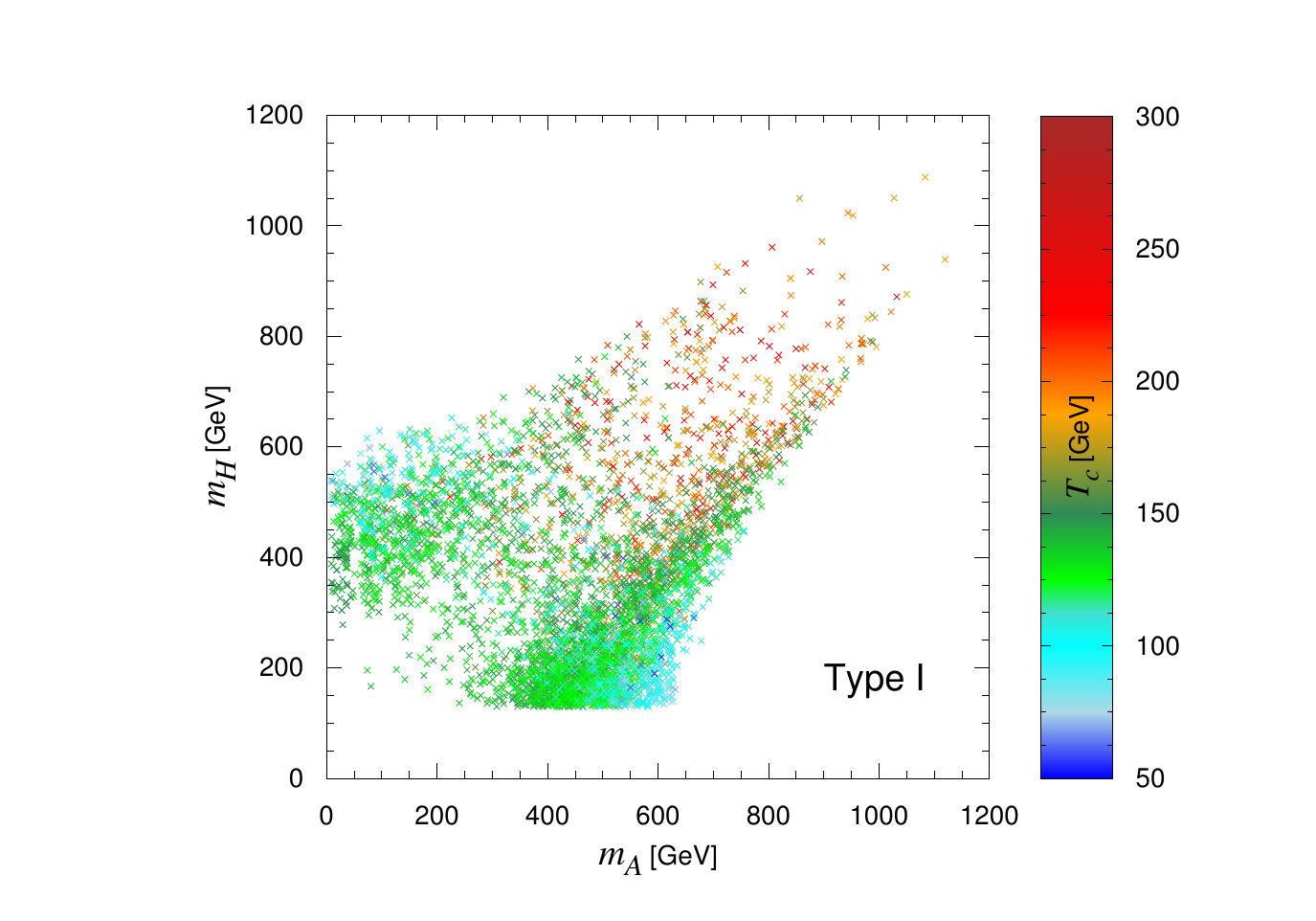}
\hspace{-24mm}
\includegraphics[width=0.57\textwidth]{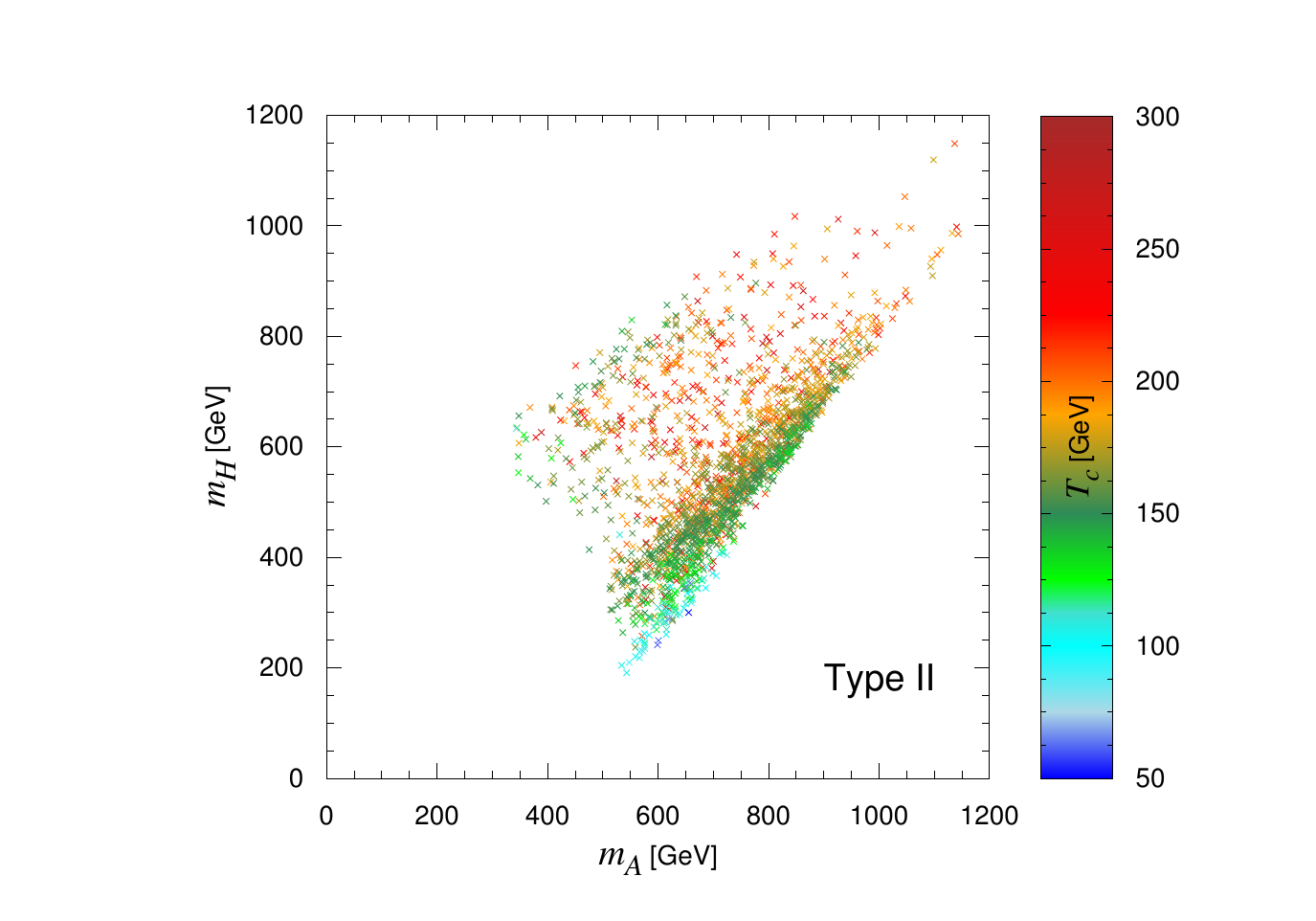}
\caption{We present the critical temperature in the mass spectrum of the model. Only $T_c \lesssim 300\gev$ points obeying the strong first order EWPT condition $\xi \geq 0.7$ are drawn. 
}
\label{fig:StrEWPT}
\end{center}
\end{figure}

An explicit dependence of the critical temperature $T_{c}$ on the mass spectrum of the three extra Higgs bosons can be visualized in Fig.~\ref{fig:StrEWPT}, where we display all points that pass the applied constraints as in Fig.~\ref{fig:vcTc} and additionally fulfill a strong first order EW phase transition (\textit{i.e.}, $\xi \geq 1$). 

%%%% Fig.10 %%%%%

\begin{figure}[t]
\begin{center}
\includegraphics[width=0.49\textwidth]{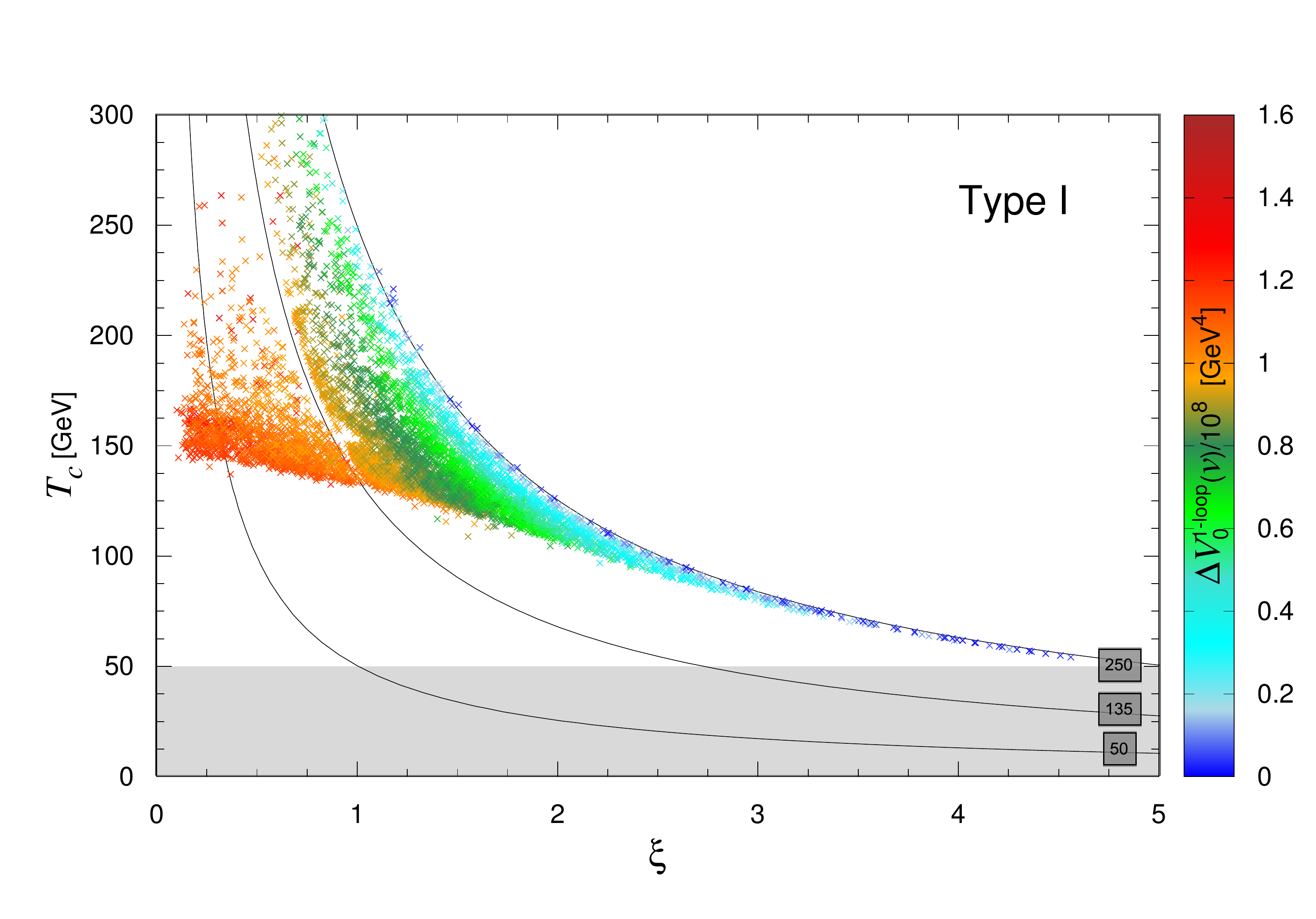}
\hspace*{-4mm}
\includegraphics[width=0.49\textwidth]{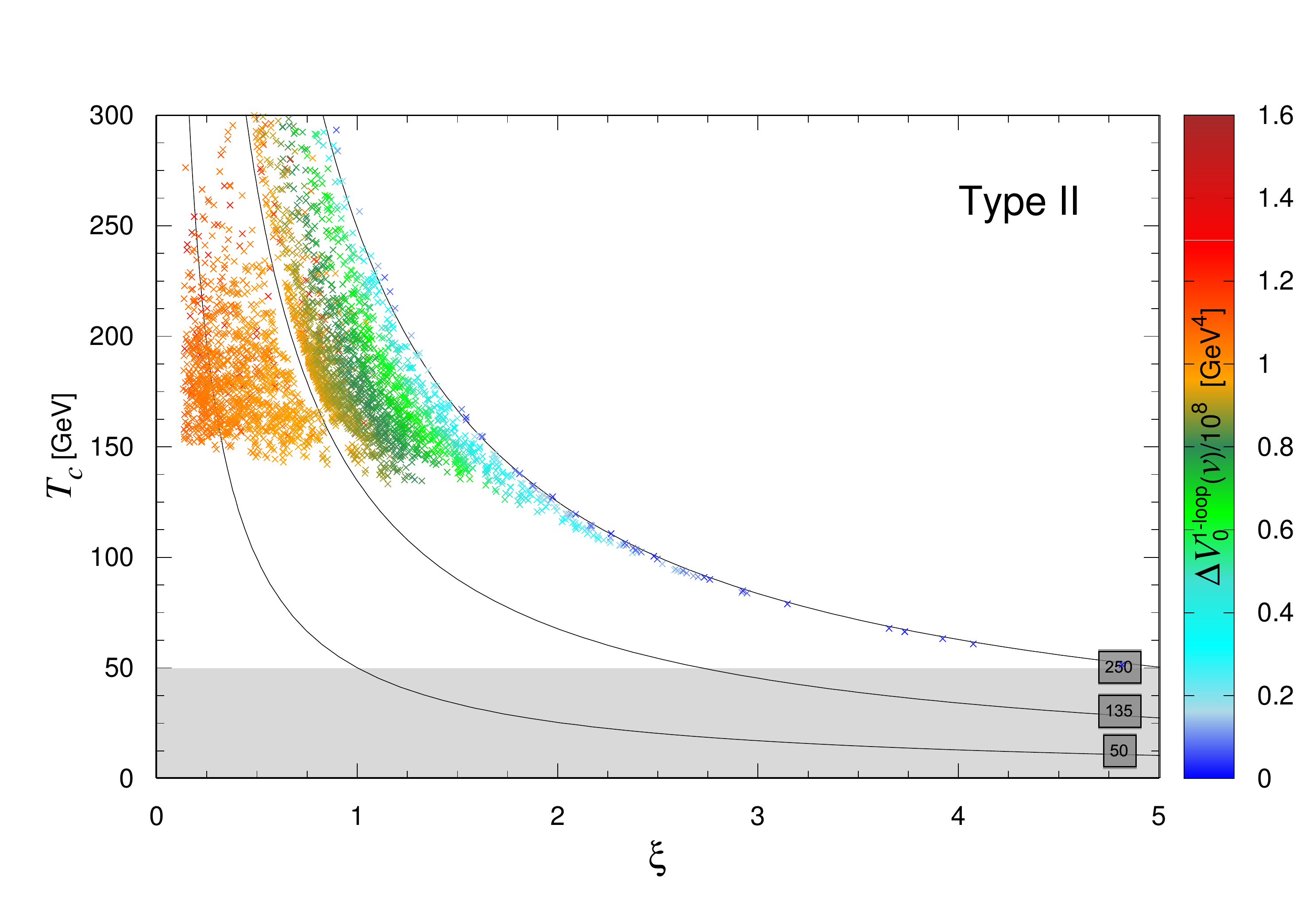}
\includegraphics[width=0.49\textwidth]{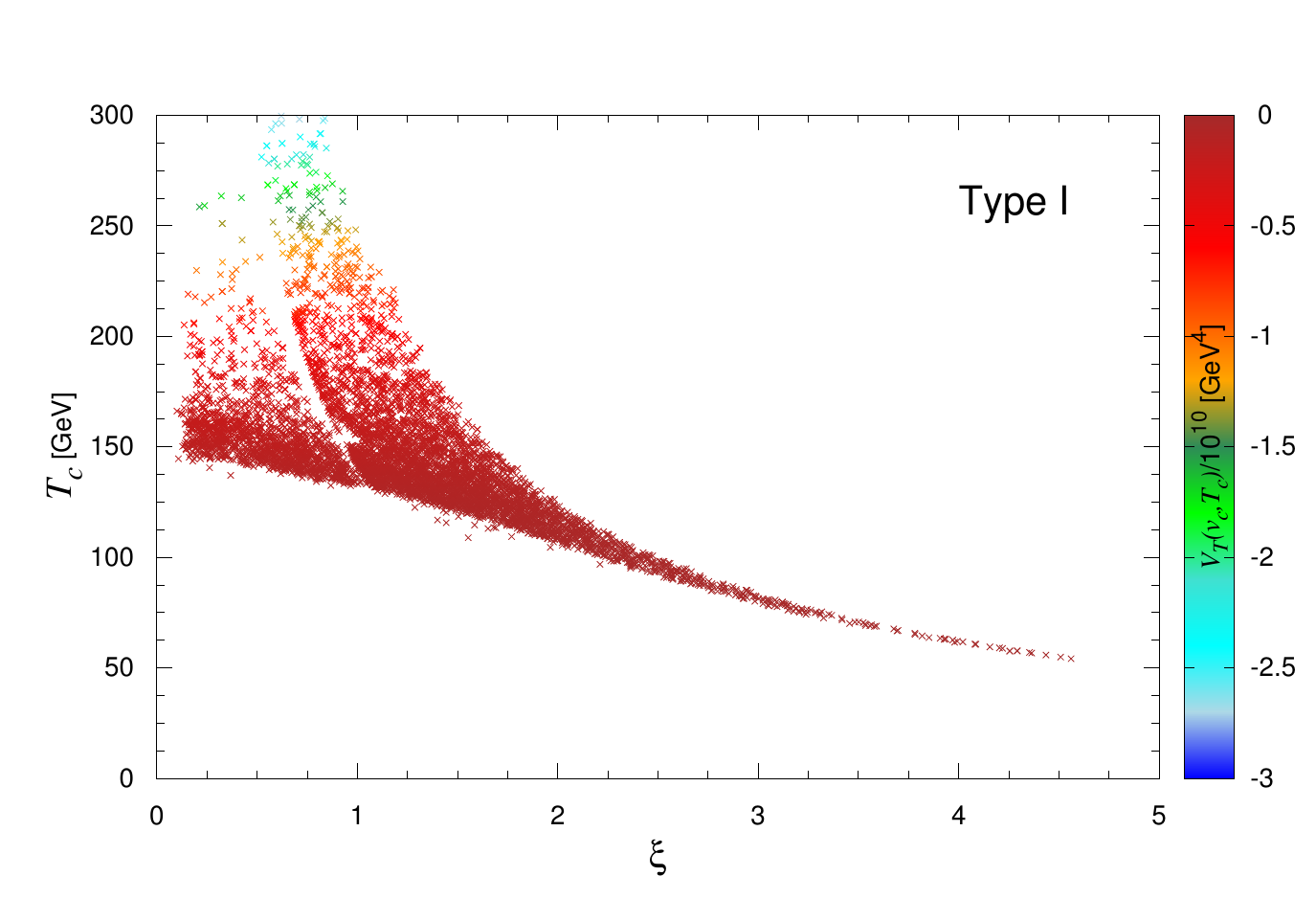}
\hspace*{-4mm}
\includegraphics[width=0.49\textwidth]{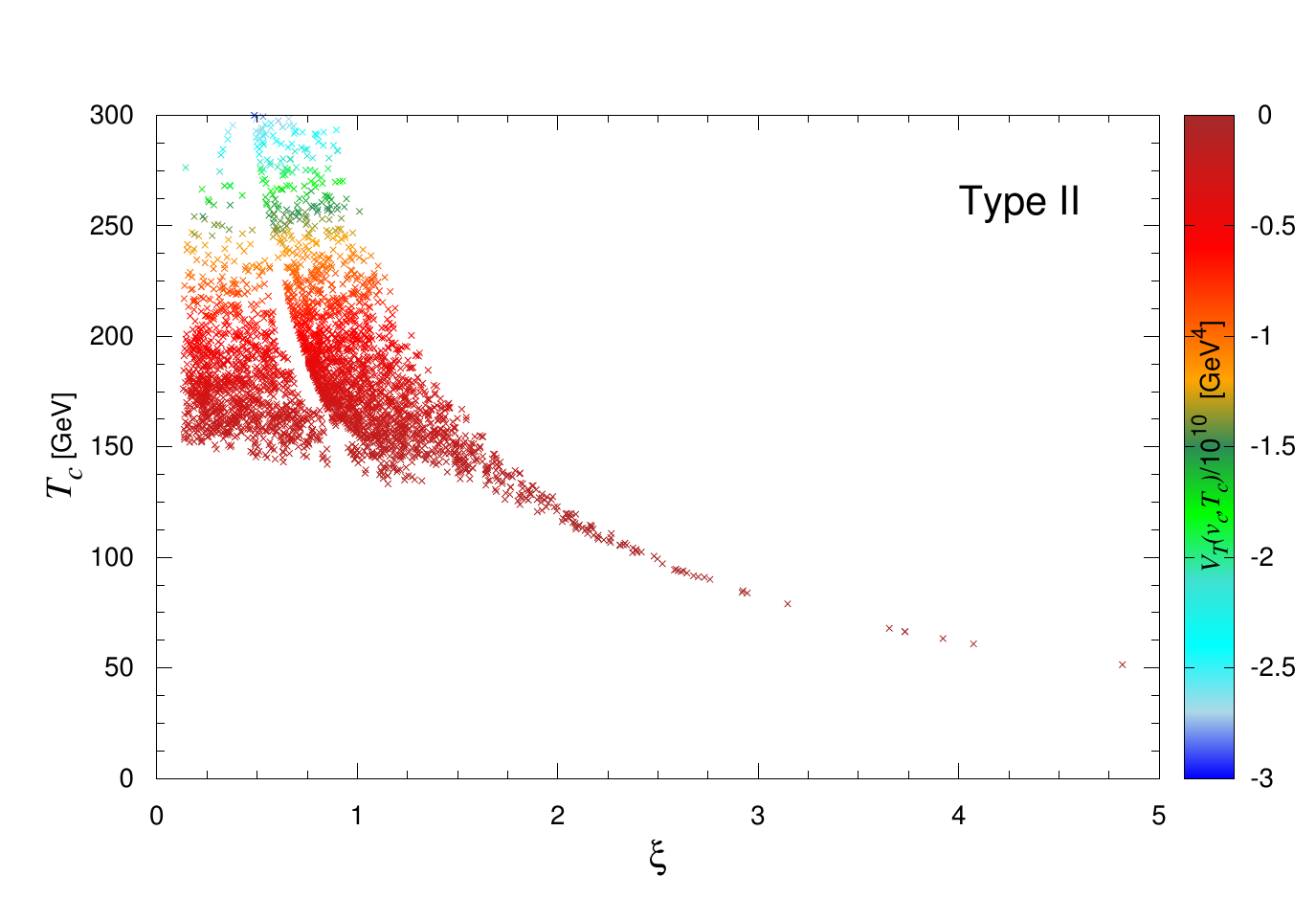}
\vspace{-5mm}
\caption{Properties of first order PT in the 2HDM. The z-axis in the upper and lower plots represents the vacuum depth of the zero temperature potential $| \Delta V^{\text{1-loop}}_{0} (v) |$ and the thermal potential in the broken vacuum at the critical temperature $V_{T} (v_{c}, T_{c})$, respectively. Only $T_c \lesssim 300\gev$ points are retained. }
\label{fig:vcTcVT}
\end{center}
\end{figure}

Having explored these SFOPT behaviors, we shall investigate the relation between critical classical field values, critical temperatures and different contributions to the effective potential in the model.
While the thermal contribution is crucial in controlling the process of vacuum tunneling, lots of attempts have been made to describe the properties of the phase (i.e. $v_{c}$ and $T_{c}$) from the effective potential at zero temperature.
A recent progress was reported in Ref.~\cite{Dorsch:2017nza} (within the framework of the CP-conserving 2HDM) that 
the strength of the phase transition is dominantly controlled by the value of $\mathcal{F}_0$, the depth of the 1-loop potential at zero temperature between the symmetry unbroken vacuum $h=0$ and the symmetry broken vacuum $h=v$ which corresponds to, in our notation,
\beq
\Delta V^{\text{1-loop}}_{0} (v) \equiv V^{\text{1-loop}}_{0} (v) - V^{\text{1-loop}}_{0} (0)
\eeq
where $V^{\text{1-loop}}_{0} (h) = V_{0} (h) + V_{\text{CW}} (h) + V_{\text{CT}} (h)$ is the full 1-loop potential at zero temperature.
Using the normalized depth $\Delta \mathcal{F}_0$ defined in~\cite{Dorsch:2017nza} one can derive an upper bound that definitely guarantees the PT to be strong, for example, $\Delta \mathcal{F}_0/\mathcal{F}_0^{\rm SM}\lesssim -0.34$ necessarily leads to $\xi \geq 1$ in the 2HDM. This, of course, can be used as an empirical test to assess the strength of the phase transition. However, a strong first order PT is still possible even though this upper bound is overflowed, in this situation the thermal potential plays a more important role for the thermal evolution of the system.
Therefore, while appreciating the advantage of this approach in simplifying the phase transition study, which allows to find regions of the parameter space where a SFOPT could be achieved, we expect a deeper comprehension by investigating not the strength $\xi$ itself, which is not an intrinsic property of the phase transition, but the characteristic quantities derived from the phase dynamics:  $v_{c}$ and $T_{c}$.
Interestingly, we find that the magnitude of $v_{c}$ increases towards the zero temperature VEV with the decrease of the vacuum depth $\Delta V^{\text{1-loop}}_{0} (v)$ independent of the value of $T_c$. 
 This is illustrated in Fig~\ref{fig:vcTcVT} and is one of the nontrivial outcomes of this work. It is naively true that  $v_{c}\simeq v$ when $|\Delta V^{\text{1-loop}}_{0} (v)|\simeq 0$, which implies that the thermal effects in the presence of extra scalars enhance the value of the effective potential at the SU(2) symmetry broken vacuum and almost do not shift the symmetry broken vacuum at the critical temperature. 
 As expected, as the vacuum depth $| \Delta V^{\text{1-loop}}_{0} (v) |$ increases, $v_{c}$ decreases towards the classical field value of $h=0$, which results in a smaller value of $\xi$ for a given $T_c$. 
In the meanwhile,  we emphasize that the precise evaluation of $v_{c}$ (and $T_{c}$), of course, requires the inclusion of the temperature-dependent part in the potential. 
The critical temperature $T_{c}$ is supposed to be more related to the thermal corrections to the effective potential, as demonstrated in the lower panels of Fig.~\ref{fig:vcTcVT}.

%%%%%%%%%%% Fig. 11

%%%% Fig.12 %%%%%
\begin{figure}[t]
\begin{center}
\includegraphics[width=0.47\textwidth]{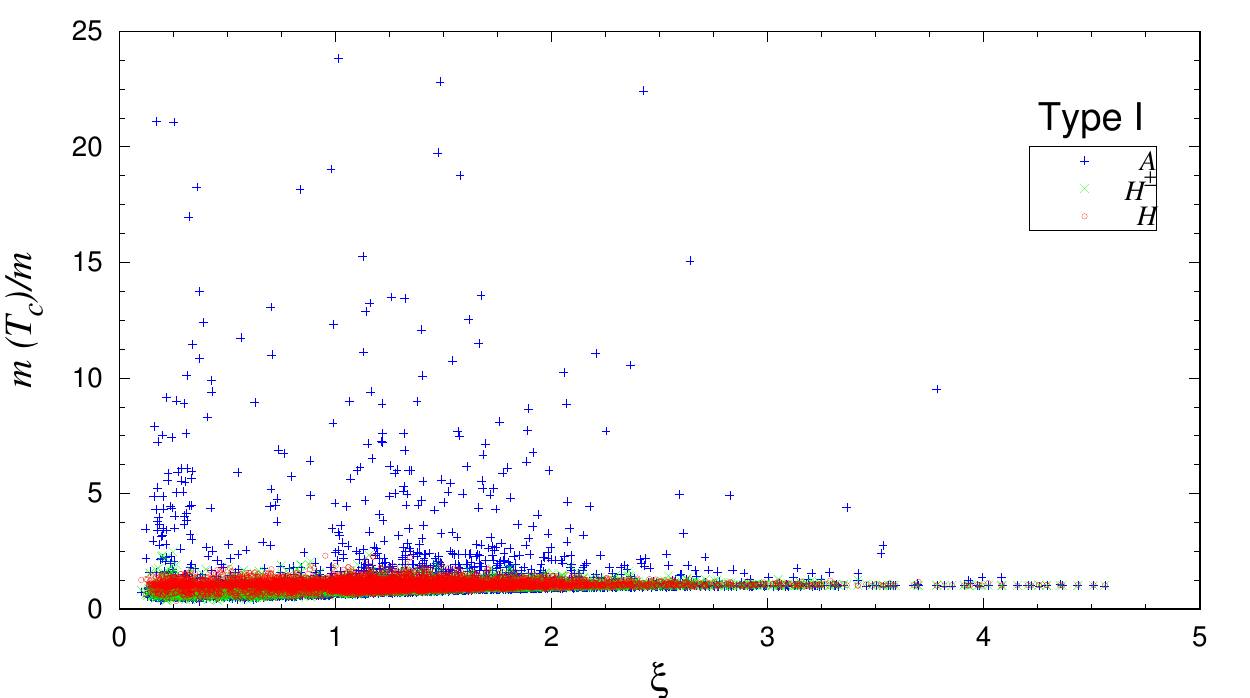}
\hspace{-4mm}
\includegraphics[width=0.47\textwidth]{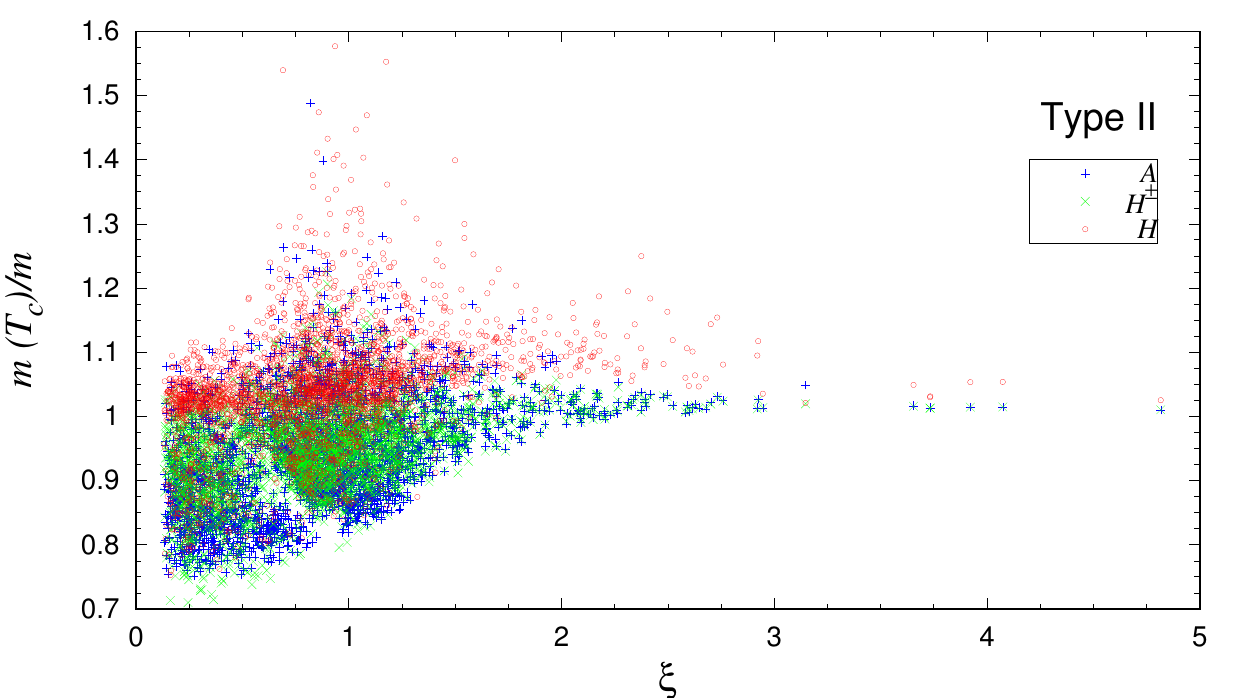}
\caption{The ratio of the thermal mass at the critical temperature to the zero-temperature mass for the three BSM states as a function of the PT strength $\xi$. Only $T_c \lesssim 300\gev$ points obeying the strong first order EWPT condition $\xi \geq 0.7$ are shown. 
}
\label{fig:thermalmass2}
\end{center}
\end{figure}

In addition to the non-thermal loop effect discussed above, the thermal effect in the presence of extra scalars is another promising source driving the SFOPT. In the 2HDM, extra BSM bosonic states are present in the plasma and induce the additional contribution to the thermal mass through the quartic couplings ($\lam_{1,2,3,4}$), see \eq{eq:thermalmass2}. Thus, if a proper cancellation between their masses and couplings is satisfied, an energy barrier can be generated so that the PT becomes strongly first order~\cite{Curtin:2016urg}. 
In order to see the importance of the thermal effect, we estimate the thermal masses for three extra Higgs bosons~\eq{eq:thermalmass} at critical temperature $T_{c}$ and present in Fig.~\ref{fig:thermalmass2} the ratio normalising the zero-temperature masses (the measured masses) as a function of the PT strength $\xi$.
Clearly, the ratio for the three states have a large variation around 1 on both sides, which means their thermal corrections can be either constructive or destructive even for the SFOPT ($\xi \geq 1$). In particular, this ratio for the CP-odd $A$ state in \typei~model can be up to $\sim 20$ owing to the presence of the extremely light $A$. While the thermal correction tends to suppress the $m_{A}$ and $m_{H^{\pm}}$ at $T_{c}$, the preference over the enhancement on the $H$ (relative to $H^{\pm}$) is still visible. 
The importance of the thermal mass maximizes at $\xi \simeq 1$ and becomes less significant as $\xi$ further grows.

\begin{figure}[t]
\begin{center}
\includegraphics[width=0.57\textwidth]{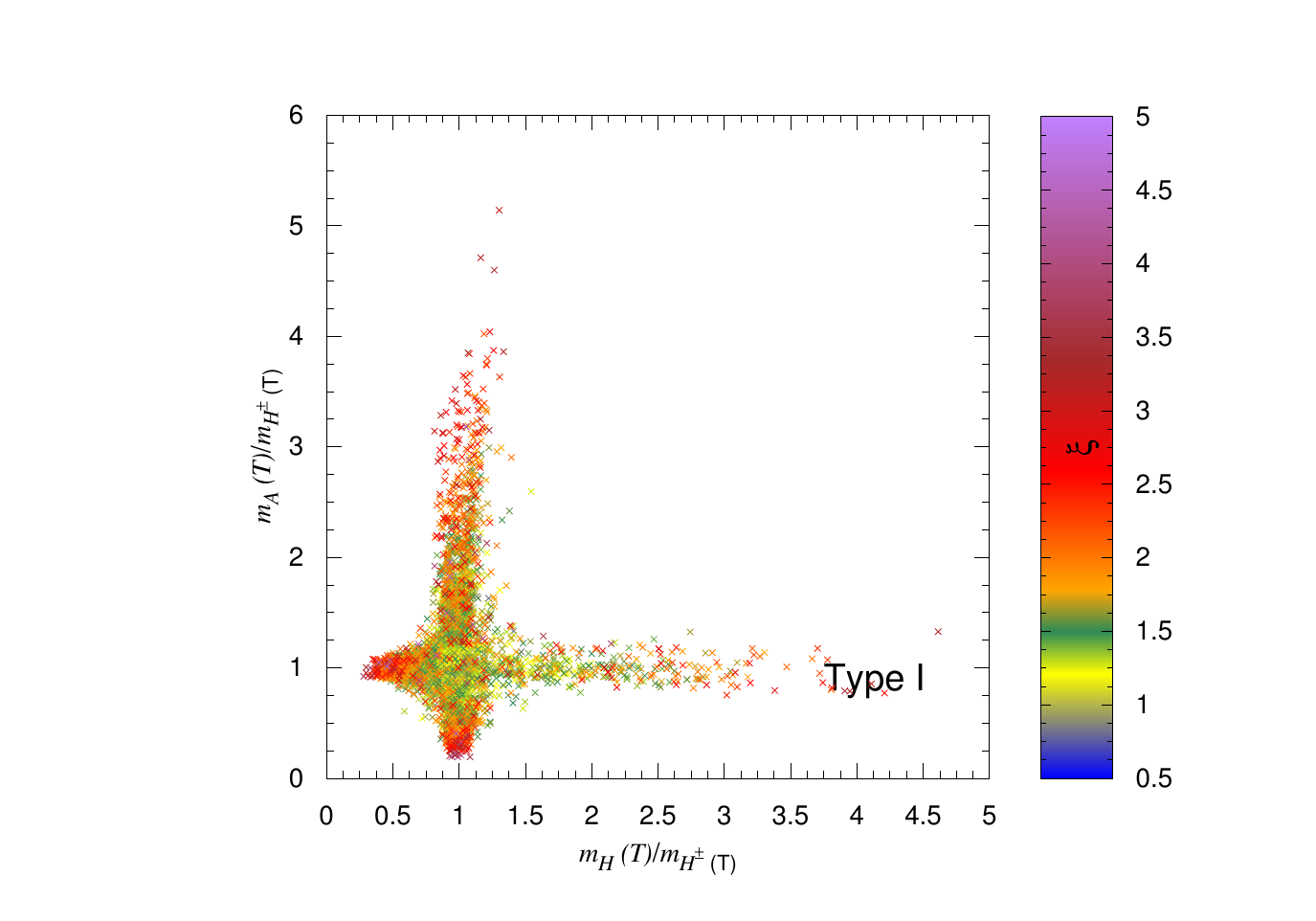}
\hspace{-24mm}
\includegraphics[width=0.57\textwidth]{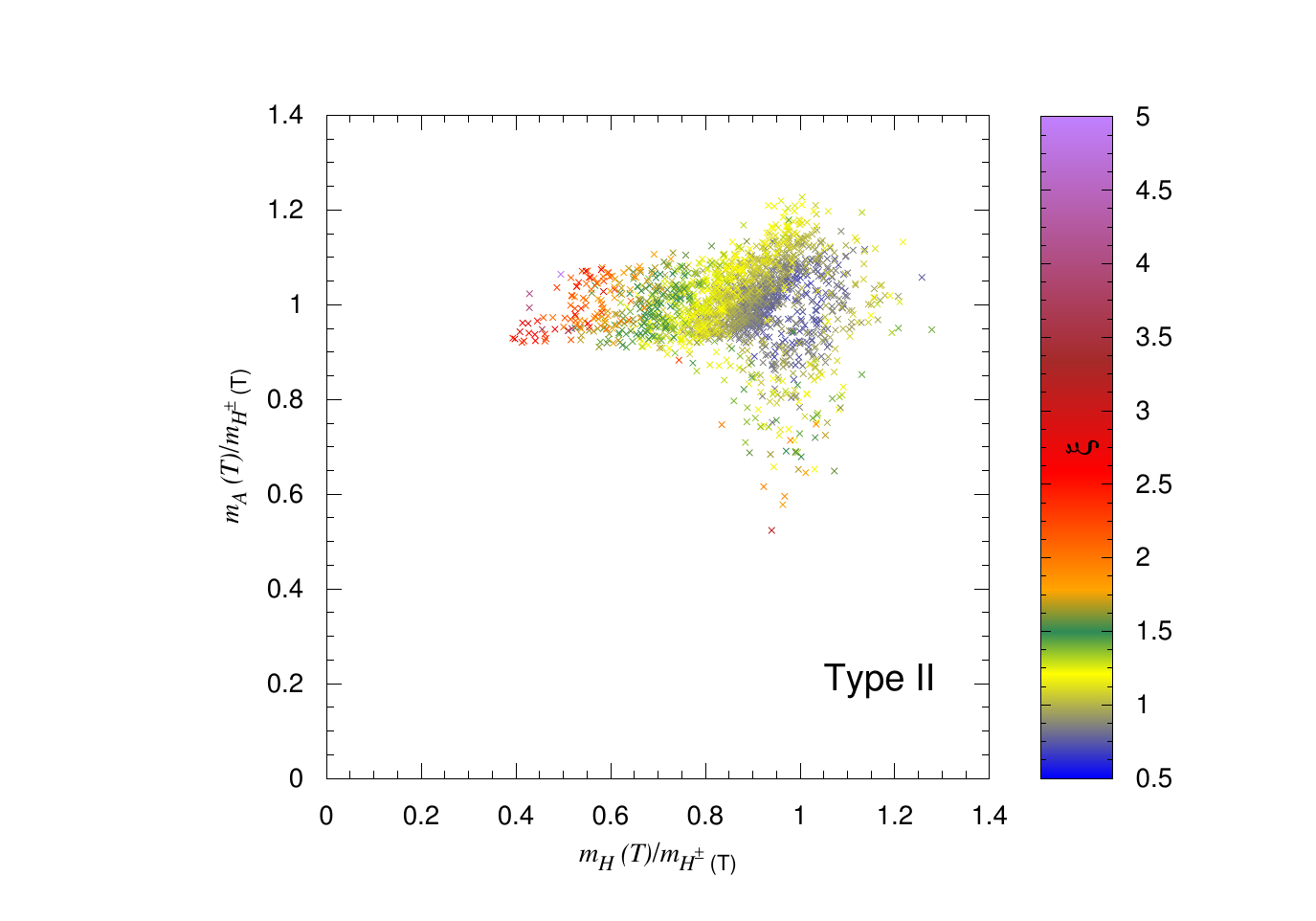}
\caption{To examine the violation of the SU(2) custodial symmetry we normalize the field-dependent mass for two neutral scalars $A$ and $H$ to the one for the charged Higgs $H^\pm$, $m_A(T_c)/m_{H^\pm}(T_c)$ and $m_H(T_c)/m_{H^\pm}(T_c)$ in the presentation. Only $T_c \lesssim 300\gev$ points obeying the strong first order EWPT condition $\xi \geq 0.7$ are shown. 
}
\label{fig:thermalmass}
\end{center}
\end{figure}

Recall that the SU(2) custodial symmetry is not severely broken at zero temperature due to the $T$ parameter in the EWPD which forces small mass difference for $|m_{H^{\pm}}-m_{A}|$ or $|m_{H^{\pm}}-m_{H}|$ or both. One may be curious whether this symmetry is broken at finite temperature. 
This is especially interesting when such symmetry plays a crucial role in selecting a particular region of parameter space. 
In general, the thermal correction to the field dependent masses might results in a shift of the symmetry of the model at finite temperature.
The particular case of interest is the $Z_2$ symmetry cases studied in Refs.~\cite{Espinosa:2011ax,Espinosa:2011eu,Cline:2017qpe} where the $Z_2$ symmetry is preserved at $T=0$ but spontaneously broken at $T\neq0$. 
To examine if the effect of thermal corrections leads to a shift of the SU(2) custodial symmetry in our model, we estimate the ratio of the thermal mass for two neutral states with respect to that for the charged state at critical temperature $T_{c}$.
The result is illustrated by Fig.~\ref{fig:thermalmass} where one can observe that the points displayed are well aligned either $m_A(T_c)/m_{H^\pm}(T_c)\simeq 1$ or $m_H(T_c)/m_{H^\pm}(T_c)\simeq 1$ with about 10-20\% departure, indicating a large violation of the SU(2) custodial symmetry is not possible at finite temperature during the SFOEWPT in the 2HDM. 

%%%% Fig.13 %%%%%

%%%%%%%%%%%%%%%%%%%%%%%%%%%
\section{Strong first order EWPT and the implications for future measurements at colliders}
\label{sec:1stcol}

\subsection{Typical mass spectra and discovery channels at LHC}

%%%% Fig.14 %%%%%
\begin{figure}[t]
\begin{center}
\includegraphics[width=0.57\textwidth]{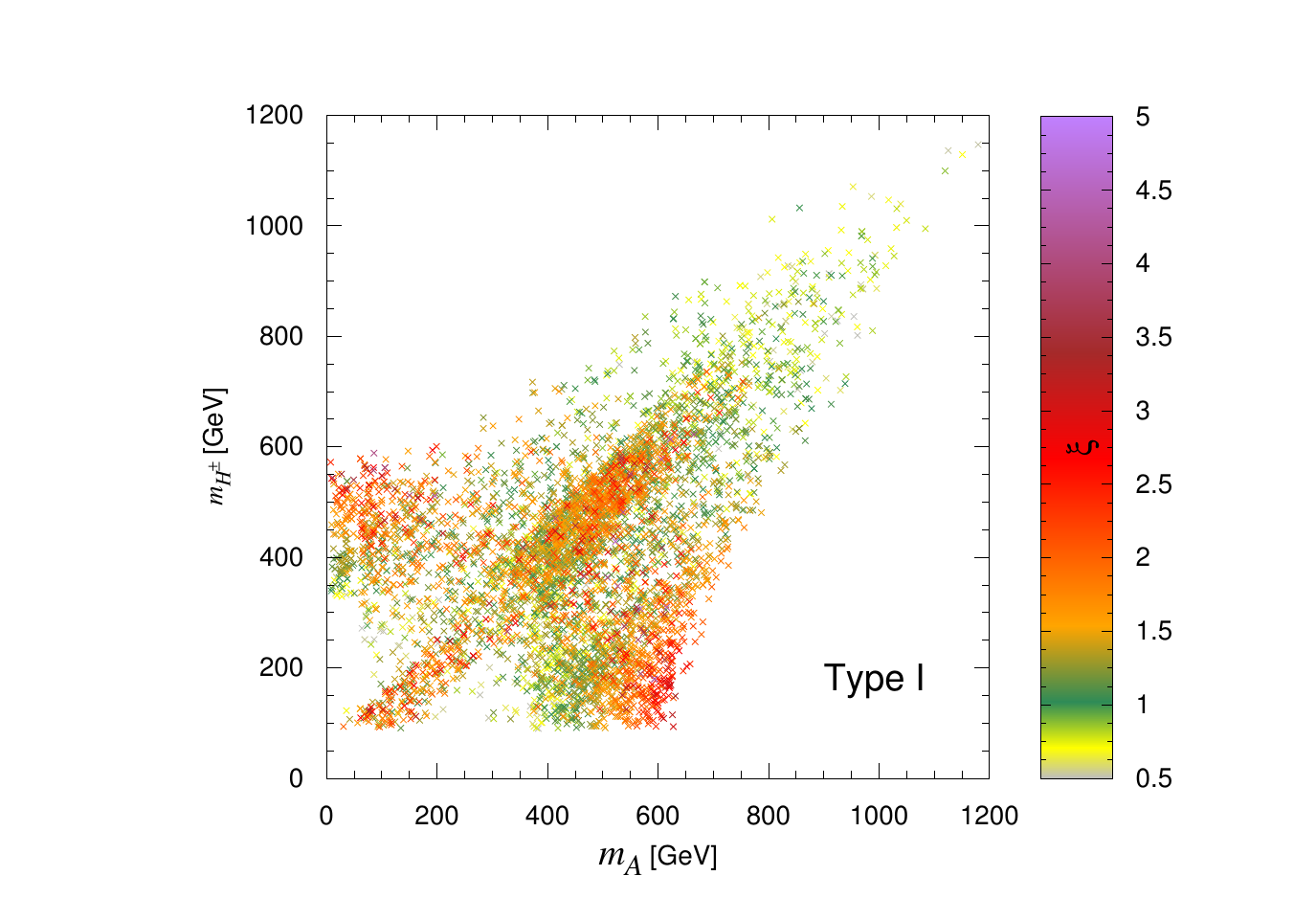}
\hspace{-24mm}
\includegraphics[width=0.57\textwidth]{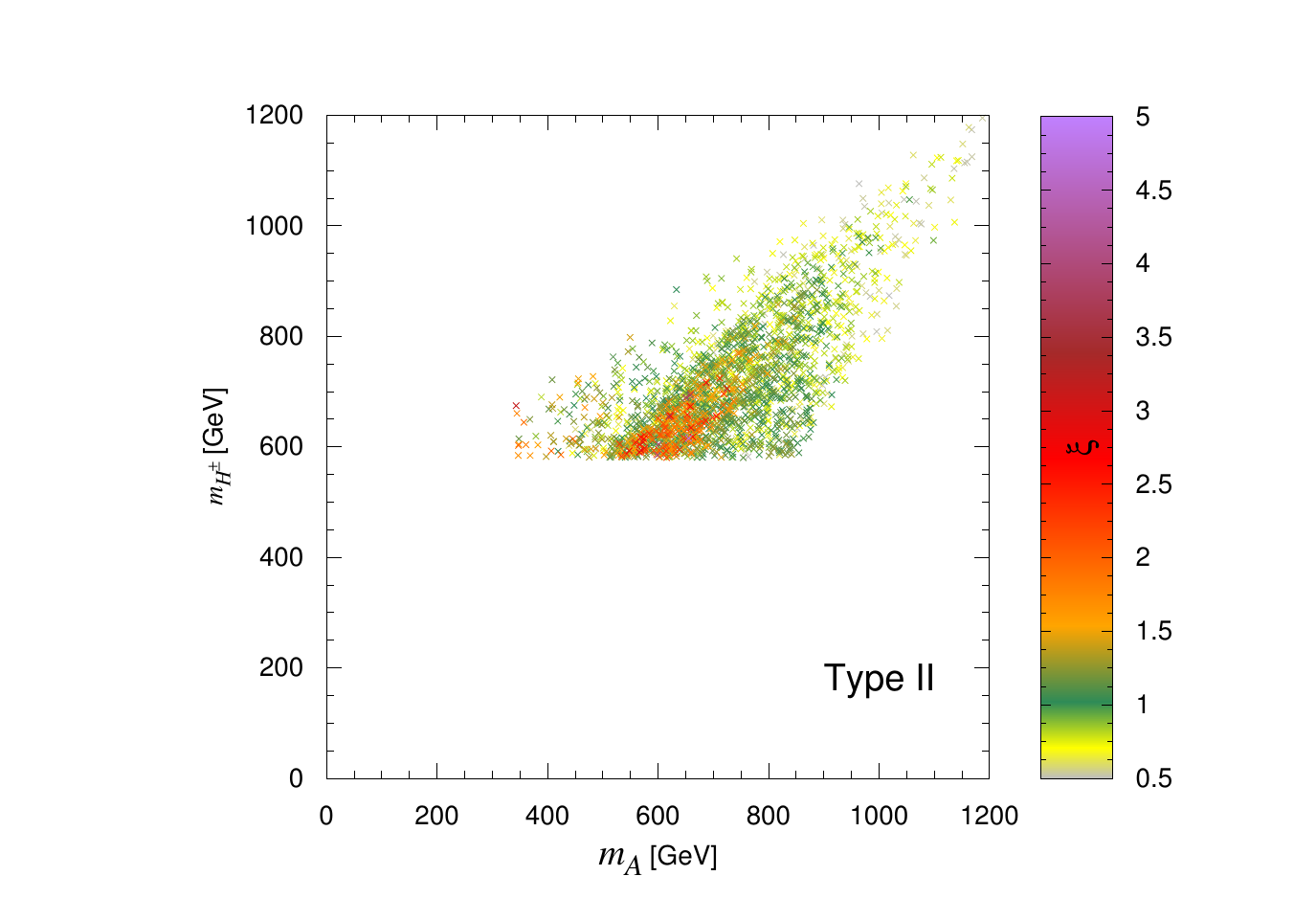}\\[-3mm]
\includegraphics[width=0.57\textwidth]{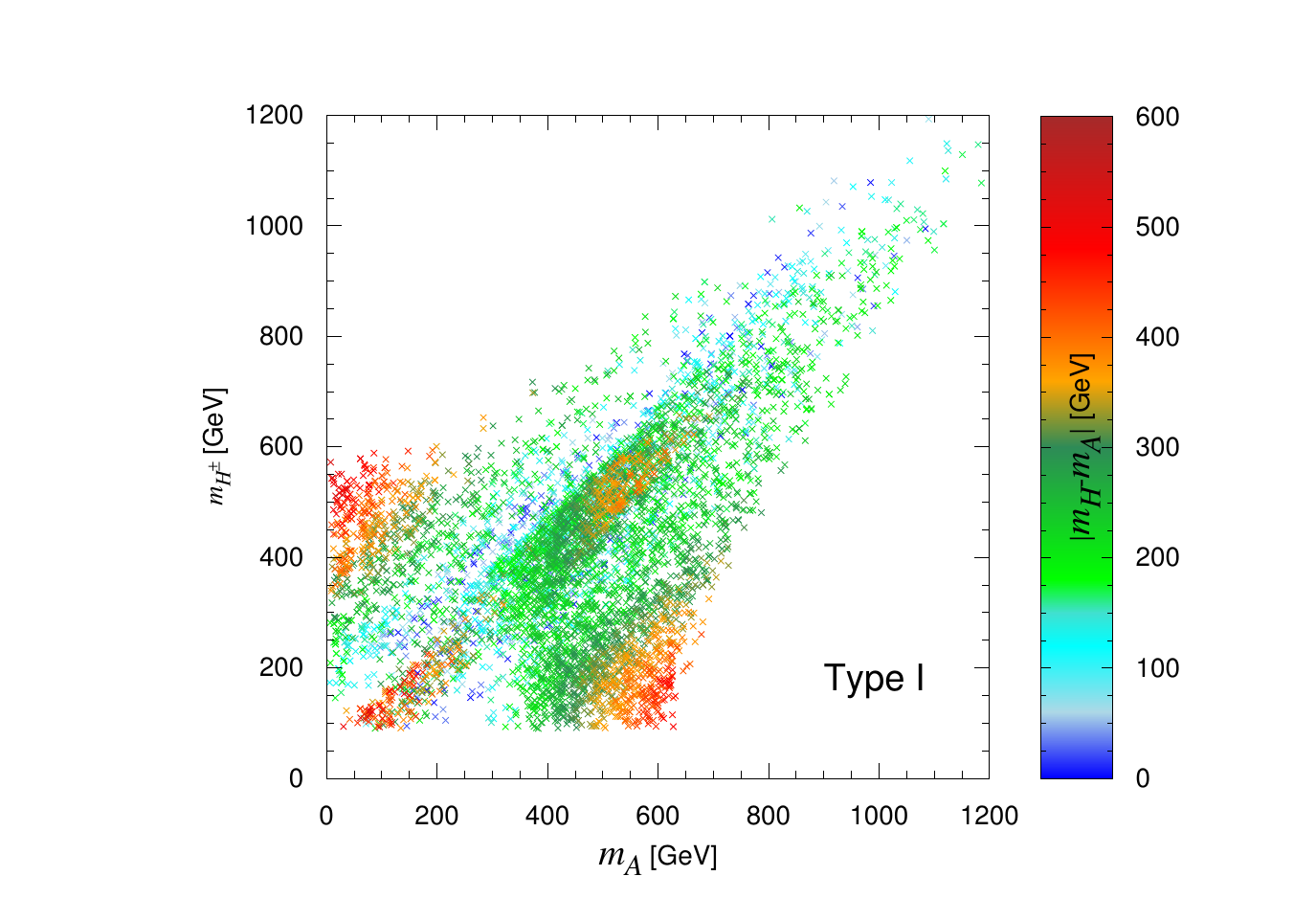}
\hspace{-24mm} 
\includegraphics[width=0.57\textwidth]{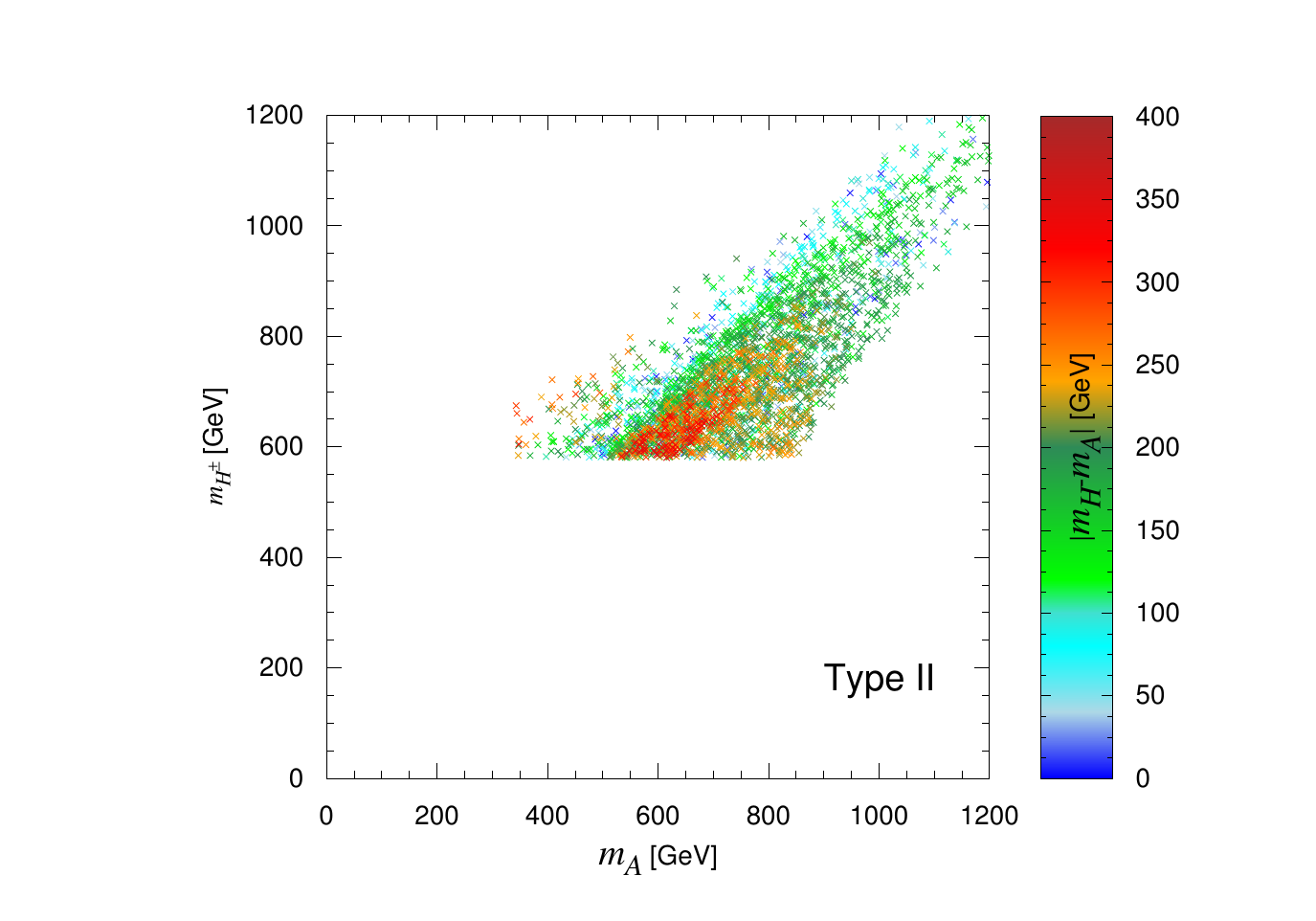}
\vspace{-3mm}
\caption{The strength of first order EWPT shown on the Higgs mass plane. Only $T_c \le 300$ GeV points are retained.}
\label{fig:vcTcMass}
\end{center}
\end{figure}

As seen from Fig.~\ref{fig:vcTc}, a SFOPT is possible in both \typei~and \typeii~models.
Then one may wonder what is the LHC Higgs phenomenology associated with a SFOPT.
To answer this question, in Fig.~\ref{fig:vcTcMass} we present in the $m_A$ versus $m_{H^{\pm}}$ (upper) and $m_H$ versus $m_{H^{\pm}}$ (lower) planes all points that pass the applied constraints as in Fig.~\ref{fig:vcTc} and additionally realize a SFOEWPT (\textit{i.e.}, $\xi \geq 0.7$).
The values of $\xi$ and the mass difference $|m_A-m_H|$ are indicated in color scale in the upper and lower panels, respectively.
We emphasize again that the EWPD, essentially the T parameter, force the mass differences between the charged Higgs boson and at least one of the extra neutral Higgs bosons to be small and strongly favor mass spectra where the masses of all new scalars are close to each other, in the decoupling limit in particular. 
This severe constraint on the mass spectra for the non-SM Higgs bosons 
leads to five benchmark scenarios achieving a SFOPT in \typei~model. They are summarized in Table~\ref{tab:masssper} where the characteristic mass spectra and the main decay modes of the heavier neutral Higgs boson ($H$ or $A$) with an estimate on the branching ratio are given in each scenario.

\begin{table}[t]
\caption{Benchmark scenarios leading to the SFOPT. Mass spectra and the main decay modes of the heavier neutral Higgs boson ($H$ or $A$) are given in each scenario.%except Scenario E where the main decays are undetermined, which strongly rely on the mass hierarchy among the three BSM Higgs states. 
The numbers in the parenthesis following each decay indicate an estimate on the branching ratios.}
\vspace{-3mm}
\begin{center}
\hspace*{-5mm}
\begin{tabular}{|c|c|c|c|c|c|}
\hline
Sce. & $m_{H}$ [GeV] & $m_{A}$ [GeV] & $m_{H^{\pm}}$ [GeV] & Type &  Main $H/A$ decays \cr
\hline
A & 130 -- 300 & 400 -- 600 & 100 -- 300 & I & $A \to W^{-}H^{+} (60\%), ZH (25\%)$ \cr
B & 400 -- 600 & 10 -- 200 & 400 -- 600 & I, II & $H \to Z A (50-75\%)$   \cr
C & 130 -- 200 & 450 -- 800 & 450 -- 800 & I, II &$A \to ZH (\sim 100\%)$  \cr
D & 400 -- 600 & 10 -- 250 & 100 -- 250 & I & $H \to W^{-}H^{+} (60\%),ZA (25\%)$  \cr
E & 300 -- 350 & 300 -- 350 & 300 -- 350 & I & $A \to Zh (\sim 100\%)$, $H \to W^{+}W^{-} (\gtrsim 40\%)$\cr
\hline
\end{tabular}
\end{center}
\label{tab:masssper}
\end{table}%

We start with the analysis in \typei~model.
First, the most widely studied mass configuration includes a pseudoscalar $A$ with mass within the range of $400-600\gev$ accompanied with $m_H \approx m_{H^{\pm}} \simeq 200\gev$~\cite{Dorsch:2013wja,Dorsch:2014qja}.
In this case, $m_{12}^2$ must be relatively small since large $m^{2}_{12}$ tends to reduce the strength of the phase transition. This leads to a special relation among the quartic couplings $\lam_{1,2,3} \simeq 0$ and $\lambda_4 \simeq - \lambda_5 \simeq 5 $, meaning that the strength of the phase transition is mainly governed by $\lambda_4$ and $\lambda_5$, see also~\cite{Dorsch:2013wja}. 
Dictated by symmetry argument, one can image that the mass spectrum consisting of a light CP-odd state and two highly mass degenerate $H$ and $H^{\pm}$ can also lead to a SFOPT, which is reflected by the existence of a bulk of red points at the upper left corner (\textit{i.e}, $m_{H^{\pm}} \simeq 400-600\gev$, $m_{A} \lesssim 200\gev$) in Fig.~\ref{fig:vcTcMass}. The situation of the model parameters is opposite due to the flip of mass hierarchy among the three BSM Higgs states. To be specific, $m_{12}$ is large as a consequence of large $m_{H}$, and $\lam_{1,2,3} \simeq 0$ and $\lambda_4 \simeq - \lambda_5 \simeq -5 $.
Likewise, a SFOPT ($\xi \geq 1$) can be also realized provided that $m_{A}$ and $m_{H^\pm}$ are close to each other, while both having a large gap relative to $m_{H}$. Strictly speaking, such condition provides two possibilities for the mass spectra: i) $m_{A} \simeq m_{H^\pm}\simeq 600\gev$ and $m_{H} \simeq 200\gev$ and ii) $m_{A} \simeq m_{H^\pm}\simeq 200\gev$ and $m_{H} \simeq 600\gev$, which correspond to two isolated red-orange points densely distributed along the diagonal line in the lower panel plot.
Deduced from Eqs~(\ref{eq:Amass}) and (\ref{eq:Hcmass}) the mass degeneracy between $A$ and $H^{\pm}$ states in this scenario restrict $\lam_{4} \simeq \lam_{5}$, while an additional coupling $\lam_{3}$ participates into the potential evolution and influences the phase transition.
Apart from these four scenarios that are visible in the low panel plot, the upper left plot in Fig.~\ref{fig:vcTcMass} demonstrates an additional possible scenario that is compatible with $\xi \geq 1$ where all
three non-SM-like Higgs bosons have similar mass scales at $300-350\gev$. This scenario was unfortunately ignored~\cite{Dorsch:2013wja,Dorsch:2014qja} or paid less attention~\cite{Basler:2016obg}.~\footnote{One might indeed have believed that large mass splitting among the non-SM Higgs bosons are a necessary condition for the requirement of a strong first-order EWPT.} It is also worth noting that in this highly degenerate scenario none of $\lam_{i}$ couplings can be close to zero if the first order PT takes place.

On the other hand, the allowed mass spectrum that is compatible with $\xi \geq 1$ in \typeii~model is quite simple. 
As explained in Sec.~\ref{sec:LHCbounds}, the combination effect from $B$-physics observables and EWPD pushes $m_{H^\pm}\gtrsim 580\gev$ and simultaneously raises the mass scale for at least one of the extra Higgs bosons. 
Consequently, many scenarios available in \typei~model are eliminated, resulting in an allowed mass spectrum that leads to a strong first-order EWPT being quite restrained: $m_{A}\simeq m_{H^{\pm}} \approx 600\gev$ and a large positive mass gap between $m_{H^\pm}$ and $m_H$ :  $m_{H^\pm} - m_H \gsim 300\gev$.

Generally speaking, requiring a SFOPT forces down the mass scale for the new scalars and the preferred ranges for all the scalar masses below $600\gev$, which coincidently approaches to the current lower bound on the charged Higgs mass strongly constrained by the latest measurement of $B \rightarrow X_s \gamma$. This means that future improvement on $B$-physics observables may decisively rule out the success of SFOPT in the \typeii~2HDM. Of course, Fig.~\ref{fig:vcTcMass} also informs us that weak first order PT (under the criterion of $\xi \simeq 0.7$) would still be possible even if no additional Higgs bosons were discovered below 1 TeV.

\begin{figure}[t]
\begin{center}
\includegraphics[width=0.5\textwidth]{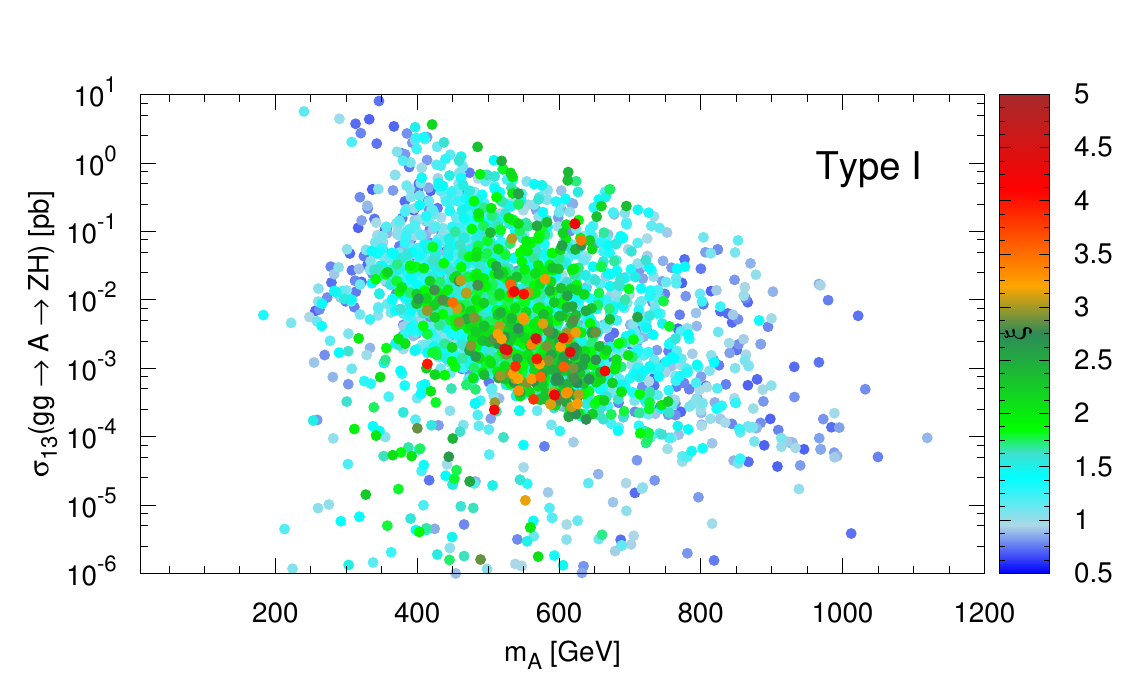}
\hspace{-3mm}
\includegraphics[width=0.5\textwidth]{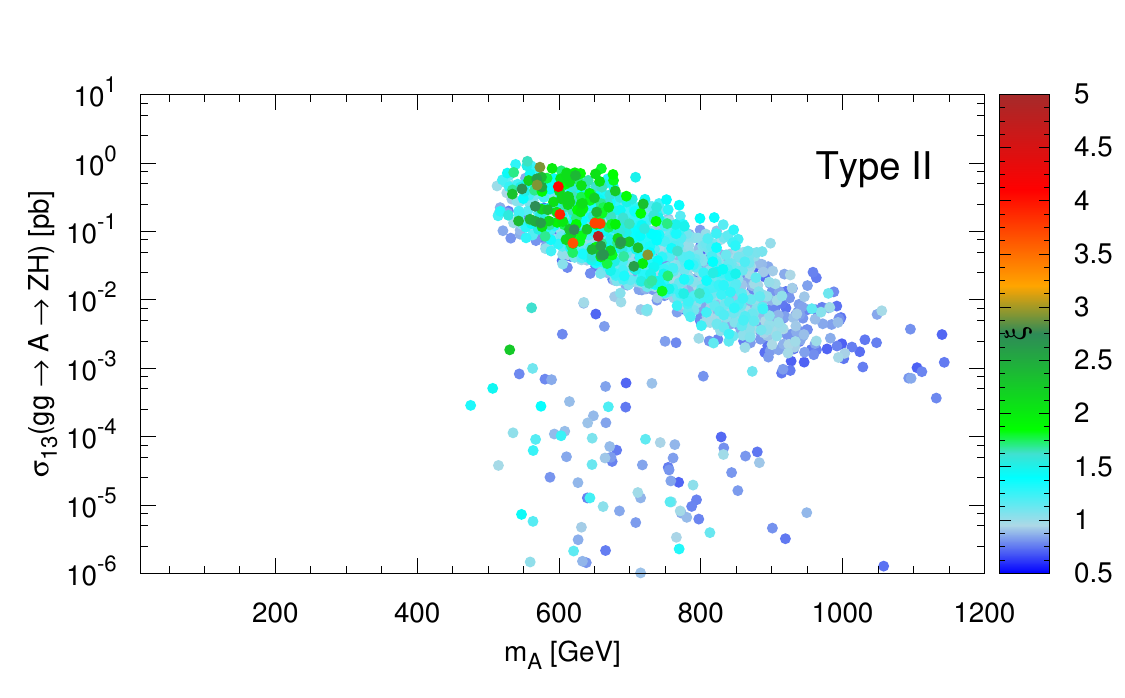}
\includegraphics[width=0.5\textwidth]{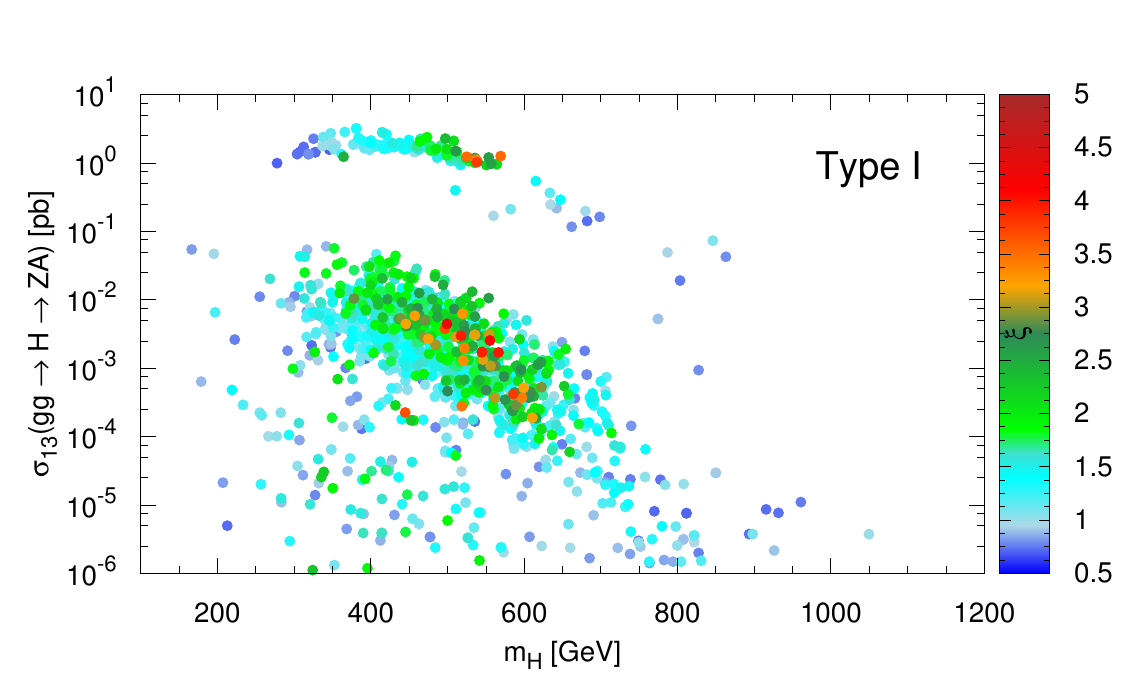}
\hspace{-3mm}
\includegraphics[width=0.5\textwidth]{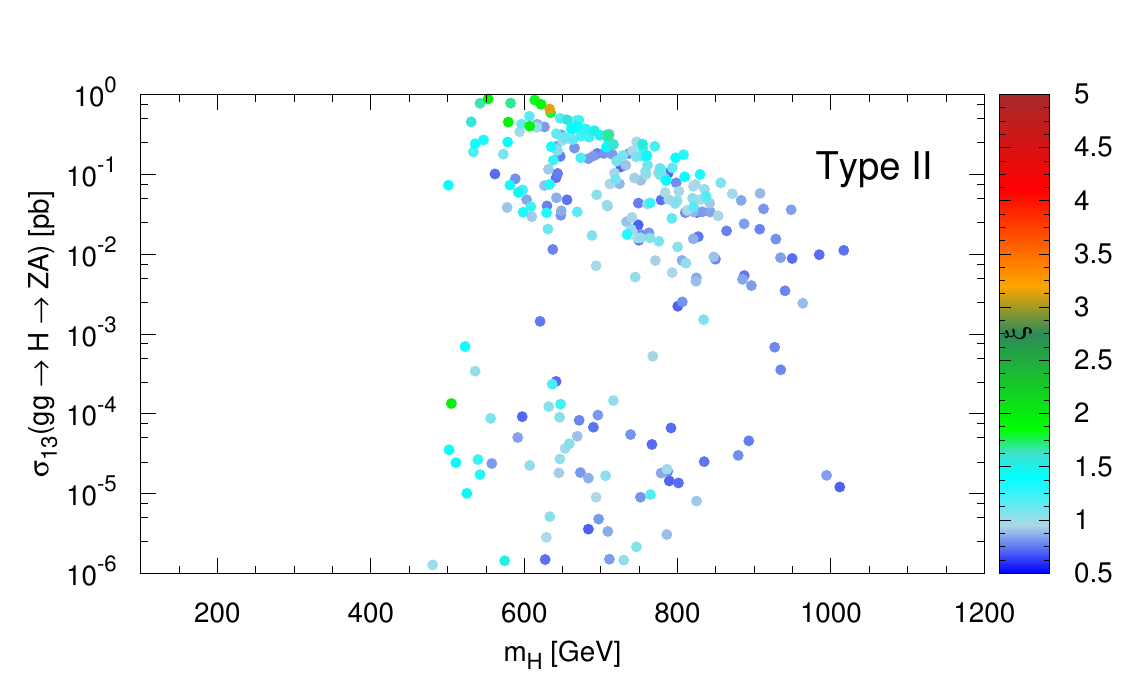}
\vspace{-0.4cm}
\caption{13 TeV cross sections at the LHC as a function of the relevant mass scale, for the $gg\to A\to ZH$ (upper panels) and $gg\to H\to ZA$ (bottom panels) channels in \typei~(left panels) and \typeii~(right panels).}
\label{fig:mxsec}
\end{center}
\end{figure}

Finally, we briefly discuss the prospects of testing the EWPT at the colliders in accordance with the mass spectrum provided above. 
In the alignment limit $\sin (\beta-\alpha) \approx 1$ we consider, the coupling $g_{hAZ}$ is vanishingly small but the coupling $g_{HAZ} \propto \sin(\beta- \alpha)$ is enhanced. Hence, the branching ratios for $A \to ZH$ and $H \to ZA$ as long as kinematically allowed can be substantially large depending on the mass spectrum in the model. 
These results point towards the observation of the $A \to ZH$ and/or $H \to ZA$ decay channels would be ``smoking gun'' signatures of 2HDMs with a SFOEWPT.~\footnote{Although our results confirm the results in the earlier literature \cite{Dorsch:2014qja}, more importantly, we clarify that the decay $A \to ZH$ is not a unique ``smoking gun'' signature of SFOPT in the 2HDM of \typei~model. This conclusion is also supported by another recent study~\cite{Basler:2016obg}.}
LHC search prospects for the former decay have been analyzed and proposed as a promising EWPT benchmark scenario in \cite{Dorsch:2014qja}, while the collider analysis looking at both decays was performed in Ref.~\cite{Coleppa:2014hxa} but not specifically aiming at the EWPT. In Fig.~\ref{fig:mxsec} we show the 13 TeV cross sections at the LHC for these two channels in the gluon-fusion production mode.  
In all cases, a cross section above the pb level can be achieved for the scenarios realizing a SFOPT. Although these signatures are characteristic ones in most of the 2HDM scenarios discussed above (see Table~\ref{tab:masssper}), no strong correlation in these channels is found between $\xi$ and the corresponding cross sections, which means that there is no guarantee to observe these decays in colliders. We leave a detailed collider analysis to future studies.

%%%% Fig.15 %%%%%

\begin{figure}[t]
\begin{center}
\includegraphics[width=0.5\textwidth]{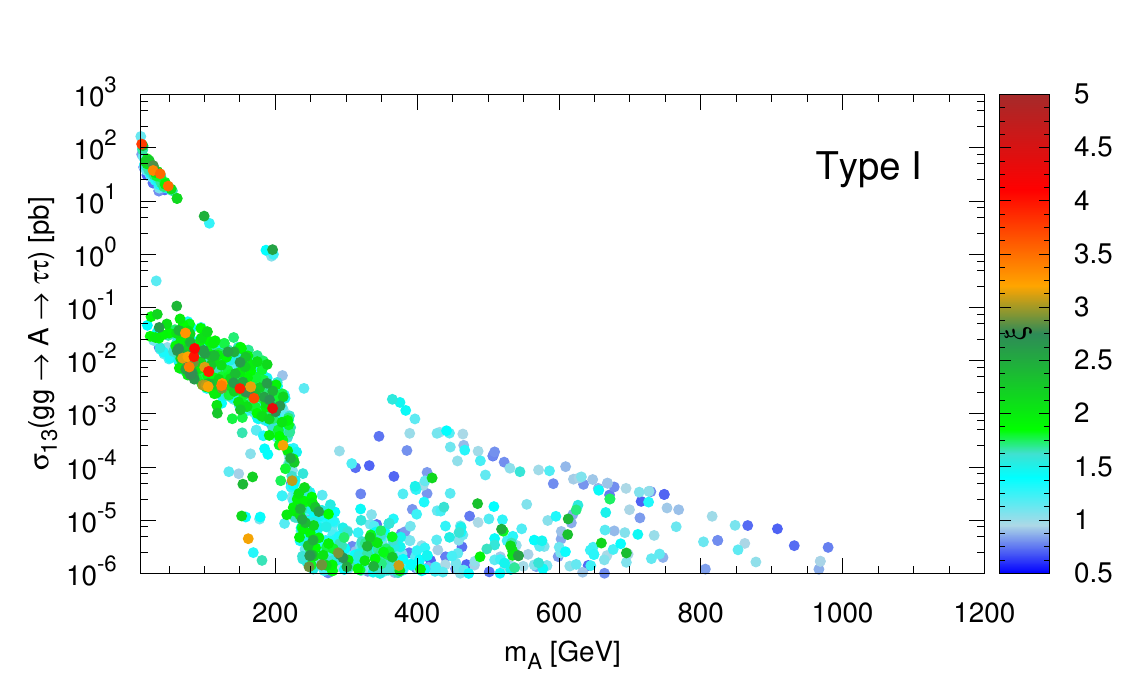}
\hspace{-3mm}
\includegraphics[width=0.5\textwidth]{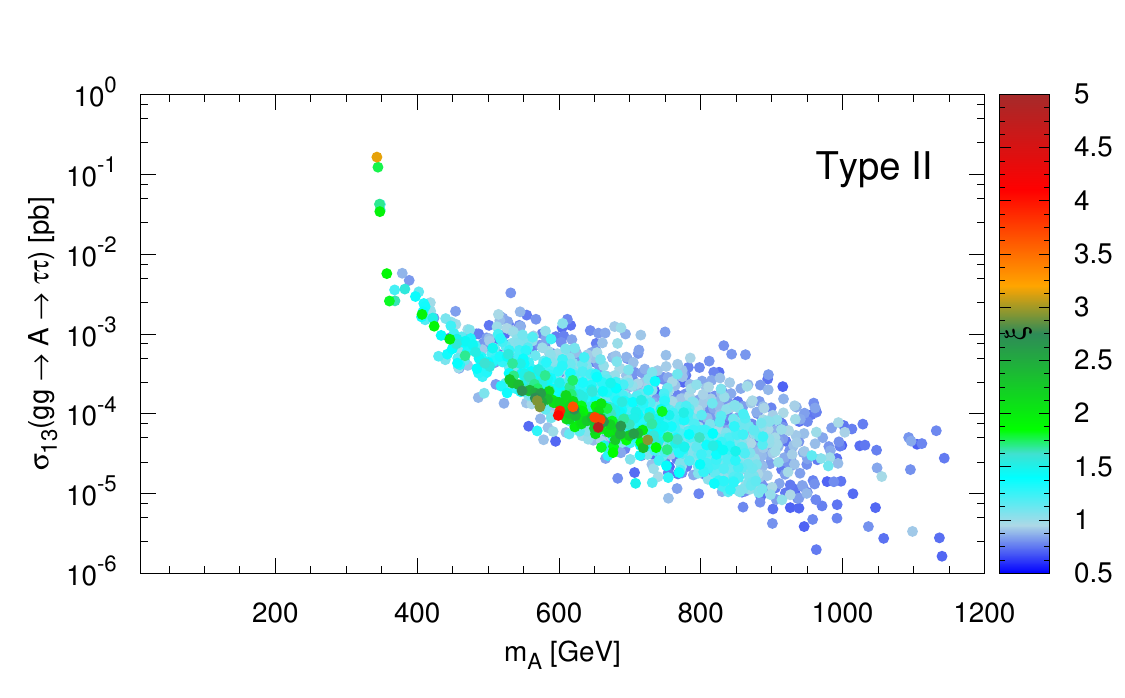}
\includegraphics[width=0.5\textwidth]{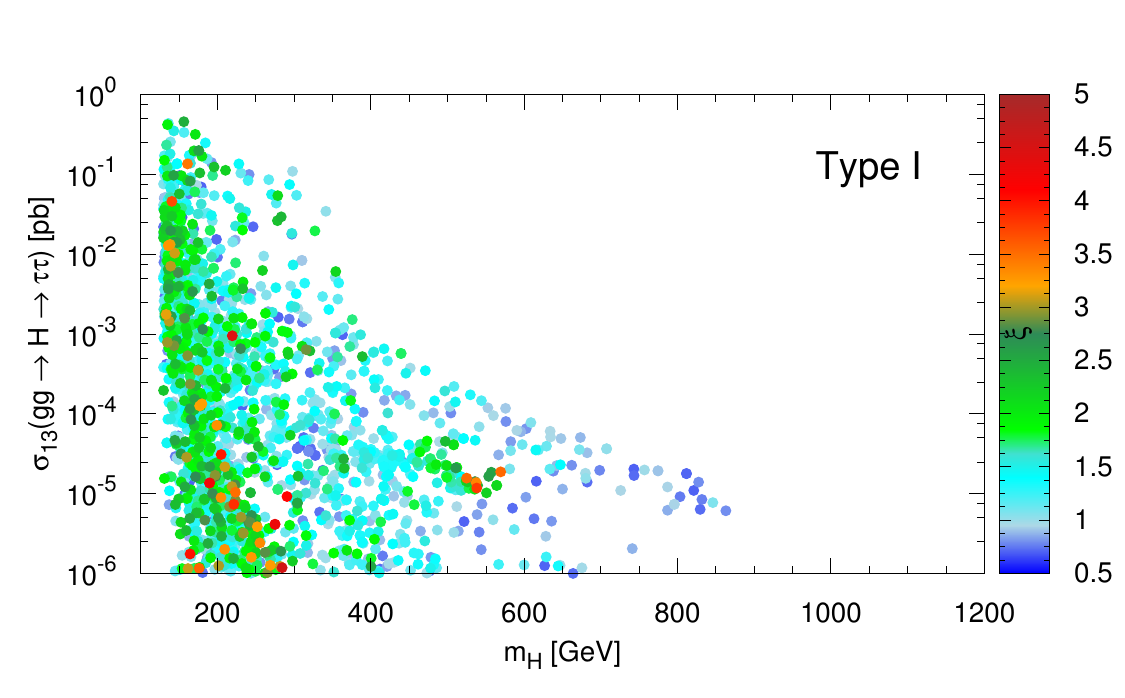}
\hspace{-3mm}
\includegraphics[width=0.5\textwidth]{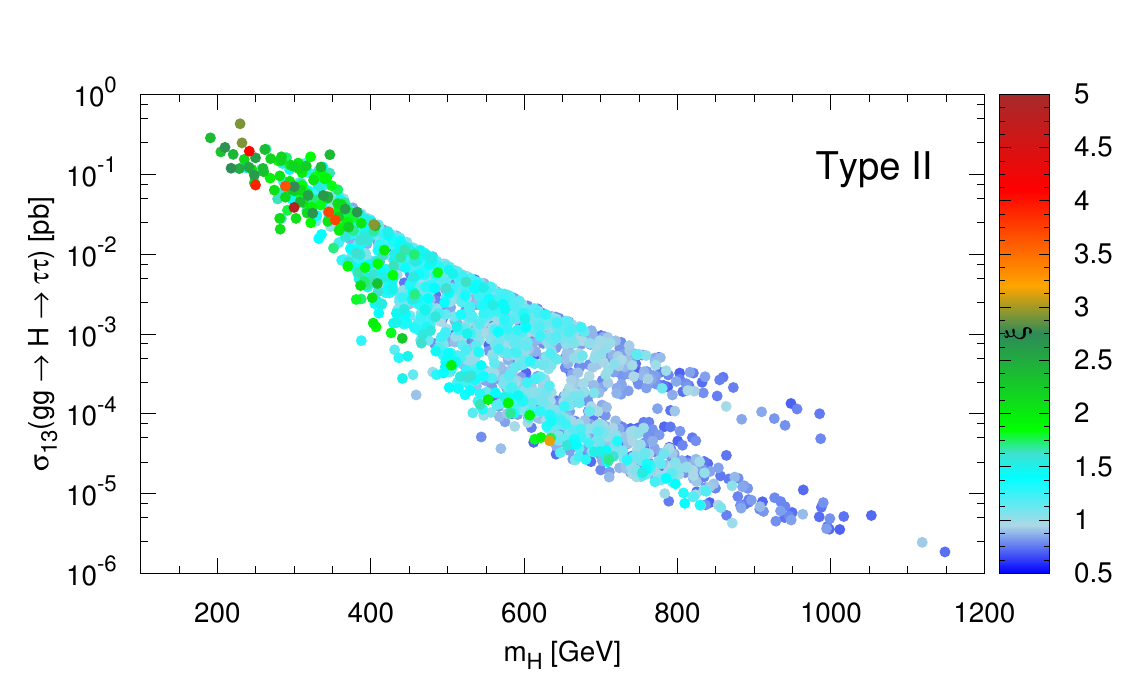}
\vspace{-0.4cm}
\caption{13 TeV cross-sections at the LHC as a function of the relevant mass scale, for the $gg\to A\to \tau\tau$ (upper panels) and $gg\to H\to \tau\tau$ (bottom panels) channels in \typei~(left panels) and \typeii~(right panels).}
\label{fig:mxsectautau}
\end{center}
\end{figure}

Searching for a new scalar resonance is performed at the LHC mostly through its decay into SM particles. These decay channels include $H\to ZZ\to 4\ell$, $H,A\to \gamma\gamma, \tau\tau, t\bar{t}$. For the purpose of testing the EWPT, it would be very useful to find channels with strong correlation to the $\xi$ value. The one served as an example here is the gluon-fusion production cross section of $A$ and $H$ in the $\tau\tau$ decay channel, which is shown in Fig.~\ref{fig:mxsectautau}.
In general, the gluon-fusion cross section in \typei~model is considerably small, so there is very little hope to ever observe $A$ or $H$ in this channel.
An exception occurs in the very light CP-odd $A$ region with cross-section as large as the level of $10-100$ pb~\cite{Bernon:2014nxa}.
Moreover in this region, $m_{A}\le 60\gev$, a few points with large $\xi$ values are observed, which could be excluded by the upcoming experimental searches in that channel.
In \typeii~the situation is different, for a given scalar mass the achievable cross-sections have a lower bound. The large $\xi$ points are located at low $m_H\lesssim 350$~GeV and have reasonably large cross-sections just below the current experimental upper limit.
In short, we estimate that a factor of 4 improvement in the search sensitivity, which is very likely to be reached, would either see an exciting signal or eliminate these points, as a result, the first order PT with strength $\xi>3$ can be fully tested at the LHC. 

%%%% Fig.16 %%%%%

%%%%%%%%%%%
\subsection{Triple Higgs couplings and the implications of the future measurements}

%%%% Fig.17 %%%%%
\begin{figure}[t]
\begin{center}
\includegraphics[width=0.5\textwidth]{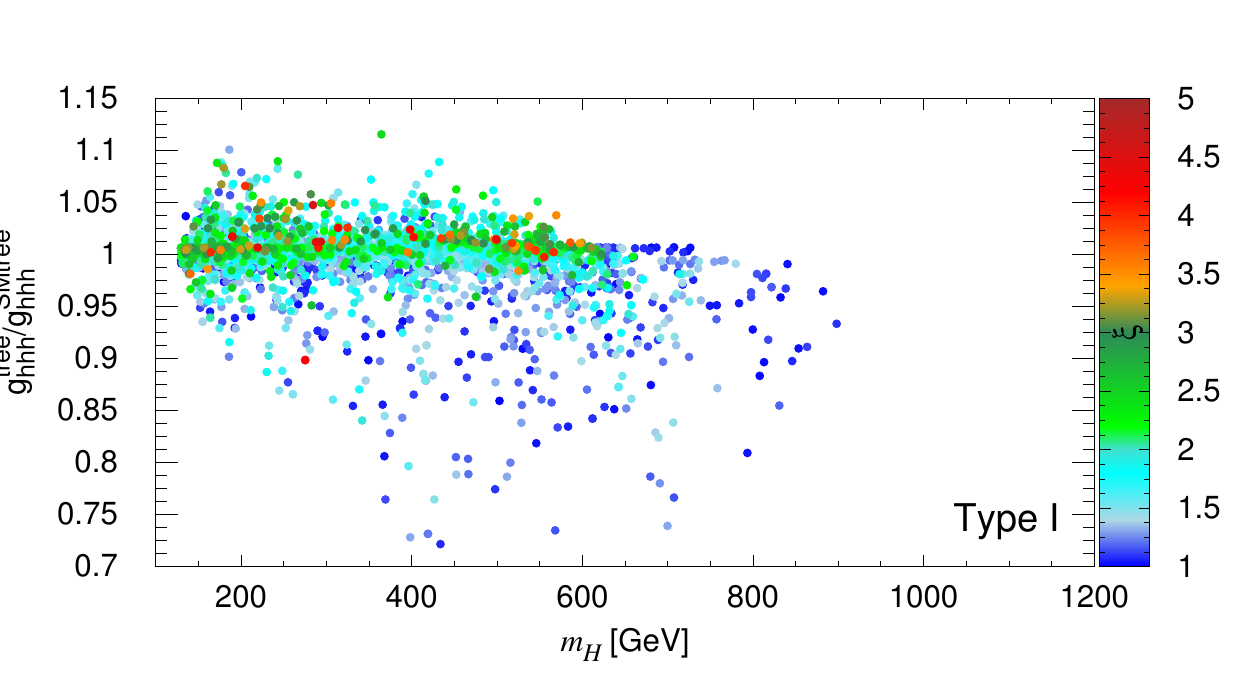}
\hspace{-3mm}
\includegraphics[width=0.5\textwidth]{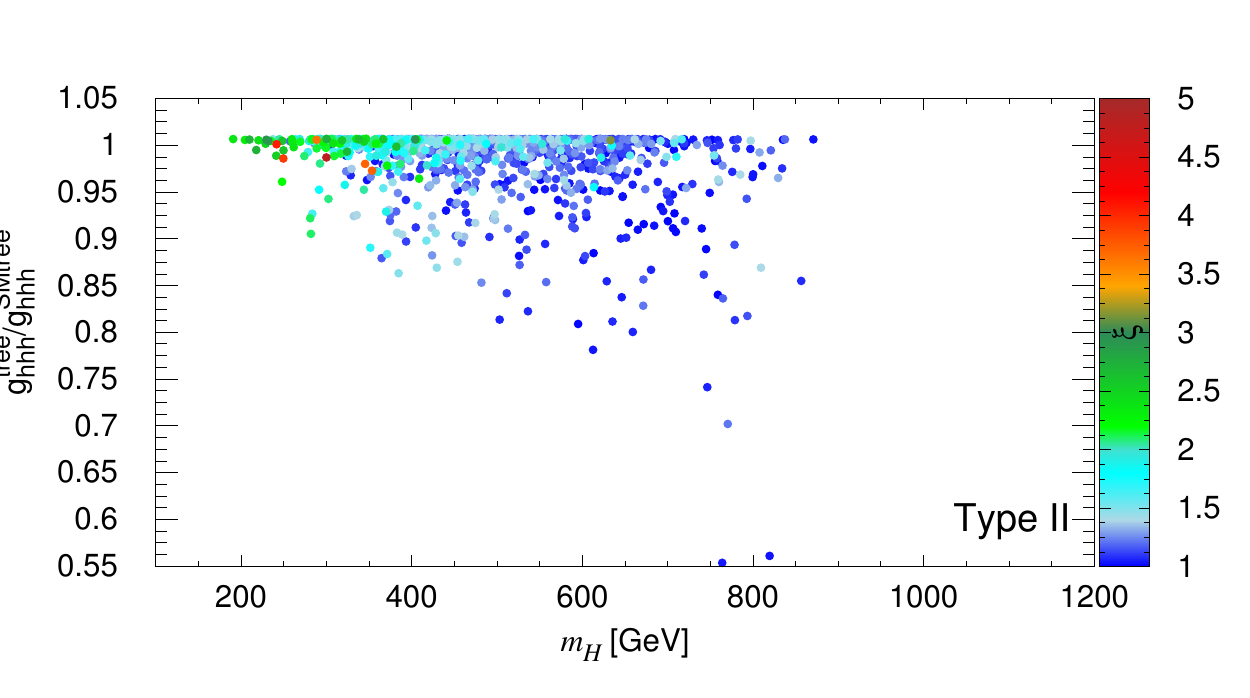}\\
\includegraphics[width=0.5\textwidth]{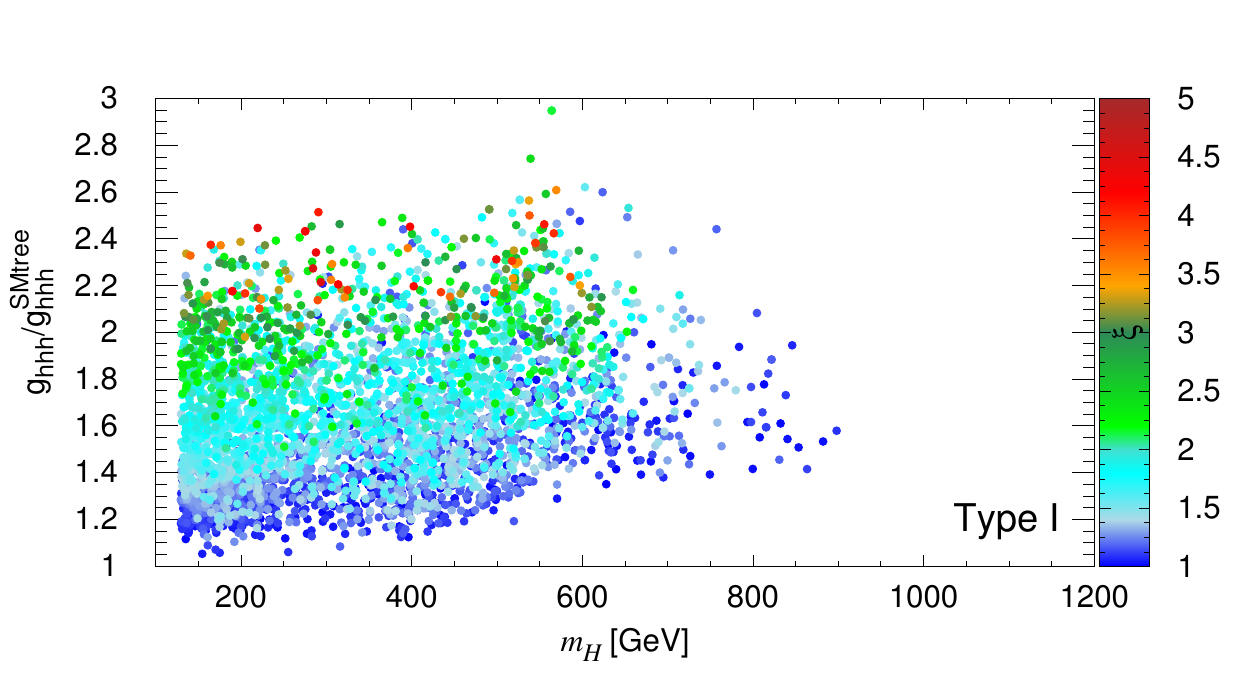}
\hspace{-3mm}
\includegraphics[width=0.5\textwidth]{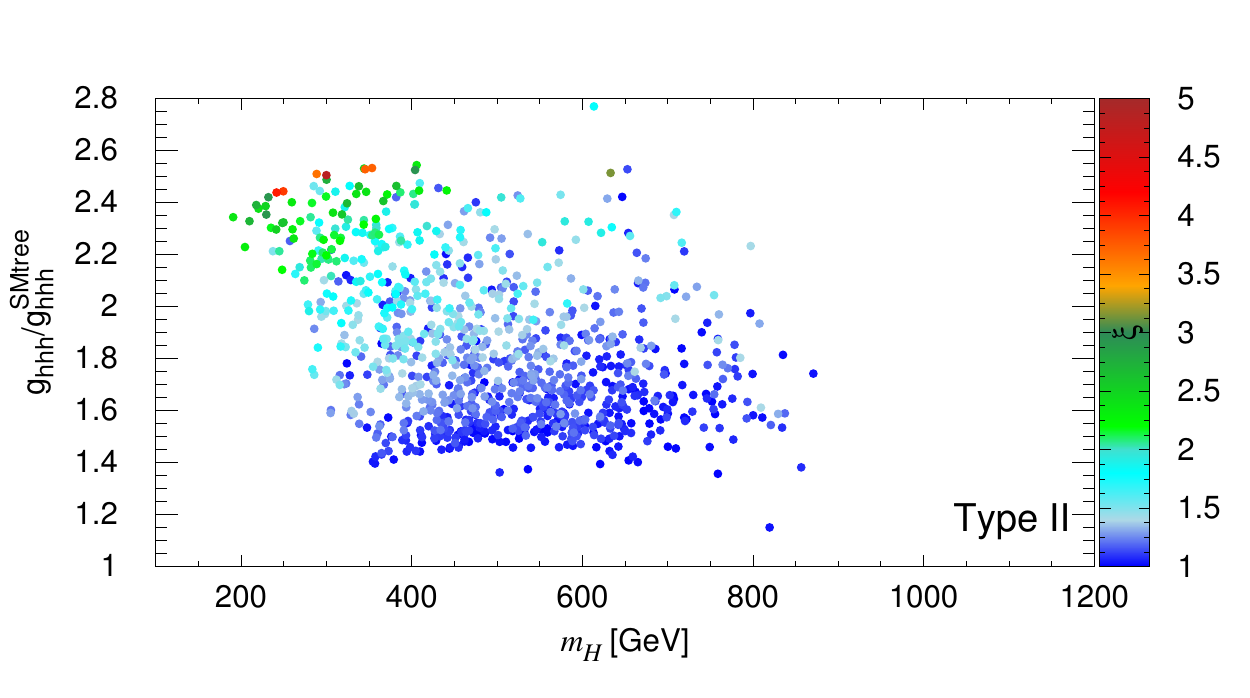}
\\[-1mm]
\caption{Triple Higgs  coupling $hhh$ at tree-level (upper) and at one-loop level (lower) normalized to the SM tree value $g^{\text {SM tree}}_{hhh}=3 m_h^2/v$. Note that tree-level Higgs self-coupling is not enhanced in \typeii. To have a better visualization only $\xi\geq 1$ points are shown.
}\label{fig:triphc}
\end{center}
\end{figure}

The scenarios that lead to the first order PT in the model have a mass spectrum below the TeV scale, as can be seen in Fig.~\ref{fig:vcTcMass} and Table~\ref{tab:masssper}. 
The presence of additional scalars that couple to the SM-like Higgs $h$ can modify the triple Higgs coupling $hhh$ at both tree-level and loop-level and thus leads to the deviation with respect to its SM value $g^{\text {SM tree}}_{hhh}$.
Moreover, such deviation can be significant near the alignment limit provided being away from the decoupling limit~\cite{Bernon:2015qea,Kanemura:2002vm}. We examine both the tree-level coupling and the one after the inclusion of the one-loop corrections. 
They are computed by taking the third derivative of the tree level potential $V_{0}$ and the one-loop potential $V_{0}+V_{\text{CW}}+V_{\text{CT}}$ with respect to $h$, respectively and shown in the upper and lower panels of Fig.~\ref{fig:triphc} (after normalizing the SM value $g^{\text {SM tree}}_{hhh}=3 m_h^2/v$). 
Focusing on the tree-level results, one can observe that the triple SM-like Higgs self-coupling $g_{hhh}$ in favor of the highly strong PT (\textit{i.e}, $\xi \gtrsim 3$) is close to its SM value $g^{\text {SM tree}}_{hhh}$, while large deviation (mostly suppression) of $g_{hhh}$ from $g^{\text {SM tree}}_{hhh}$ is possible for the weakly strong PT (\textit{i.e}, $\xi \lesssim 1.5$).
Another transparent observation is that the $hhh$ coupling at tree-level cannot be enhanced in \typei~(for $m_{H} \gtrsim 600\gev$) and \typeii~models, see Ref.~\cite{Bernon:2015qea} for analytical understanding of these features.
However, we stress that this conclusion will be dramatically changed when the one-loop corrections to the $hhh$ coupling are taken into account.
As shown in the lower panel plots, the coupling $g_{hhh}$ at one-loop level are absolutely enhanced in both models and the largest normalized coupling $g_{hhh}/g^{\text {SM tree}}_{hhh}$ can be about 2.5, corresponding to $\sim\!\!150\%$ enhancement. 
This allow us to conclude that the strong PT ($\xi \geq 1$) in the 2HDM typically induces the enhancement on the $hhh$ coupling.
Next, we would like to quantitatively explore the relation between the phase transition strength and the content of the derivation the triple Higgs coupling.
In general, the loop-level $hhh$ coupling exhibits a larger deviation with increased strength of the phase transition. 
Whereas, the tree-level $hhh$ coupling shown in the upper panel plots does not display such a proportionality behavior. 
This dramatic change implies that the loop corrections coming from the CW potential and counter-terms are important in general when the phase transition is of strong first order and can even be dominant over the tree-level contribution in the case of the extremely strong phase transition. 
It is also apparent in Fig.~\ref{fig:triphc} that the highly strong PT induces a substantial enhancement on the $hhh$ coupling. In contrast, the $hhh$ coupling normalized to the SM tree value can vary from $\sim1$ to 2.5 for the weakly strong PT of $\xi=1-2$.
This means that large triple Higgs coupling $hhh$ is a necessary but not sufficient condition of realizing the highly strong PT. 
For instance, if the deviation is smaller than 100\%, then possibility of the highly strong PT ($\xi \gtrsim 3~(2.5)$ in \typei~(II)) will be eliminated.
As a result, the size of the triple Higgs coupling $hhh$ derives an upper bound on the achievable value of $\xi$. In some sense, this is phenomenologically useful because we have built a connection between the phase transition involving the thermal contribution and a measurable observable at zero temperature. Therefore, the measurement of the triple Higgs coupling could be an indirect approach of probing the phase transition at colliders.

Experimentally, the deviation of the triple Higgs coupling can be detected at both lepton colliders (i.e., ILC~\cite{Asner:2013psa}, CEPC~\cite{CEPC-SPPCStudyGroup:2015csa} and FCC-ee\cite{Gomez-Ceballos:2013zzn} ) and hadron colliders such as LHC and SppC~\cite{Arkani-Hamed:2015vfh}.  
At hadron colliders, the resonant Higgs pair production is promising while special attention needs to be paid when the heavier CP-even state $H$ produces a destructive interference with the SM top box diagram process~\cite{Huang:2015tdv}.
Upon the sensitivity of 50\% supposed to be achieved at the HL-LHC, a large amount of the (nearly entire) parameter space in \typei~(II) model leading to strong PT can be probed through the di-Higgs production into $b\bar{b}\gamma\gamma$ and $b\bar{b}W^+W^-$ channels in the ultimate operation of LHC Run-2~\cite{Profumo:2014opa,Huang:2017jws,Kotwal:2016tex}.
In our case, $g_{hhh}$ has the same sign as the SM value and hence results in the destructive interference between the $s$-channel $h$-mediator triangle diagram and the top box diagrams of the $gg\to hh$ production process. This implies increasing $g_{hhh}$ will decrease the production cross section~\cite{Shao:2013bz}.
Previous studies demonstrated that when $g_{hhh}\simeq2.45g_{hhh}^{\text{SM tree}}$ an exact cancellation between these two diagrams is accomplished at the threshold of the di-Higgs invariant mass $m_{hh}=2m_t$ ~\cite{Barger:2013jfa,Frederix:2014hta,Huang:2015tdv}. 
Due to the low acceptance at LHC for large $g_{hhh}$, a cut $m_{hh}<2 m_{t}$ is imposed~\cite{Barger:2013jfa,Huang:2015tdv}. MVA analysis of Ref.~\cite{Barger:2013jfa} shows that, for the parameter space leading to the SFOPT (presented in Fig.~\ref{fig:triphc}), the observation significance in the $b\bar{b}\gamma\gamma$ channel with the integrated luminosity of 3 ab$^{-1}$ at 14 TeV would decrease from 10 to 4 in both \typei~and \typeii~models.
In measuring the triple Higgs coupling $hhh$ the lepton machines are typically more powerful, using the Higgs associated process $e^+e^-\rightarrow Z^{*} \to Zh^{*}(hh)$. The better designed sensitivities at the CEPC~\cite{CEPC-SPPCStudyGroup:2015csa}, FCC-ee\cite{Gomez-Ceballos:2013zzn} and ILC1000 are roughly 20-30\%.
This indicates that almost the full parameter spaces that are compatible with $\xi \geq 1$, particularly for $m_H \gtrsim 500\gev$, are within the future detection reach.

%%%% Fig.18 %%%%%
\begin{figure}[t]
\begin{center}
\includegraphics[width=0.5\textwidth]{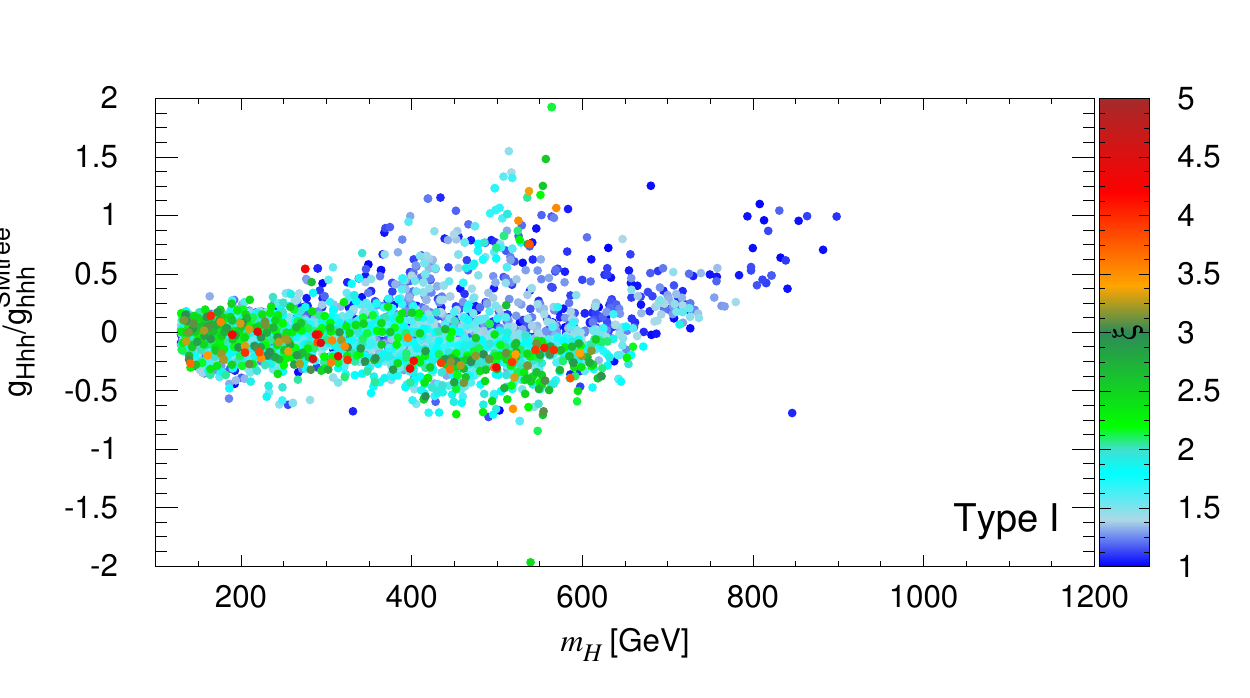}
\hspace{-3mm}
\includegraphics[width=0.5\textwidth]{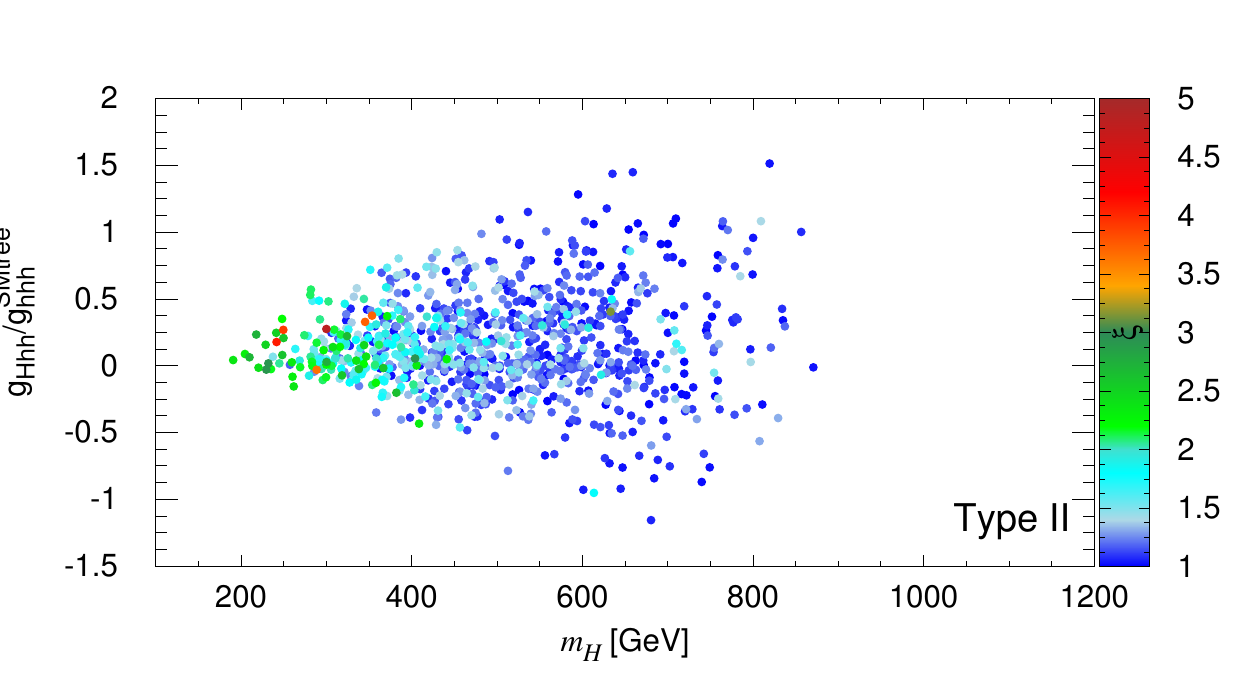}
\\[-1mm]
\caption{Triple Higgs  coupling $Hhh$ at one-loop level normalized to the SM tree value $g^{\text {SM tree}}_{hhh}=3 m_h^2/v$. In contrast to the $hhh$ coupling, the one-loop corrections to the $Hhh$ coupling are vanishingly small near the alignment limit. To have a better visualization only $\xi\geq1$ points are shown.
}\label{fig:triphhc}
\end{center}
\end{figure}

The other Higgs self-coupling of interest is the $Hhh$ coupling $g_{Hhh}$, which is also relevant to the Higgs pair production through the s-channel $H$ mediator triangle diagram.
The $Hhh$ coupling at one-loop as a function of $m_{H}$ is depicted in Fig.~\ref{fig:triphhc}.
In contrast to the $hhh$ coupling, the one-loop corrections to $Hhh$ coupling are vanishingly small near the alignment limit. We thus do not show the tree-level result.
It is important to mention that the $Hhh$ coupling can be significant even in the alignment limit, which can be observed in 
Fig.~\ref{fig:triphhc}. For instance, the $Hhh$ coupling is about $\pm(30-50)\%$ of the tree-level SM $hhh$ coupling for the highly strong PT ($\xi \geq 3$) and can be even comparable with or larger than $g^{\text{SM tree}}_{hhh}$ as the PT is weakly strong ($\xi \approx 1-2$).
Notably, the obtained $Hhh$ coupling $g_{Hhh}$ in the successful SFOEWPT scenarios can have either the same sign as or the opposite sign to the coupling $g_{hhh}$. The consequence of the sign flip of the $g_{Hhh}$ will affect the s-channel $H$-resonant triangle diagram contribution to the $gg \to hh$ process, whose amplitude is proportional to the product of $g_{Hhh}$ and $g_{Ht\bar{t}}=C^{H}_{U} y_{t}$, resulting in a change on the $m_{hh}$ lineshape due to the interference between the triangle diagram of the signal and the continuum top box diagram.
When the interference is destructive, special attention needs to be paid~\cite{Huang:2015tdv}.
The study of Ref.~\cite{Barger:2014taa} indicates that most of our SFOPT points can be detected at $5\sigma$ significance provided that $g_{Hhh}\times g_{Ht\bar{t}}> 300\gev$ at 14 TeV with integrated luminosity of 3 ab$^{-1}$ using the $b\bar{b}\gamma\gamma$ channel.

%-------------------------------------------------------------------------------
\section{Conclusions and Outlook}
\label{sec:Con}
%-------------------------------------------------------------------------------

Taking into account theoretical and up-to-date experimental constraints, we studied the electroweak phase transition in the framework of the CP-conserving 2HDM of \typei~and \typeii~models near the alignment limit. The thermal potential was expressed in terms of modified Bessel functions, which allows for a fast numerical evaluation and high precision compared to the simpler high/low temperature approximations.  
While both 1-stage and 2-stage phase transitions were shown to be realized within the 2HDM, in this paper we focused on scenarios leading to 1-stage phase transitions at electroweak scale, for which first order and the second order phase transitions are distinguished.

We analyzed the properties of the first order phase transition, observing that the field value of the electroweak symmetry breaking vacuum at the critical temperature is strongly related to the vacuum depth of the 1-loop potential at zero temperature, while the critical temperature reflecting the size of the thermal effect is characterized by the temperature-dependent potential.
In general, the critical temperature $T_{c}$ tends to be higher as the BSM states becomes heavier, and on the other hand $T_{c}$ can be down to $\sim~100\gev$ when at least one light BSM Higgs bosons present in the spectrum. We have also observed that the thermal correction to the mass is important in driving a SFOPT.

The strength of the transition, a key property for the electroweak baryogenesis mechanism, depends largely on the allowed mass spectrum. 
Requiring a SFOPT with $\xi \geq 1$ forces down the mass scale for the new scalars and the preferred ranges for all the scalar masses below $600\gev$.
We demonstrate that SFOPT (\textit{i.e.}, $\xi \geq 1$) required for baryogenesis is possible in both \typei~and \typeii~models. 
In \typei~model, SFOPT is achievable in the parameter space where a large mass splitting is present between two neutral Higgs bosons such as $m_H \gg m_{A}$ and $m_A \gg m_{H}$. In either case, the charged Higgs mass is close to either $m_{H}$ or $m_{A}$ required by the EWPD. 
The mass spectrum among the extra Higgs bosons in the \typeii~model is, on the contrary, strongly constrained due to flavor observables, which push the mass of the charged Higgs above $\sim 600 \gev$. As a result, scenarios leading to a SFOPT in \typeii~are $m_{H^{\pm}} \simeq m_H \gg m_{A}$ and $m_{H^{\pm}} \simeq  m_A \gg m_{H}$.
In view of large mass splitting between $H$ and $A$, both $pp\rightarrow H\rightarrow ZA$ and $pp\rightarrow A\rightarrow ZH$ can be ``smoking gun'' collider signatures related to a SFOPT in the 2HDMs as the cross sections via gluon-fusion production in these two channels predicted for SFOPT points are typically up to $\sim 1$~pb.
%depending on whether the background is sufficiently suppressed after employing the cut when performing the analysis, due to the fact that the cross sections via gluon-fusion production in these two channels predicted for SFOPT points are typically below $\sim 1$~pb.
%Cross-sections for selected channels at the 13 TeV LHC have been discussed.  
In addition to large mass splitting, SFOPT can also take place in \typei~even if all the masses of the three extra Higgs bosons ($A$, $H$ and $H^{\pm}$) are degenerate around 350~GeV. Such scenario leads to potentially testable consequences through the $A\to Zh$ decay channel at colliders. 

Following the analysis of the benchmark scenarios, we investigated the implications of a SFOEWPT on the LHC Higgs phenomenology. 
Various characteristic collider signatures at the 13 TeV LHC have been identified, among which the gluon-fusion production cross section of $A$ and $H$ in the $\tau\tau$ decay channel displays a correlation with the PT strength $\xi$. 
It turns out that new physics searches at collider machines can provide an indirect channel to examine the EWPT scenarios.
Finally, we verify that an enhancement on the triple Higgs coupling $hhh$ (including loop corrections) is a typical signature of the SFOPT driven by the additional doublet. The PT with larger strength is associated with larger deviation of the loop-level triple Higgs coupling $hhh$ with respect to the SM value, which can help to enhance an energy barrier. Meanwhile, we notice that the other triple Higgs coupling $g_{Hhh}$ can also be comparable with the triple Higgs coupling in the SM for SFOPT so that the search for the heavy neutral Higgs $H$ through the $gg\to hh$ process is possible for small $\tan\beta$ since the top Yukawa coupling of the $H$ is proportional to $\cot\beta$.

We leave for future work the interplay of gravitational waves signals and testable colliders signatures for SFOPT benchmark scenarios presented in this paper.
This success would build a link between early Universe cosmology and collider detection, which could provide additional constraints in the allowed parameter space of the 2HDM. We believe that such connection will have a significant physical value and serves as a useful guide for collider search strategies.
%It is also worthwhile to study the role of the CP-violating phase in the EWBG mechanism during the Universe cooling down.

%###################################################################

\section*{ACKNOWLEDGMENTS}

We would like to thank M.~Trott and J.~Cline for useful discussions and communication. We also appreciate N.~Chen for careful reading and comments on the manuscript.
JB is supported by the Collaborative Research Fund (CRF) under Grant No. HUKST4/CRF/13G. He also
thanks the LPSC Grenoble for support for a research stay during which part of this work was performed. The work of LGB is partially supported by the National Natural Science Foundation of China (under Grant No. 11605016), Basic Science Research Program through the National Research Foundation of Korea (NRF) funded by the Ministry of Education, Science and Technology (Grant No. NRF- 2016R1A2B4008759), and Korea Research Fellowship Program through the National Research Foundation of Korea (NRF) funded by the Ministry of Science, and I. C. T (Grant No. 2017H1D3A1A01014046). Y.J. acknowledges generous support by the Villum Fonden and the Discovery center. Y.J. also thanks for financial support provided by Chongqing University for multiple visits at different stages during the completion of this paper.

%%%%%%%%%%%%
\appendix
%%%%%%%%%%%%%%%%%%%%%%%%%%%%%%%%%%%%%%%%%%%%%%%%%%%%%%%%%%%%%%%%%%%%%%%%%%%%%%%%%%%%%%%%%%%%%%%%%%%%%%%%%%%%%%%%%%%%%%%%%%%%%%%%%%%%%%%%%%%%
\clearpage
%%%%%%%%%%%%%%%%%
\section{Thermal mass for SM gauge bosons}
\label{sec:thermass}
The thermal masses of the gauge bosons are more complicated. 
Only the longitudinal components receive corrections. The expressions for 
these in the SM can be found in Ref.~\cite{Carrington:1991hz}, 
\begin{eqnarray}
\Pi _{W^{\pm}}^{L} &=&\frac{11}{6}g^{2}T^{2},~\Pi _{W^{\pm}}^{T}=0  \notag \\
\Pi _{W^{3}}^{L} &=&\frac{11}{6}g^{2}T^{2},~\Pi _{W^{3}}^{T}=0   \\
\Pi _{A}^{L} &=&\frac{11}{6}g'^{2}T^{2} \notag
\end{eqnarray}%
where the script $L$ ($T$) denotes the longitudinal (transversal) mode. 
Their contributions from the extra Higgs doublet are easy to be included
\begin{align}
\Delta \Pi _{W^{\pm}}^{L} & = \Delta \Pi _{W^{3}}^{L} =\frac{1}{6}g^{2}T^{2}, \\
\Delta \Pi _{A}^{L} & =\frac{1}{6}g'^{2}T^{2},
\end{align}

Adding them together, for the longitudinally polarized $W$ boson, the result is 
\beq
M_{W^{\pm}_L}^2 = {1 \over 4} g^2 (h^2_1+h^2_2) + 2 g^2 T^2.
\eeq 
This includes contributions from gauge boson self-interactions, two Higgs doublets and 
all three fermion families. 
The masses of the longitudinal $Z$ and $A$ are determined by diagonalizing the matrix 
\beq
\frac{1}{4}(h^2_1+h^2_2)
\begin{pmatrix}
g^2& -g g^\prime \\
-g g^\prime & g^{\prime 2}
\end{pmatrix}
+
\begin{pmatrix}
2 g^2 T^2 & 0 \\
0 & 2 g^{\prime 2} T^2 
\end{pmatrix}.
\eeq
The eigenvalues can be written as
\beq
M_{Z_L,\gamma_L}^2 = \frac{1}{8} (g^2+g'^2) (h^2_1+h^2_2) + (g^2 + g^{\prime 2} )T^2 \pm \Delta, 
\eeq
where 
\beq
\Delta^2 =\frac{1}{64}  (g^2 + g^{\prime 2} )^2(h_{1}^2 + h_{2}^2+8T^2)^2
- g^2 g^{\prime 2} T^2 ( h_{1}^2 + h_{2}^2 + 4 T^2). 
\eeq

\bibliography{ewpt2hdm_jhep_rev}

\providecommand{\href}[2]{#2}\begingroup\raggedright\begin{thebibliography}{100}

\bibitem{Aad:2012tfa}
{\bf ATLAS} Collaboration, G.~Aad et~al., {\it {Observation of a new particle
  in the search for the Standard Model Higgs boson with the ATLAS detector at
  the LHC}},  {\em Phys. Lett.} {\bf B716} (2012) 1--29,
  [\href{http://arxiv.org/abs/1207.7214}{{\tt arXiv:1207.7214}}].

\bibitem{Chatrchyan:2012xdj}
{\bf CMS} Collaboration, S.~Chatrchyan et~al., {\it {Observation of a new boson
  at a mass of 125 GeV with the CMS experiment at the LHC}},  {\em Phys. Lett.}
  {\bf B716} (2012) 30--61, [\href{http://arxiv.org/abs/1207.7235}{{\tt
  arXiv:1207.7235}}].

\bibitem{Sakharov:1967dj}
A.~D. Sakharov, {\it {Violation of CP Invariance, c Asymmetry, and Baryon
  Asymmetry of the Universe}},  {\em Pisma Zh. Eksp. Teor. Fiz.} {\bf 5} (1967)
  32--35. [Usp. Fiz. Nauk161,61(1991)].

\bibitem{Kuzmin:1985mm}
V.~A. Kuzmin, V.~A. Rubakov, and M.~E. Shaposhnikov, {\it {On the Anomalous
  Electroweak Baryon Number Nonconservation in the Early Universe}},  {\em
  Phys. Lett.} {\bf 155B} (1985) 36.

\bibitem{DOnofrio:2014rug}
M.~D'Onofrio, K.~Rummukainen, and A.~Tranberg, {\it {Sphaleron Rate in the
  Minimal Standard Model}},  {\em Phys. Rev. Lett.} {\bf 113} (2014), no.~14
  141602, [\href{http://arxiv.org/abs/1404.3565}{{\tt arXiv:1404.3565}}].

\bibitem{Kajantie:1996mn}
K.~Kajantie, M.~Laine, K.~Rummukainen, and M.~E. Shaposhnikov, {\it {Is there a
  hot electroweak phase transition at m(H) larger or equal to m(W)?}},  {\em
  Phys. Rev. Lett.} {\bf 77} (1996) 2887--2890,
  [\href{http://arxiv.org/abs/hep-ph/9605288}{{\tt hep-ph/9605288}}].

\bibitem{Kajantie:1996qd}
K.~Kajantie, M.~Laine, K.~Rummukainen, and M.~E. Shaposhnikov, {\it {A
  Nonperturbative analysis of the finite T phase transition in SU(2) x U(1)
  electroweak theory}},  {\em Nucl. Phys.} {\bf B493} (1997) 413--438,
  [\href{http://arxiv.org/abs/hep-lat/9612006}{{\tt hep-lat/9612006}}].

\bibitem{Csikor:1998eu}
F.~Csikor, Z.~Fodor, and J.~Heitger, {\it {Endpoint of the hot electroweak
  phase transition}},  {\em Phys. Rev. Lett.} {\bf 82} (1999) 21--24,
  [\href{http://arxiv.org/abs/hep-ph/9809291}{{\tt hep-ph/9809291}}].

\bibitem{Aoki:1999fi}
Y.~Aoki, F.~Csikor, Z.~Fodor, and A.~Ukawa, {\it {The Endpoint of the first
  order phase transition of the SU(2) gauge Higgs model on a four-dimensional
  isotropic lattice}},  {\em Phys. Rev.} {\bf D60} (1999) 013001,
  [\href{http://arxiv.org/abs/hep-lat/9901021}{{\tt hep-lat/9901021}}].

\bibitem{Morrissey:2012db}
D.~E. Morrissey and M.~J. Ramsey-Musolf, {\it {Electroweak baryogenesis}},
  {\em New J. Phys.} {\bf 14} (2012) 125003,
  [\href{http://arxiv.org/abs/1206.2942}{{\tt arXiv:1206.2942}}].

\bibitem{Jiang:2015cwa}
M.~Jiang, L.~Bian, W.~Huang, and J.~Shu, {\it {Impact of a complex singlet:
  Electroweak baryogenesis and dark matter}},  {\em Phys. Rev.} {\bf D93}
  (2016), no.~6 065032, [\href{http://arxiv.org/abs/1502.07574}{{\tt
  arXiv:1502.07574}}].

\bibitem{Lee:1974jb}
T.~D. Lee, {\it {CP Nonconservation and Spontaneous Symmetry Breaking}},  {\em
  Phys. Rept.} {\bf 9} (1974) 143--177. [,124(1974)].

\bibitem{Inoue:2014nva}
S.~Inoue, M.~J. Ramsey-Musolf, and Y.~Zhang, {\it {CP-violating phenomenology
  of flavor conserving two Higgs doublet models}},  {\em Phys. Rev.} {\bf D89}
  (2014), no.~11 115023, [\href{http://arxiv.org/abs/1403.4257}{{\tt
  arXiv:1403.4257}}].

\bibitem{Bian:2017jpt}
L.~Bian, N.~Chen, and Y.~Zhang, {\it {CP violation effects in the diphoton
  spectrum of heavy scalars}},  {\em Phys. Rev.} {\bf D96} (2017), no.~9
  095008, [\href{http://arxiv.org/abs/1706.09425}{{\tt arXiv:1706.09425}}].

\bibitem{Bian:2017wfv}
L.~Bian, H.-K. Guo, and J.~Shu, {\it {Gravitational Waves, baryon asymmetry of
  the universe and electric dipole moment in the CP-violating NMSSM}},
  \href{http://arxiv.org/abs/1704.02488}{{\tt arXiv:1704.02488}}.

\bibitem{Bian:2016zba}
L.~Bian and N.~Chen, {\it {Cancellation mechanism in the predictions of
  electric dipole moments}},  {\em Phys. Rev.} {\bf D95} (2017), no.~11 115029,
  [\href{http://arxiv.org/abs/1608.07975}{{\tt arXiv:1608.07975}}].

\bibitem{Bian:2014zka}
L.~Bian, T.~Liu, and J.~Shu, {\it {Cancellations Between Two-Loop Contributions
  to the Electron Electric Dipole Moment with a CP-Violating Higgs Sector}},
  {\em Phys. Rev. Lett.} {\bf 115} (2015) 021801,
  [\href{http://arxiv.org/abs/1411.6695}{{\tt arXiv:1411.6695}}].

\bibitem{Chupp:2017rkp}
T.~Chupp, P.~Fierlinger, M.~Ramsey-Musolf, and J.~Singh, {\it {Electric Dipole
  Moments of the Atoms, Molecules, Nuclei and Particles}},
  \href{http://arxiv.org/abs/1710.02504}{{\tt arXiv:1710.02504}}.

\bibitem{Dorsch:2013wja}
G.~C. Dorsch, S.~J. Huber, and J.~M. No, {\it {A strong electroweak phase
  transition in the 2HDM after LHC8}},  {\em JHEP} {\bf 10} (2013) 029,
  [\href{http://arxiv.org/abs/1305.6610}{{\tt arXiv:1305.6610}}].

\bibitem{Dorsch:2014qja}
G.~C. Dorsch, S.~J. Huber, K.~Mimasu, and J.~M. No, {\it {Echoes of the
  Electroweak Phase Transition: Discovering a second Higgs doublet through $A_0
  \rightarrow ZH_0$}},  {\em Phys. Rev. Lett.} {\bf 113} (2014), no.~21 211802,
  [\href{http://arxiv.org/abs/1405.5537}{{\tt arXiv:1405.5537}}].

\bibitem{Basler:2016obg}
P.~Basler, M.~Krause, M.~Muhlleitner, J.~Wittbrodt, and A.~Wlotzka, {\it
  {Strong First Order Electroweak Phase Transition in the CP-Conserving 2HDM
  Revisited}},  {\em JHEP} {\bf 02} (2017) 121,
  [\href{http://arxiv.org/abs/1612.04086}{{\tt arXiv:1612.04086}}].

\bibitem{Cline:1996mga}
J.~M. Cline and P.-A. Lemieux, {\it {Electroweak phase transition in two Higgs
  doublet models}},  {\em Phys. Rev.} {\bf D55} (1997) 3873--3881,
  [\href{http://arxiv.org/abs/hep-ph/9609240}{{\tt hep-ph/9609240}}].

\bibitem{Fromme:2006cm}
L.~Fromme, S.~J. Huber, and M.~Seniuch, {\it {Baryogenesis in the two-Higgs
  doublet model}},  {\em JHEP} {\bf 11} (2006) 038,
  [\href{http://arxiv.org/abs/hep-ph/0605242}{{\tt hep-ph/0605242}}].

\bibitem{Cline:2011mm}
J.~M. Cline, K.~Kainulainen, and M.~Trott, {\it {Electroweak Baryogenesis in
  Two Higgs Doublet Models and B meson anomalies}},  {\em JHEP} {\bf 11} (2011)
  089, [\href{http://arxiv.org/abs/1107.3559}{{\tt arXiv:1107.3559}}].

\bibitem{Dorsch:2016nrg}
G.~C. Dorsch, S.~J. Huber, T.~Konstandin, and J.~M. No, {\it {A Second Higgs
  Doublet in the Early Universe: Baryogenesis and Gravitational Waves}},  {\em
  JCAP} {\bf 1705} (2017), no.~05 052,
  [\href{http://arxiv.org/abs/1611.05874}{{\tt arXiv:1611.05874}}].

\bibitem{Haarr:2016qzq}
A.~Haarr, A.~Kvellestad, and T.~C. Petersen, {\it {Disfavouring Electroweak
  Baryogenesis and a hidden Higgs in a CP-violating Two-Higgs-Doublet Model}},
  \href{http://arxiv.org/abs/1611.05757}{{\tt arXiv:1611.05757}}.

\bibitem{Basler:2017uxn}
P.~Basler, M.~Mühlleitner, and J.~Wittbrodt, {\it {The CP-Violating 2HDM in
  Light of a Strong First Order Electroweak Phase Transition and Implications
  for Higgs Pair Production}},  \href{http://arxiv.org/abs/1711.04097}{{\tt
  arXiv:1711.04097}}.

\bibitem{Bernon:2015qea}
J.~Bernon, J.~F. Gunion, H.~E. Haber, Y.~Jiang, and S.~Kraml, {\it
  {Scrutinizing the alignment limit in two-Higgs-doublet models:
  m$_h$=125??GeV}},  {\em Phys. Rev.} {\bf D92} (2015), no.~7 075004,
  [\href{http://arxiv.org/abs/1507.00933}{{\tt arXiv:1507.00933}}].

\bibitem{Bernon:2015wef}
J.~Bernon, J.~F. Gunion, H.~E. Haber, Y.~Jiang, and S.~Kraml, {\it
  {Scrutinizing the alignment limit in two-Higgs-doublet models. II.
  m$_H$=125??GeV}},  {\em Phys. Rev.} {\bf D93} (2016), no.~3 035027,
  [\href{http://arxiv.org/abs/1511.03682}{{\tt arXiv:1511.03682}}].

\bibitem{Grzadkowski:2016szj}
B.~Grzadkowski, O.~M. Ogreid, and P.~Osland, {\it {Spontaneous CP violation in
  the 2HDM: physical conditions and the alignment limit}},  {\em Phys. Rev.}
  {\bf D94} (2016), no.~11 115002, [\href{http://arxiv.org/abs/1609.04764}{{\tt
  arXiv:1609.04764}}].

\bibitem{Kamionkowski:1993fg}
M.~Kamionkowski, A.~Kosowsky, and M.~S. Turner, {\it {Gravitational radiation
  from first order phase transitions}},  {\em Phys. Rev.} {\bf D49} (1994)
  2837--2851, [\href{http://arxiv.org/abs/astro-ph/9310044}{{\tt
  astro-ph/9310044}}].

\bibitem{Caprini:2015zlo}
C.~Caprini et~al., {\it {Science with the space-based interferometer eLISA. II:
  Gravitational waves from cosmological phase transitions}},  {\em JCAP} {\bf
  1604} (2016), no.~04 001, [\href{http://arxiv.org/abs/1512.06239}{{\tt
  arXiv:1512.06239}}].

\bibitem{Kudoh:2005as}
H.~Kudoh, A.~Taruya, T.~Hiramatsu, and Y.~Himemoto, {\it {Detecting a
  gravitational-wave background with next-generation space interferometers}},
  {\em Phys. Rev.} {\bf D73} (2006) 064006,
  [\href{http://arxiv.org/abs/gr-qc/0511145}{{\tt gr-qc/0511145}}].

\bibitem{Lee:2004we}
C.~Lee, V.~Cirigliano, and M.~J. Ramsey-Musolf, {\it {Resonant relaxation in
  electroweak baryogenesis}},  {\em Phys. Rev.} {\bf D71} (2005) 075010,
  [\href{http://arxiv.org/abs/hep-ph/0412354}{{\tt hep-ph/0412354}}].

\bibitem{Riotto:1998zb}
A.~Riotto, {\it {The More relaxed supersymmetric electroweak baryogenesis}},
  {\em Phys. Rev.} {\bf D58} (1998) 095009,
  [\href{http://arxiv.org/abs/hep-ph/9803357}{{\tt hep-ph/9803357}}].

\bibitem{Moore:2000wx}
G.~D. Moore, {\it {Electroweak bubble wall friction: Analytic results}},  {\em
  JHEP} {\bf 03} (2000) 006, [\href{http://arxiv.org/abs/hep-ph/0001274}{{\tt
  hep-ph/0001274}}].

\bibitem{Konstandin:2004gy}
T.~Konstandin, T.~Prokopec, and M.~G. Schmidt, {\it {Kinetic description of
  fermion flavor mixing and CP-violating sources for baryogenesis}},  {\em
  Nucl. Phys.} {\bf B716} (2005) 373--400,
  [\href{http://arxiv.org/abs/hep-ph/0410135}{{\tt hep-ph/0410135}}].

\bibitem{Li:2008ez}
Y.~Li, S.~Profumo, and M.~Ramsey-Musolf, {\it {Bino-driven Electroweak
  Baryogenesis with highly suppressed Electric Dipole Moments}},  {\em Phys.
  Lett.} {\bf B673} (2009) 95--100, [\href{http://arxiv.org/abs/0811.1987}{{\tt
  arXiv:0811.1987}}].

\bibitem{Cline:1997vk}
J.~M. Cline, M.~Joyce, and K.~Kainulainen, {\it {Supersymmetric electroweak
  baryogenesis in the WKB approximation}},  {\em Phys. Lett.} {\bf B417} (1998)
  79--86, [\href{http://arxiv.org/abs/hep-ph/9708393}{{\tt hep-ph/9708393}}].
  [Erratum: Phys. Lett.B448,321(1999)].

\bibitem{Hindmarsh:2013xza}
M.~Hindmarsh, S.~J. Huber, K.~Rummukainen, and D.~J. Weir, {\it {Gravitational
  waves from the sound of a first order phase transition}},  {\em Phys. Rev.
  Lett.} {\bf 112} (2014) 041301, [\href{http://arxiv.org/abs/1304.2433}{{\tt
  arXiv:1304.2433}}].

\bibitem{Hindmarsh:2015qta}
M.~Hindmarsh, S.~J. Huber, K.~Rummukainen, and D.~J. Weir, {\it {Numerical
  simulations of acoustically generated gravitational waves at a first order
  phase transition}},  {\em Phys. Rev.} {\bf D92} (2015), no.~12 123009,
  [\href{http://arxiv.org/abs/1504.03291}{{\tt arXiv:1504.03291}}].

\bibitem{Glashow:1976nt}
S.~L. Glashow and S.~Weinberg, {\it {Natural Conservation Laws for Neutral
  Currents}},  {\em Phys. Rev.} {\bf D15} (1977) 1958.

\bibitem{Paschos:1976ay}
E.~A. Paschos, {\it {Diagonal Neutral Currents}},  {\em Phys. Rev.} {\bf D15}
  (1977) 1966.

\bibitem{Davidson:2005cw}
S.~Davidson and H.~E. Haber, {\it {Basis-independent methods for the
  two-Higgs-doublet model}},  {\em Phys. Rev.} {\bf D72} (2005) 035004,
  [\href{http://arxiv.org/abs/hep-ph/0504050}{{\tt hep-ph/0504050}}]. [Erratum:
  Phys. Rev.D72,099902(2005)].

\bibitem{Gunion:2002zf}
J.~F. Gunion and H.~E. Haber, {\it {The CP conserving two Higgs doublet model:
  The Approach to the decoupling limit}},  {\em Phys. Rev.} {\bf D67} (2003)
  075019, [\href{http://arxiv.org/abs/hep-ph/0207010}{{\tt hep-ph/0207010}}].

\bibitem{Branco:2011iw}
G.~C. Branco, P.~M. Ferreira, L.~Lavoura, M.~N. Rebelo, M.~Sher, and J.~P.
  Silva, {\it {Theory and phenomenology of two-Higgs-doublet models}},  {\em
  Phys. Rept.} {\bf 516} (2012) 1--102,
  [\href{http://arxiv.org/abs/1106.0034}{{\tt arXiv:1106.0034}}].

\bibitem{Aoki:2009ha}
M.~Aoki, S.~Kanemura, K.~Tsumura, and K.~Yagyu, {\it {Models of Yukawa
  interaction in the two Higgs doublet model, and their collider
  phenomenology}},  {\em Phys. Rev.} {\bf D80} (2009) 015017,
  [\href{http://arxiv.org/abs/0902.4665}{{\tt arXiv:0902.4665}}].

\bibitem{Grinstein:2015rtl}
B.~Grinstein, C.~W. Murphy, and P.~Uttayarat, {\it {One-loop corrections to the
  perturbative unitarity bounds in the CP-conserving two-Higgs doublet model
  with a softly broken $ {\mathrm{\mathbb{Z}}}_2 $ symmetry}},  {\em JHEP} {\bf
  06} (2016) 070, [\href{http://arxiv.org/abs/1512.04567}{{\tt
  arXiv:1512.04567}}].

\bibitem{Akeroyd:2000wc}
A.~G. Akeroyd, A.~Arhrib, and E.-M. Naimi, {\it {Note on tree level unitarity
  in the general two Higgs doublet model}},  {\em Phys. Lett.} {\bf B490}
  (2000) 119--124, [\href{http://arxiv.org/abs/hep-ph/0006035}{{\tt
  hep-ph/0006035}}].

\bibitem{Ginzburg:2005dt}
I.~F. Ginzburg and I.~P. Ivanov, {\it {Tree-level unitarity constraints in the
  most general 2HDM}},  {\em Phys. Rev.} {\bf D72} (2005) 115010,
  [\href{http://arxiv.org/abs/hep-ph/0508020}{{\tt hep-ph/0508020}}].

\bibitem{Gerard:2007kn}
J.~M. Gerard and M.~Herquet, {\it {A Twisted custodial symmetry in the
  two-Higgs-doublet model}},  {\em Phys. Rev. Lett.} {\bf 98} (2007) 251802,
  [\href{http://arxiv.org/abs/hep-ph/0703051}{{\tt hep-ph/0703051}}].

\bibitem{Haber:2010bw}
H.~E. Haber and D.~O'Neil, {\it {Basis-independent methods for the
  two-Higgs-doublet model III: The CP-conserving limit, custodial symmetry, and
  the oblique parameters S, T, U}},  {\em Phys. Rev.} {\bf D83} (2011) 055017,
  [\href{http://arxiv.org/abs/1011.6188}{{\tt arXiv:1011.6188}}].

\bibitem{Belle:2016ufb}
{\bf Belle} Collaboration, A.~Abdesselam et~al., {\it {Measurement of the
  inclusive $B\to X_{s+d} \gamma$ branching fraction, photon energy spectrum
  and HQE parameters}},  in {\em {Proceedings, 38th International Conference on
  High Energy Physics (ICHEP 2016): Chicago, IL, USA, August 3-10, 2016}},
  2016.
\newblock \href{http://arxiv.org/abs/1608.02344}{{\tt arXiv:1608.02344}}.

\bibitem{Misiak:2017bgg}
M.~Misiak and M.~Steinhauser, {\it {Weak radiative decays of the B meson and
  bounds on $M_{H^\pm }$ in the Two-Higgs-Doublet Model}},  {\em Eur. Phys. J.}
  {\bf C77} (2017), no.~3 201, [\href{http://arxiv.org/abs/1702.04571}{{\tt
  arXiv:1702.04571}}].

\bibitem{Bernon:2015hsa}
J.~Bernon and B.~Dumont, {\it {Lilith: a tool for constraining new physics from
  Higgs measurements}},  {\em Eur. Phys. J.} {\bf C75} (2015), no.~9 440,
  [\href{http://arxiv.org/abs/1502.04138}{{\tt arXiv:1502.04138}}].

\bibitem{Aaboud:2017sjh}
{\bf ATLAS} Collaboration, M.~Aaboud et~al., {\it {Search for additional heavy
  neutral Higgs and gauge bosons in the ditau final state produced in 36
  fb$^{-1}$ of $pp$ collisions at $\sqrt{s}$ = 13 TeV with the ATLAS
  detector}},  \href{http://arxiv.org/abs/1709.07242}{{\tt arXiv:1709.07242}}.

\bibitem{CMS:2016pkt}
{\bf CMS} Collaboration, C.~Collaboration, {\it {Search for a neutral MSSM
  Higgs boson decaying into $\tau \tau$ at 13 TeV}}, .

\bibitem{Khachatryan:2016are}
{\bf CMS} Collaboration, V.~Khachatryan et~al., {\it {Search for neutral
  resonances decaying into a Z boson and a pair of b jets or $\tau$ leptons}},
  {\em Phys. Lett.} {\bf B759} (2016) 369--394,
  [\href{http://arxiv.org/abs/1603.02991}{{\tt arXiv:1603.02991}}].

\bibitem{Aad:2015wra}
{\bf ATLAS} Collaboration, G.~Aad et~al., {\it {Search for a CP-odd Higgs boson
  decaying to Zh in pp collisions at $\sqrt{s} = 8$ TeV with the ATLAS
  detector}},  {\em Phys. Lett.} {\bf B744} (2015) 163--183,
  [\href{http://arxiv.org/abs/1502.04478}{{\tt arXiv:1502.04478}}].

\bibitem{ATLAS:2017nxi}
{\bf ATLAS} Collaboration, T.~A. collaboration, {\it {Search for heavy
  resonances decaying to a $W$ or $Z$ boson and a Higgs boson in final states
  with leptons and $b$-jets in 36.1~fb$^{-1}$ of $pp$ collision data at $\sqrt
  s = 13$~TeV with the ATLAS detector}}, .

\bibitem{DiazCruz:1992uw}
J.~L. Diaz-Cruz and A.~Mendez, {\it {Vacuum alignment in multiscalar models}},
  {\em Nucl. Phys.} {\bf B380} (1992) 39--50.

\bibitem{Ferreira:2004yd}
P.~M. Ferreira, R.~Santos, and A.~Barroso, {\it {Stability of the tree-level
  vacuum in two Higgs doublet models against charge or CP spontaneous
  violation}},  {\em Phys. Lett.} {\bf B603} (2004) 219--229,
  [\href{http://arxiv.org/abs/hep-ph/0406231}{{\tt hep-ph/0406231}}]. [Erratum:
  Phys. Lett.B629,114(2005)].

\bibitem{Barroso:2005sm}
A.~Barroso, P.~M. Ferreira, and R.~Santos, {\it {Charge and CP symmetry
  breaking in two Higgs doublet models}},  {\em Phys. Lett.} {\bf B632} (2006)
  684--687, [\href{http://arxiv.org/abs/hep-ph/0507224}{{\tt hep-ph/0507224}}].

\bibitem{Barroso:2006pa}
A.~Barroso, P.~M. Ferreira, R.~Santos, and J.~P. Silva, {\it {Stability of the
  normal vacuum in multi-Higgs-doublet models}},  {\em Phys. Rev.} {\bf D74}
  (2006) 085016, [\href{http://arxiv.org/abs/hep-ph/0608282}{{\tt
  hep-ph/0608282}}].

\bibitem{Barroso:2007rr}
A.~Barroso, P.~M. Ferreira, and R.~Santos, {\it {Neutral minima in two-Higgs
  doublet models}},  {\em Phys. Lett.} {\bf B652} (2007) 181--193,
  [\href{http://arxiv.org/abs/hep-ph/0702098}{{\tt hep-ph/0702098}}].

\bibitem{Ginzburg:2007jn}
I.~F. Ginzburg and K.~A. Kanishev, {\it {Different vacua in 2HDM}},  {\em Phys.
  Rev.} {\bf D76} (2007) 095013, [\href{http://arxiv.org/abs/0704.3664}{{\tt
  arXiv:0704.3664}}].

\bibitem{Ginzburg:2010wa}
I.~F. Ginzburg, K.~A. Kanishev, M.~Krawczyk, and D.~Sokolowska, {\it {Evolution
  of Universe to the present inert phase}},  {\em Phys. Rev.} {\bf D82} (2010)
  123533, [\href{http://arxiv.org/abs/1009.4593}{{\tt arXiv:1009.4593}}].

\bibitem{Patel:2013zla}
H.~H. Patel, M.~J. Ramsey-Musolf, and M.~B. Wise, {\it {Color Breaking in the
  Early Universe}},  {\em Phys. Rev.} {\bf D88} (2013), no.~1 015003,
  [\href{http://arxiv.org/abs/1303.1140}{{\tt arXiv:1303.1140}}].

\bibitem{Ramsey-Musolf:2017tgh}
M.~J. Ramsey-Musolf, G.~White, and P.~Winslow, {\it {Color Breaking
  Baryogenesis}},  \href{http://arxiv.org/abs/1708.07511}{{\tt
  arXiv:1708.07511}}.

\bibitem{Coleman:1973jx}
S.~R. Coleman and E.~J. Weinberg, {\it {Radiative Corrections as the Origin of
  Spontaneous Symmetry Breaking}},  {\em Phys. Rev.} {\bf D7} (1973)
  1888--1910.

\bibitem{Camargo-Molina:2013qva}
J.~E. Camargo-Molina, B.~O'Leary, W.~Porod, and F.~Staub, {\it
  {$\mathbf{Vevacious}$: A Tool For Finding The Global Minima Of One-Loop
  Effective Potentials With Many Scalars}},  {\em Eur. Phys. J.} {\bf C73}
  (2013), no.~10 2588, [\href{http://arxiv.org/abs/1307.1477}{{\tt
  arXiv:1307.1477}}].

\bibitem{Quiros:1999jp}
M.~Quiros, {\it {Finite temperature field theory and phase transitions}},  in
  {\em {Proceedings, Summer School in High-energy physics and cosmology:
  Trieste, Italy, June 29-July 17, 1998}}, pp.~187--259, 1999.
\newblock \href{http://arxiv.org/abs/hep-ph/9901312}{{\tt hep-ph/9901312}}.

\bibitem{Delaunay:2007wb}
C.~Delaunay, C.~Grojean, and J.~D. Wells, {\it {Dynamics of Non-renormalizable
  Electroweak Symmetry Breaking}},  {\em JHEP} {\bf 04} (2008) 029,
  [\href{http://arxiv.org/abs/0711.2511}{{\tt arXiv:0711.2511}}].

\bibitem{Bernon:2014nxa}
J.~Bernon, J.~F. Gunion, Y.~Jiang, and S.~Kraml, {\it {Light Higgs bosons in
  Two-Higgs-Doublet Models}},  {\em Phys. Rev.} {\bf D91} (2015), no.~7 075019,
  [\href{http://arxiv.org/abs/1412.3385}{{\tt arXiv:1412.3385}}].

\bibitem{Khachatryan:2015baw}
{\bf CMS} Collaboration, V.~Khachatryan et~al., {\it {Search for a low-mass
  pseudoscalar Higgs boson produced in association with a $b\bar{b}$ pair in pp
  collisions at $\sqrt{s} =$ 8 TeV}},  {\em Phys. Lett.} {\bf B758} (2016)
  296--320, [\href{http://arxiv.org/abs/1511.03610}{{\tt arXiv:1511.03610}}].

\bibitem{Dolan:1973qd}
L.~Dolan and R.~Jackiw, {\it {Symmetry Behavior at Finite Temperature}},  {\em
  Phys. Rev.} {\bf D9} (1974) 3320--3341.

\bibitem{Anderson:1991zb}
G.~W. Anderson and L.~J. Hall, {\it {The Electroweak phase transition and
  baryogenesis}},  {\em Phys. Rev.} {\bf D45} (1992) 2685--2698.

\bibitem{Jain:2017sqm}
B.~Jain, S.~J. Lee, and M.~Son, {\it {On the Validity of the Effective
  Potential and the Precision of Higgs Self Couplings}},
  \href{http://arxiv.org/abs/1709.03232}{{\tt arXiv:1709.03232}}.

\bibitem{Carrington:1991hz}
M.~E. Carrington, {\it {The Effective potential at finite temperature in the
  Standard Model}},  {\em Phys. Rev.} {\bf D45} (1992) 2933--2944.

\bibitem{Arnold:1992rz}
P.~B. Arnold and O.~Espinosa, {\it {The Effective potential and first order
  phase transitions: Beyond leading-order}},  {\em Phys. Rev.} {\bf D47} (1993)
  3546, [\href{http://arxiv.org/abs/hep-ph/9212235}{{\tt hep-ph/9212235}}].
  [Erratum: Phys. Rev.D50,6662(1994)].

\bibitem{Blinov:2015vma}
N.~Blinov, S.~Profumo, and T.~Stefaniak, {\it {The Electroweak Phase Transition
  in the Inert Doublet Model}},  {\em JCAP} {\bf 1507} (2015), no.~07 028,
  [\href{http://arxiv.org/abs/1504.05949}{{\tt arXiv:1504.05949}}].

\bibitem{Parwani:1991gq}
R.~R. Parwani, {\it {Resummation in a hot scalar field theory}},  {\em Phys.
  Rev.} {\bf D45} (1992) 4695, [\href{http://arxiv.org/abs/hep-ph/9204216}{{\tt
  hep-ph/9204216}}]. [Erratum: Phys. Rev.D48,5965(1993)].

\bibitem{Land:1992sm}
D.~Land and E.~D. Carlson, {\it {Two stage phase transition in two Higgs
  models}},  {\em Phys. Lett.} {\bf B292} (1992) 107--112,
  [\href{http://arxiv.org/abs/hep-ph/9208227}{{\tt hep-ph/9208227}}].

\bibitem{Hammerschmitt:1994fn}
A.~Hammerschmitt, J.~Kripfganz, and M.~G. Schmidt, {\it {Baryon asymmetry from
  a two stage electroweak phase transition?}},  {\em Z. Phys.} {\bf C64} (1994)
  105--110, [\href{http://arxiv.org/abs/hep-ph/9404272}{{\tt hep-ph/9404272}}].

\bibitem{Patel:2011th}
H.~H. Patel and M.~J. Ramsey-Musolf, {\it {Baryon Washout, Electroweak Phase
  Transition, and Perturbation Theory}},  {\em JHEP} {\bf 07} (2011) 029,
  [\href{http://arxiv.org/abs/1101.4665}{{\tt arXiv:1101.4665}}].

\bibitem{Wainwright:2011qy}
C.~Wainwright, S.~Profumo, and M.~J. Ramsey-Musolf, {\it {Gravity Waves from a
  Cosmological Phase Transition: Gauge Artifacts and Daisy Resummations}},
  {\em Phys. Rev.} {\bf D84} (2011) 023521,
  [\href{http://arxiv.org/abs/1104.5487}{{\tt arXiv:1104.5487}}].

\bibitem{Garny:2012cg}
M.~Garny and T.~Konstandin, {\it {On the gauge dependence of vacuum transitions
  at finite temperature}},  {\em JHEP} {\bf 07} (2012) 189,
  [\href{http://arxiv.org/abs/1205.3392}{{\tt arXiv:1205.3392}}].

\bibitem{Moore:1998swa}
G.~D. Moore, {\it {Measuring the broken phase sphaleron rate
  nonperturbatively}},  {\em Phys. Rev.} {\bf D59} (1999) 014503,
  [\href{http://arxiv.org/abs/hep-ph/9805264}{{\tt hep-ph/9805264}}].

\bibitem{Patel:2014}
``{Public talk at ACFI Higgs Portal Workshop}.''
  \url{https://www.physics.umass.edu/acfi/sites/acfi/files/slides/2014_acfi-higgs_0.pdf}.

\bibitem{Ginzburg:2009dp}
I.~F. Ginzburg, I.~P. Ivanov, and K.~A. Kanishev, {\it {The Evolution of vacuum
  states and phase transitions in 2HDM during cooling of Universe}},  {\em
  Phys. Rev.} {\bf D81} (2010) 085031,
  [\href{http://arxiv.org/abs/0911.2383}{{\tt arXiv:0911.2383}}].

\bibitem{Ginzburg:2004vp}
I.~F. Ginzburg and M.~Krawczyk, {\it {Symmetries of two Higgs doublet model and
  CP violation}},  {\em Phys. Rev.} {\bf D72} (2005) 115013,
  [\href{http://arxiv.org/abs/hep-ph/0408011}{{\tt hep-ph/0408011}}].

\bibitem{Jiang:2017}
Y.~Jiang, ``Higgs boson physics beyond the standard model (ph.d thesis), uc
  davis, 2015..''
  \url{http://particle.physics.ucdavis.edu/hefti/members/lib/exe/fetch.php?media=students:bsmhiggs_final.pdf}.

\bibitem{Dorsch:2017nza}
G.~C. Dorsch, S.~J. Huber, K.~Mimasu, and J.~M. No, {\it {The Higgs Vacuum
  Uplifted: Revisiting the Electroweak Phase Transition with a Second Higgs
  Doublet}},  \href{http://arxiv.org/abs/1705.09186}{{\tt arXiv:1705.09186}}.

\bibitem{Curtin:2016urg}
D.~Curtin, P.~Meade, and H.~Ramani, {\it {Thermal Resummation and Phase
  Transitions}},  \href{http://arxiv.org/abs/1612.00466}{{\tt
  arXiv:1612.00466}}.

\bibitem{Espinosa:2011ax}
J.~R. Espinosa, T.~Konstandin, and F.~Riva, {\it {Strong Electroweak Phase
  Transitions in the Standard Model with a Singlet}},  {\em Nucl. Phys.} {\bf
  B854} (2012) 592--630, [\href{http://arxiv.org/abs/1107.5441}{{\tt
  arXiv:1107.5441}}].

\bibitem{Espinosa:2011eu}
J.~R. Espinosa, B.~Gripaios, T.~Konstandin, and F.~Riva, {\it {Electroweak
  Baryogenesis in Non-minimal Composite Higgs Models}},  {\em JCAP} {\bf 1201}
  (2012) 012, [\href{http://arxiv.org/abs/1110.2876}{{\tt arXiv:1110.2876}}].

\bibitem{Cline:2017qpe}
J.~M. Cline, K.~Kainulainen, and D.~Tucker-Smith, {\it {Electroweak
  baryogenesis from a dark sector}},  {\em Phys. Rev.} {\bf D95} (2017), no.~11
  115006, [\href{http://arxiv.org/abs/1702.08909}{{\tt arXiv:1702.08909}}].

\bibitem{Coleppa:2014hxa}
B.~Coleppa, F.~Kling, and S.~Su, {\it {Exotic Decays Of A Heavy Neutral Higgs
  Through HZ/AZ Channel}},  {\em JHEP} {\bf 09} (2014) 161,
  [\href{http://arxiv.org/abs/1404.1922}{{\tt arXiv:1404.1922}}].

\bibitem{Kanemura:2002vm}
S.~Kanemura, S.~Kiyoura, Y.~Okada, E.~Senaha, and C.~P. Yuan, {\it {New physics
  effect on the Higgs selfcoupling}},  {\em Phys. Lett.} {\bf B558} (2003)
  157--164, [\href{http://arxiv.org/abs/hep-ph/0211308}{{\tt hep-ph/0211308}}].

\bibitem{Asner:2013psa}
D.~M. Asner et~al., {\it {ILC Higgs White Paper}},  in {\em {Proceedings, 2013
  Community Summer Study on the Future of U.S. Particle Physics}}, 2013.
\newblock \href{http://arxiv.org/abs/1310.0763}{{\tt arXiv:1310.0763}}.

\bibitem{CEPC-SPPCStudyGroup:2015csa}
C.-S.~S. Group, {\it {CEPC-SPPC Preliminary Conceptual Design Report. 1.
  Physics and Detector}}, .

\bibitem{Gomez-Ceballos:2013zzn}
{\bf TLEP Design Study Working Group} Collaboration, M.~Bicer et~al., {\it
  {First Look at the Physics Case of TLEP}},  {\em JHEP} {\bf 01} (2014) 164,
  [\href{http://arxiv.org/abs/1308.6176}{{\tt arXiv:1308.6176}}].

\bibitem{Arkani-Hamed:2015vfh}
N.~Arkani-Hamed, T.~Han, M.~Mangano, and L.-T. Wang, {\it {Physics
  opportunities of a 100 TeV proton–proton collider}},  {\em Phys. Rept.}
  {\bf 652} (2016) 1--49, [\href{http://arxiv.org/abs/1511.06495}{{\tt
  arXiv:1511.06495}}].

\bibitem{Huang:2015tdv}
P.~Huang, A.~Joglekar, B.~Li, and C.~E.~M. Wagner, {\it {Probing the
  Electroweak Phase Transition at the LHC}},  {\em Phys. Rev.} {\bf D93}
  (2016), no.~5 055049, [\href{http://arxiv.org/abs/1512.00068}{{\tt
  arXiv:1512.00068}}].

\bibitem{Profumo:2014opa}
S.~Profumo, M.~J. Ramsey-Musolf, C.~L. Wainwright, and P.~Winslow, {\it
  {Singlet-catalyzed electroweak phase transitions and precision Higgs boson
  studies}},  {\em Phys. Rev.} {\bf D91} (2015), no.~3 035018,
  [\href{http://arxiv.org/abs/1407.5342}{{\tt arXiv:1407.5342}}].

\bibitem{Huang:2017jws}
T.~Huang, J.~M. No, L.~Pernié, M.~Ramsey-Musolf, A.~Safonov, M.~Spannowsky,
  and P.~Winslow, {\it {Resonant di-Higgs boson production in the $b{\bar b}WW$
  channel: Probing the electroweak phase transition at the LHC}},  {\em Phys.
  Rev.} {\bf D96} (2017), no.~3 035007,
  [\href{http://arxiv.org/abs/1701.04442}{{\tt arXiv:1701.04442}}].

\bibitem{Kotwal:2016tex}
A.~V. Kotwal, M.~J. Ramsey-Musolf, J.~M. No, and P.~Winslow, {\it
  {Singlet-catalyzed electroweak phase transitions in the 100 TeV frontier}},
  {\em Phys. Rev.} {\bf D94} (2016), no.~3 035022,
  [\href{http://arxiv.org/abs/1605.06123}{{\tt arXiv:1605.06123}}].

\bibitem{Shao:2013bz}
D.~Y. Shao, C.~S. Li, H.~T. Li, and J.~Wang, {\it {Threshold resummation
  effects in Higgs boson pair production at the LHC}},  {\em JHEP} {\bf 07}
  (2013) 169, [\href{http://arxiv.org/abs/1301.1245}{{\tt arXiv:1301.1245}}].

\bibitem{Barger:2013jfa}
V.~Barger, L.~L. Everett, C.~B. Jackson, and G.~Shaughnessy, {\it {Higgs-Pair
  Production and Measurement of the Triscalar Coupling at LHC(8,14)}},  {\em
  Phys. Lett.} {\bf B728} (2014) 433--436,
  [\href{http://arxiv.org/abs/1311.2931}{{\tt arXiv:1311.2931}}].

\bibitem{Frederix:2014hta}
R.~Frederix, S.~Frixione, V.~Hirschi, F.~Maltoni, O.~Mattelaer, P.~Torrielli,
  E.~Vryonidou, and M.~Zaro, {\it {Higgs pair production at the LHC with NLO
  and parton-shower effects}},  {\em Phys. Lett.} {\bf B732} (2014) 142--149,
  [\href{http://arxiv.org/abs/1401.7340}{{\tt arXiv:1401.7340}}].

\bibitem{Barger:2014taa}
V.~Barger, L.~L. Everett, C.~B. Jackson, A.~D. Peterson, and G.~Shaughnessy,
  {\it {New physics in resonant production of Higgs boson pairs}},  {\em Phys.
  Rev. Lett.} {\bf 114} (2015), no.~1 011801,
  [\href{http://arxiv.org/abs/1408.0003}{{\tt arXiv:1408.0003}}].

\end{thebibliography}\endgroup

\end{document}